\documentclass[review,5p]{elsarticle}

\journal{Physica D}
\usepackage{amsfonts}
\usepackage{amsmath}
\usepackage{amssymb}
\usepackage{graphicx}
\usepackage{dcolumn}
\usepackage{amscd}
\usepackage{graphicx}

\usepackage{color}
\usepackage{subfigure}

\numberwithin{equation}{section}

\newcommand{\comment}[1]{}
\usepackage{xcolor}

 \newcommand{\ihat}{\hat{\textbf{\i}}}
\newcommand{\jhat}{\hat{\textbf{\j}}}
 
\begin{document}

\begin{frontmatter}

\title{Nonlinear Optical Waveguide Lattices: Asymptotic Analysis, Solitons, and Topological Insulators} 

\author{Mark J. Ablowitz}
\ead{mark.ablowitz@colorado.edu}
\address{Department of Applied Mathematics, University of Colorado, Campus Box 526, Boulder, Colorado, USA}

\author{Justin T. Cole}
\ead{jcole13@uccs.edu}
\address{Department of Mathematics, University of Colorado, Colorado Springs, Colorado, USA}

\begin{abstract}

In recent years, there has been considerable interest in the study of wave
propagation in nonlinear {photonic} 
lattices.  The interplay between nonlinearity
and periodicity has led researchers to manipulate light and discover  new and
interesting  phenomena such as new classes of localized modes, usually referred
to as solitons {and novel {surface states} 
that propagate robustly.} 
A field where both nonlinearity and periodicity arises
naturally is nonlinear optics. But there are other areas where waves
propagating on background lattices play an important role{,} including photonic
crystal fibers and Bose--Einstein condensation.  In this {review article} 
the propagation of wave envelopes in {one and} two-dimensional periodic lattices associated with additional potential
in {the nonlinear Schr\"odinger (NLS) equation,} 
termed lattice NLS equations, are studied. A discrete {reduction,} 
known as the tight-binding approximation, is employed in order to find the linear dispersion relation and the equations governing nonlinear
discrete envelopes for two-dimensional simple periodic lattices and two-dimensional non-simple honeycomb lattices. In the limit under which the
envelopes vary slowly{,} 
continuous envelope equations are derived from the
discrete system. The coefficients of the linear evolution system are related to
the dispersion relation in both the discrete and 
continuous cases.  For
simple lattices, the continuous systems are  {NLS} 
type equations. In honeycomb lattices, in certain cases, the continuous system {is} 
found to be nonlinear Dirac equations. {Finally, it is possible to realize 
{so-called topological insulator {systems} in} an optical waveguide setting. The modes {supported by} 
these systems are associated with spectral topological invariants and{,} 
 {remarkably{,} can propagate without backscatter from lattice defects.}}


\end{abstract}

\end{frontmatter}




\section{Introduction}
\label{intro_sec}

In nonlinear optics, periodic structures that have been carefully  studied
are arrays of coupled nonlinear optical waveguides. {These waveguides typically consist of media with higher refractive indices that {tend} 
to confine and steer light beams.} The first theoretical
prediction of discrete solitons in an optical waveguide array was reported by
Christodoulides and Joseph \cite{ChJo88}. Many properties of such
discrete solitons were subsequently  studied cf.
\cite{Kevrekidis2001,AbMu03}. However, after the theoretical
prediction of \cite{ChJo88}{,} it was almost a decade until self-trapping
of light in 
nonlinear waveguide array was experimentally observed
\cite{Eisenberg1998}.  

\begin{figure}
\centering
\centerline{\scalebox{0.25}{\includegraphics{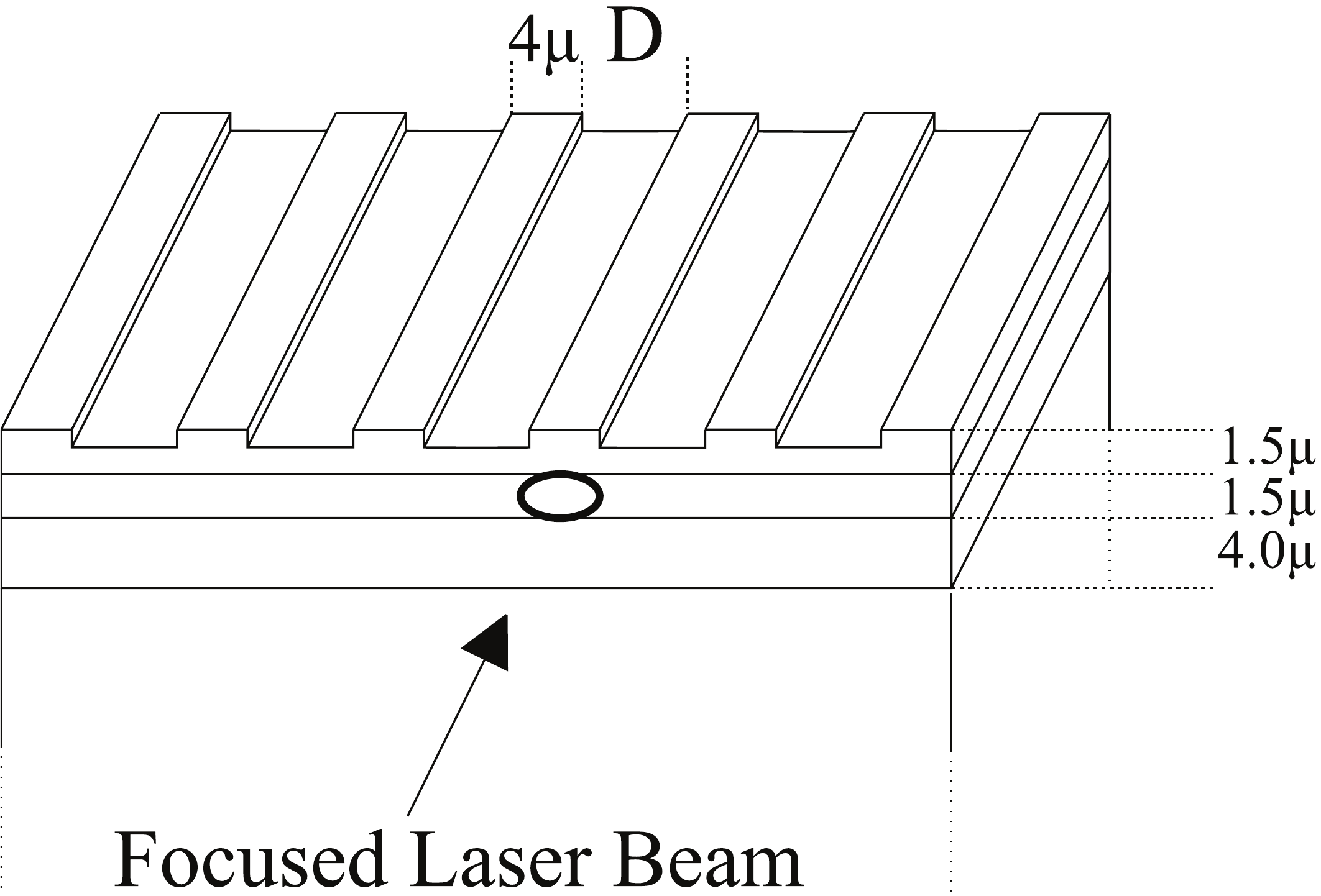}}}
\caption{Schematic illustrating the coupled waveguide array used in \cite{Eisenberg1998}. Reprinted figure with permission from \cite{Eisenberg1998}, copyright (1998) by the American Physical Society.}
\label{WGuide}
\end{figure}

\begin{figure}
\centering
\includegraphics[scale=.25]{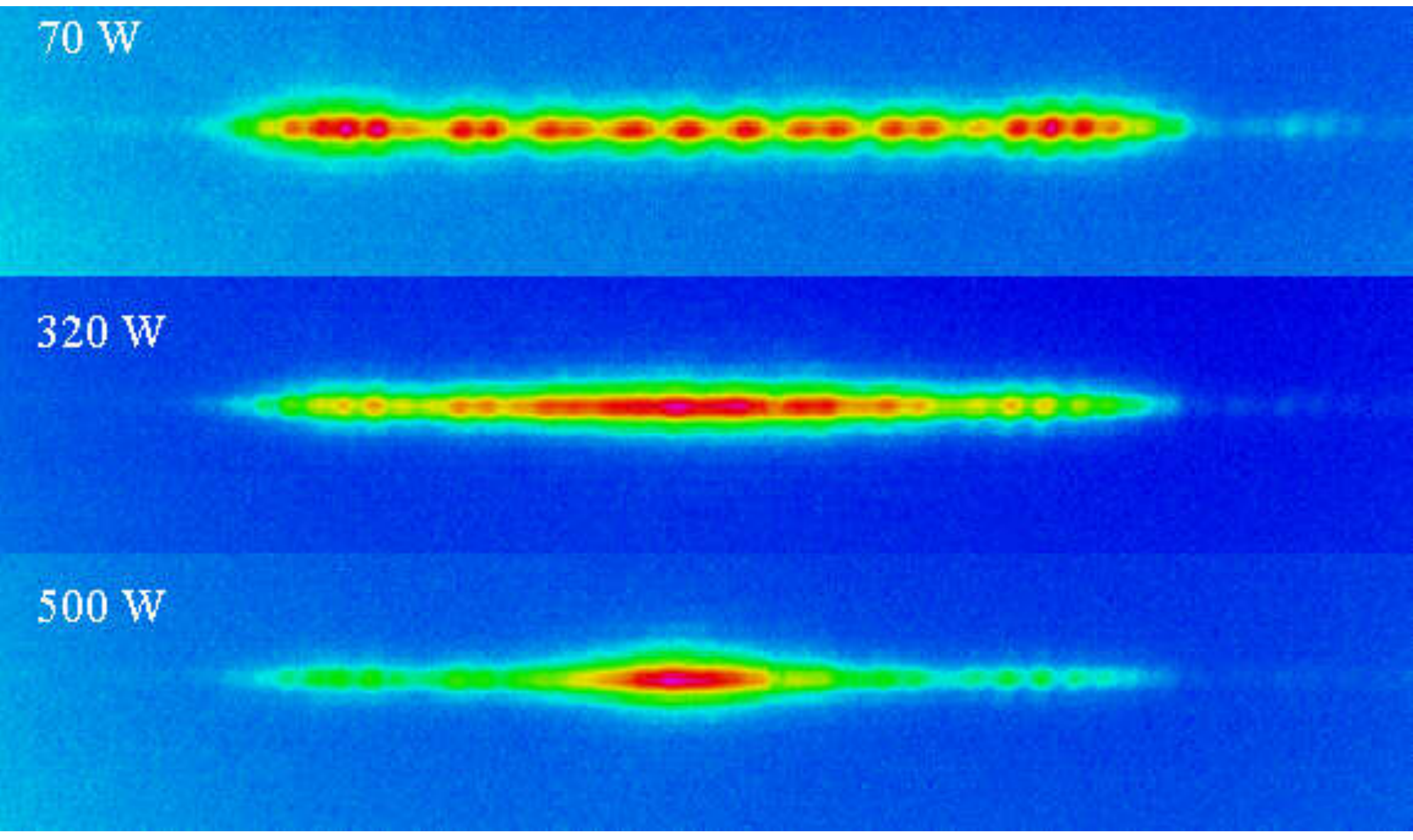}
\caption{Output field measured at the output facet of the waveguide \cite{Eisenberg1998}.
Input power: (top) low, (middle) medium, and (bottom) high. A solitary wave forms at high power. Reprinted figure with permission from \cite{Eisenberg1998}, copyright (1998) by the American Physical Society.}
\label{wguidebeam}
\end{figure}

{Early on{,} it was difficult }
to fabricate specialized materials with fixed geometry {at such} 
small scales. This has been largely overcome by {optical and etching techniques.} 
A schematic 
{illustrating} the coupled waveguide configuration {used in \cite{Eisenberg1998}} is
given in Fig.~\ref {WGuide}{. The array consists of} 
approximately 40 waveguide ``ridges" 
{that are $4~ \mu$m wide and $0.95~ \mu$m deep;} 
the longitudinal propagation length of the waveguide {is} 
$6 $ mm. 
An input laser beam {is} 
{injected} at the central location of the waveguides. {The results of the experiment are shown in Fig.~\ref{wguidebeam}.} At low power the beam {diffracts;} 
at moderate power the beam {begins} 
to self{-}focus. {Finally, a}t high power
the beam strongly {self-focuses} 
and a {highly localized} soliton beam {is} 
observed.

A few years later{,} 
a new method of creating optical 
periodic lattices in photosensitive materials using optical induction {was proposed} 
{\cite{Efremidis2002}.} 
Soon afterwards, 
{using this `all optical' technique} {two-dimensional (2D)} 
periodic lattices were
created and 2D solitons were observed and studied
{\cite{Fleischer2003PRL,Fleischer2003Nature}}. {These solitons are sometimes termed `gap' solitons because they are found in the frequency gaps of the underlying periodic wave spectrum.}
This area has attracted considerable
interest from {engineers, physicists, and mathematicians.} 
{Subsequently, m}any novel types of localized modes{,} {e.g. solitons}{,} have been predicted theoretically and demonstrated
experimentally. Examples include dipole solitons \cite{Yang04OL}, vortex
solitons \cite{Neshev04PRL}, soliton trains \cite{Chen07}{,} etc.

\begin{figure}
{\scalebox{0.2}{\includegraphics{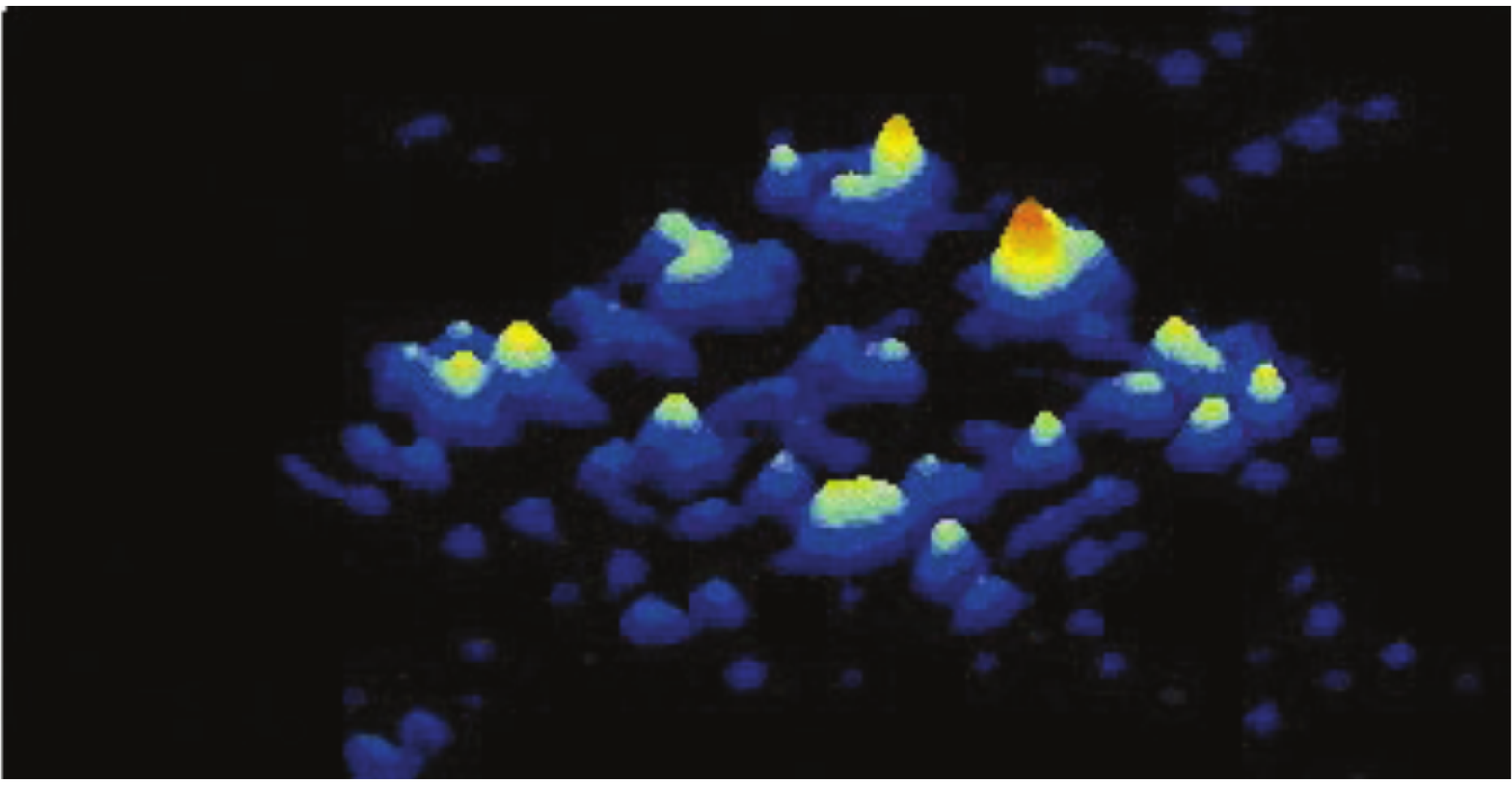}}}
{\scalebox{0.2}{\includegraphics{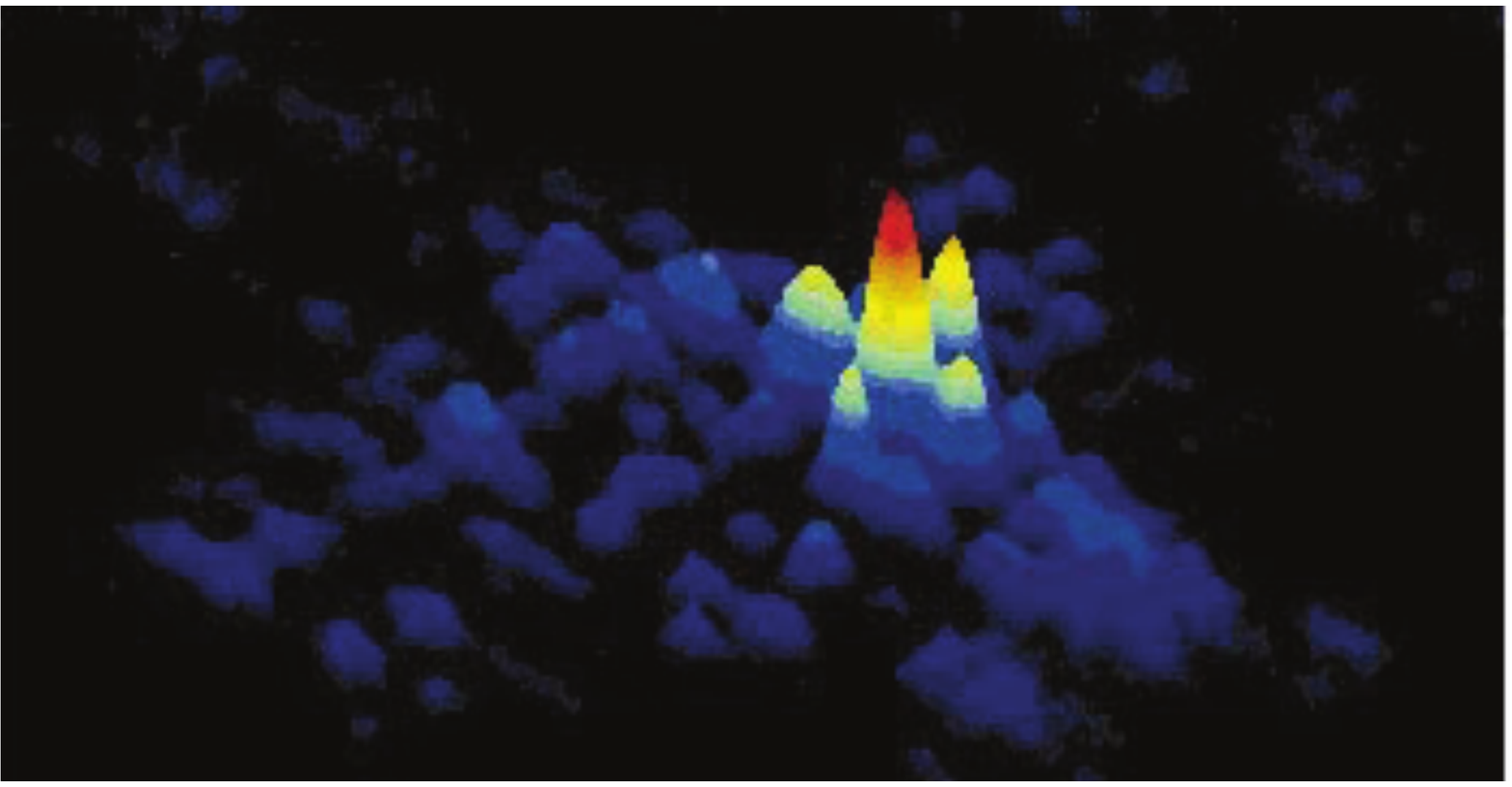}}}
\centerline{\scalebox{1.5}{\includegraphics{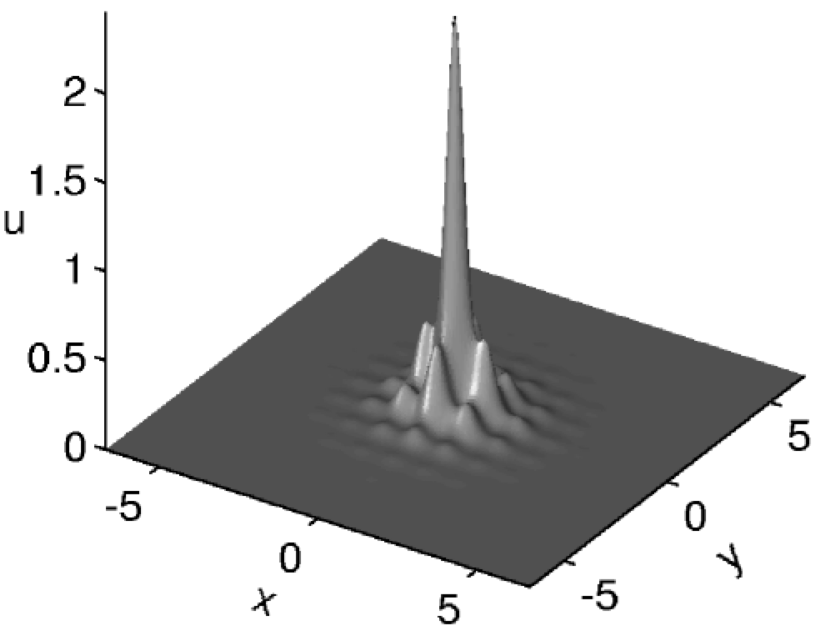}}}
\caption{(Top) Output intensity measurements obtained in a nonlinear waveguide array. Left, low input power; Right, high input power. Reprinted by permission from Springer Nature: Nature \cite{Fleischer2003Nature}, copyright (2003). (Bottom) Numerically obtained lattice soliton at high power. Reprinted figure with permission from \cite{Efremidis}., copyright (2003) by the American Physical Society.}
\label{exptlattice}
\end{figure}

The experimental results {of \cite{Fleischer2003Nature}} are depicted in Fig.~\ref{exptlattice}. Here, as with
the one dimensional configuration, at low input power the beam diffracts and at
high input power the beam self-focuses and a localized structure is seen to
emerge; i.e., a soliton is formed.

\begin{figure}
\centerline{\scalebox{0.35}{\includegraphics{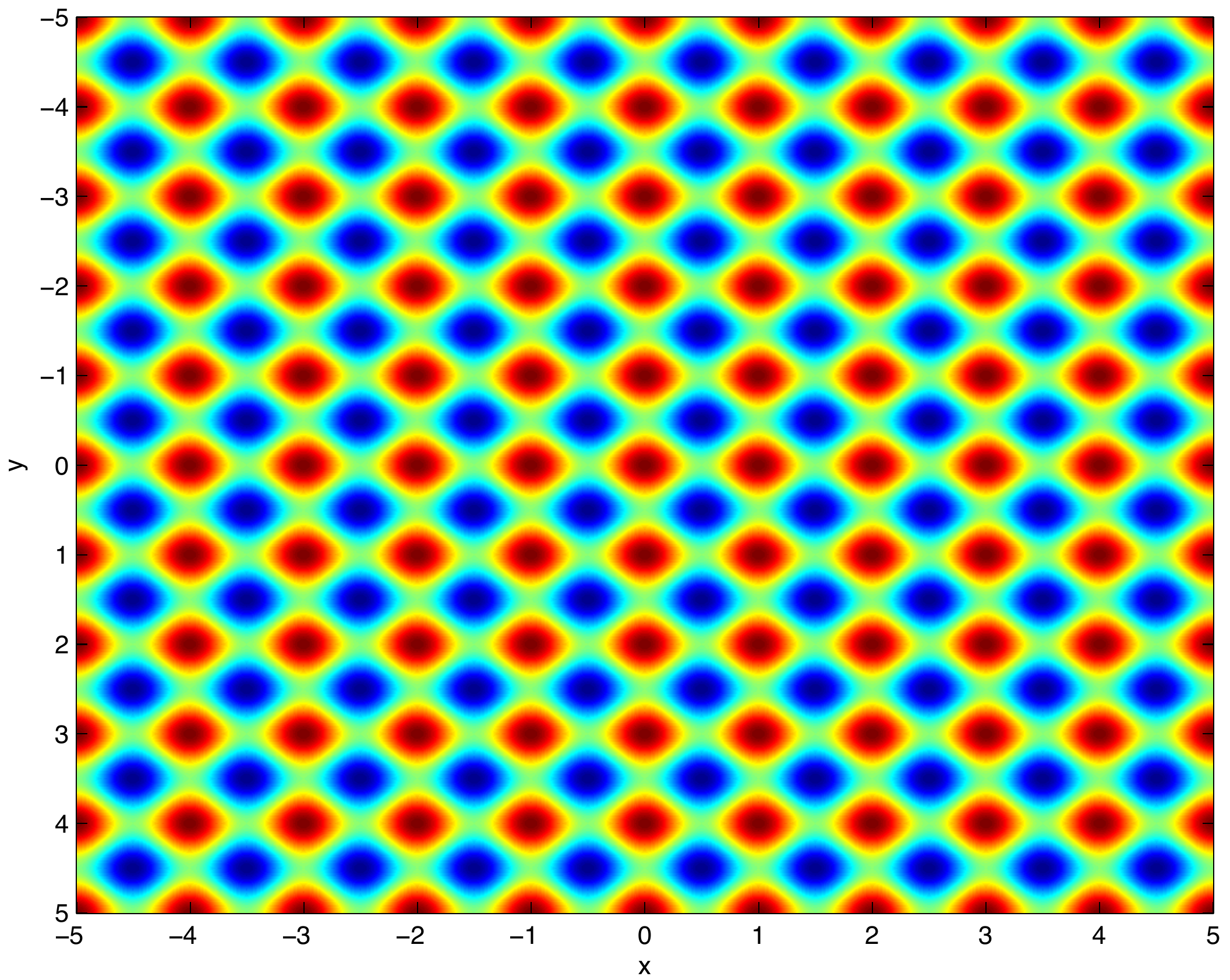}}}
\caption{$V(x,y)= V_0(\cos^2\pi x+\cos^2\pi y); V_0=1$} 
\label{2drectlattice}
\end{figure}

Researchers usually model the above phenomena by  a general 2D lattice
nonlinear Schr\"odinger (NLS) equation, written in dimensionless form:
\[
i\psi_z + \nabla^2 \psi - V(\mathbf{r})\psi + f(|\psi|^2) \psi= 0,
\]
where {$\nabla^2 = \partial_x^2 + \partial_y^2$, $\mathbf{r}=(x,y)$ is the transverse spatial dimensions,  and $z$ (the direction of propagation) behaves like a temporal variable. The effective refractive index is split into two parts: a linear periodic potential $V(\mathbf{r})$, 
and an intensity-dependent nonlinear term $f(|\psi|^2)$. The  most commonly studied type of nonlinearity is cubic, i.e.
$ f(|\psi|^2) = \sigma |\psi|^2 $ where $\sigma$ is constant.} 
This model 
describes light propagation
in a periodic Kerr nonlinear medium as well as in Bose--Einstein condensates
trapped in a 2D optical lattice {\cite{Pethick2008}}. 
{A representation of a typical cross-section of two-dimensional {rectangular} lattice  potential, $V(x,y),$
is given} in Fig. \ref {2drectlattice}. Here the maxima and minima
play the role of the `ridges and valleys' in the waveguide {(high and low refractive index)}. 

{We note that} the {photonic} material employed in 
experiments
\cite{Fleischer2003PRL,Fleischer2003Nature} {used} 
so-called photorefractive media, 
not Kerr
media{; {in this case,} 
the nonlinearity/potential} 
is usually modeled by saturable nonlinear media 
\[- V(\mathbf{r})+ f(|\psi|^2) \rightarrow -  (1 - W({\bf r}) + \sigma  |\psi|^2)^{-1}.\]
Despite their different forms,  saturable nonlinearity reduces to a cubic/Kerr nonlinearity  in the small refractive index limit, 
{$|W({\bf r})+ \sigma |\psi|^2| \ll 1$, where 
\[-  (1 - W({\bf r}) + \sigma  |\psi|^2)^{-1}\approx  ( -1 -  W({\bf r}) +\sigma   |\psi|^2) \]
and $V(\mathbf{r}) = -1 -  W({\bf r})$.
In  {\cite{Efremidis,Efremidis2002}}{,} this type of saturable lattice was used numerically  and shown to yield solitons at high  {input power.}


After these results  in 
{2D} periodic lattices were reported,
many novel localized structures were predicted theoretically and demonstrated
experimentally. Examples include dipole solitons, vortex solitons, soliton
trains, cf.
\cite{Yang04OL,Neshev04PRL,Chen07,Kivshar06_PRL,Kivshar06_OL}.
Similarly{,} in condensed matter physics{,} ultracold atoms,  Bose--Einstein
condensates {(BECs)} can be trapped in a periodic optical lattice {which is described by a lattice NL equation, also known as the Gross–Pitaevskii equation}. The
experimental observation of gap solitons {in BECs} was 
reported {in} \cite{Morsch2006}{,} 
{and vortices theoretically predicted in \cite{Kivshar04_PRL}}. With observations and theory in different fields,
the study of related phenomena such as localized modes  and their properties
has gained significant scientific interest.

Background lattice periodicity alone leads to interesting mathematical
investigations. An important feature follows from what it is often termed Bloch
theory {\cite{Odeh_64}.}  Namely{, the} associated {spectrum} 
has multi-band structure. Bands are regions
that support bounded, quasi-periodic, eigenmodes. Between two {adjacent} 
bands, there can {exist} 
a gap where
bounded linear eigenmodes do not exist. Analogous to Fourier modes, Bloch modes
can propagate in a periodic linear waveguide; here different Bloch modes admit
different dynamics that, in turn, do not influence each other because of the
superposition principle. 

Nonlinearity can change the eigenmodes associated with
band structures. The allowed regions where modes can propagate can be extended
by 
nonlinearity {into the band gaps.} So{,} in the 
gap {region} where linear bounded modes {do not propagate, i.e. are {forbidden},} 
there can exist nonlinear bounded 
eigenmodes{. Localized nonlinear gap modes are known as {\it band gap solitons}.} 
The dynamics {can} 
become more interesting with
nonlinearities; {for example,} 
in BECs, nonlinear Bloch oscillations, nonlinear
Landau--Zenner tunneling etc. have been reported; cf. \cite{Morsch2006}. In
optics, conical diffraction that was thought to be a linear phenomena is also
exhibited in nonlinear honeycomb lattices \cite{Ablowitz2009a}. In addition,
since the superposition principle does not hold when nonlinearity is {present}, 
different Bloch modes may interfere each other. Energy can spread among
these linear Bloch modes and new Bloch modes may be produced due to
interference--as seen in  supercontinuum generation \cite{Manela06,Dong08}.
{Asymptotic descriptions can be obtained via multiple-scales approach, as in \cite{Dohnal2009a,Dohnal2009,Ilan2010} and  \cite{Pelinovsky2011}.}

The geometric distribution of local minima of the potentials, also called
{\it{sites}}, can be used to classify the potentials. These sites are the positions of
the potential wells. In optics, they have increased  refractive index and the
electromagnetic field is attracted to these {regions.} 
The distribution of these
sites greatly influence the properties of the associated dynamics/waves.
Discrete one dimensional evolution equations on 1D lattices were studied by
the so-called Wannier function approach in (cf.
\cite{Kevrekidis2001,Kevrekidis2002w}). However, there are significant
differences {that occur among} 
2D periodic lattices. 

{First,} 
we will divide 2D periodic lattices into two groups: simple and
non-simple {stationary ($z$-independent)} lattices. Simple lattices only have one site in a basic unit cell
while non-simple lattices 
have more than one site per cell. Examples of
simple lattices are rectangular and triangular lattices. A {well-known} 
non-simple
lattice is the honeycomb hexagonal lattice that has two sites in a unit cell
and breaks up into two triangular sublattices. 
Due to the underlying symmetries in the
honeycomb lattice,  we will see below  that the dispersion relation of the
associated Bloch theory may have isolated degenerate points where two
dispersion surfaces touch each other. These 
are called Dirac points and
near these points the dispersion surface has {a} conical structure. {It was {rigorously} proven in \cite{FW2012} that dispersion surfaces touch each other at Dirac points.}

The evolution
of {a} 
Bloch mode envelope in the neighborhood of these points is governed by
nonlinear Dirac systems \cite{Ablowitz2009a}. There are interesting phenomena
associated with the Dirac system. An example in optics is  conical
diffraction-- where a narrow beam transforms into bright expanding rings, see
\cite{Segev07prl,Berry_Jeffrey,Ablowitz2009a}. Honeycomb lattices also admit
various types of band gap solitons that like other 2D periodic lattices is
due to the effect of nonlinearity; cf \cite{Kevrekidis2002}. Another important
application is the material graphene that has a honeycomb lattice structure,
see \cite{Geim_2007}. In BECs, honeycomb background lattices may also lead to
interesting phenomena, see \cite{Haddad}.

{From the field of beam propagation in waveguide arrays, naturally came a way to realize a type of system {(or media)} known as {a} {\it topological insulator} (TI) in a  photonic setting. Topological insulators have their origins in  condensed matter physics, and in particular the quantum Hall effect \cite{vonKlitzing1980,Thouless1982,Haldane1988}.  The first discussion of a TI in an electromagnetic system 
can be traced back to the seminal work of Haldane and Raghu \cite{Haldane2008}. The {experimental} realization of a TI in {an electromagnetic system with {anisotropic} permeability; i.e. in}
a magneto-optic system occurred soon afterward \cite{Wang2009}. The {realization} of a TI in a photonic system came a few years later \cite{Rechtsman2013} and notably did not require an external magnetic field. Instead, researchers induced an {\it effective} magnetic field by fabricating waveguides that helically-varied in the direction of beam propagation. {These studies assumed the wave propagation was linear.}}
{Interest in the field of topological insulators in electromagnetic systems has expanded considerably {since this earlier research}-- see {e.g.} \cite{Lu2014,Ozawa2019}.}

\begin{figure}
{\scalebox{0.9}{\includegraphics{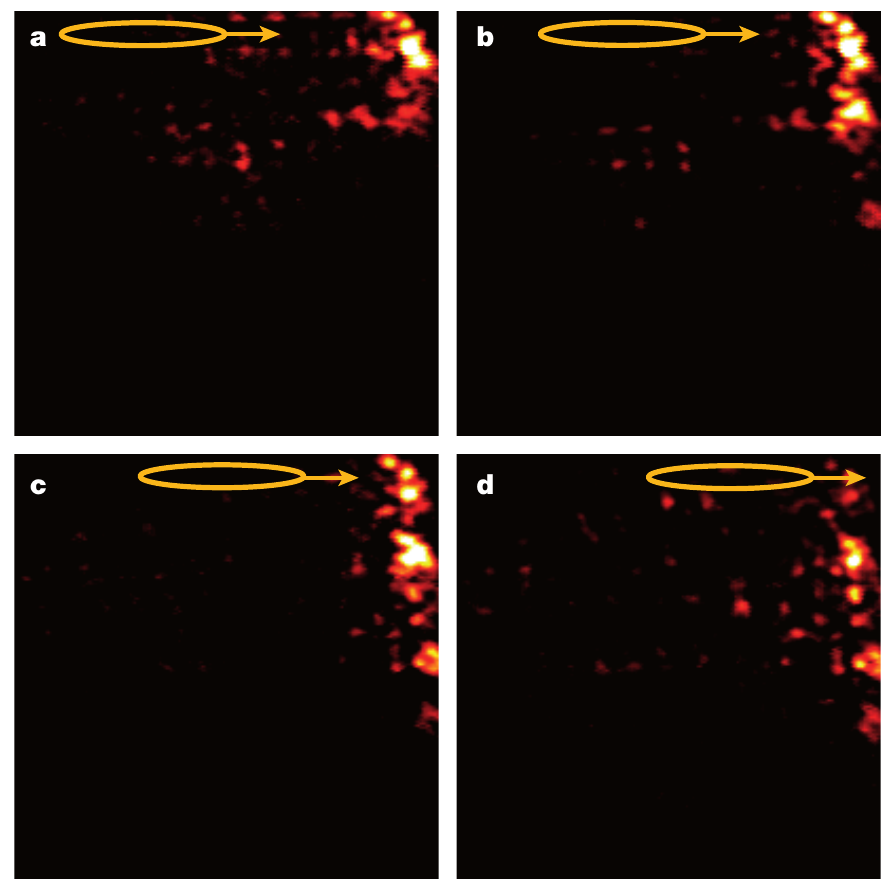}}}
\caption{Experimental results from a Floquet topological insulator in a helically-driven waveguide array \cite{Rechtsman2013}. (a-d) Yellow ellipses shows input beam location, output beam is shown in heat map. Collectively, the snapshots show a unidirectional edge mode which does not backscatter at corners. Reprinted by permission from Springer Nature: Nature \cite{Rechtsman2013}, copyright (2013).}
\label{exptlatticeA}
\end{figure}

{In a topological insulator system, linear wave propagation is possible at frequencies lying in  band gaps of the spectrum, 
{typically when they propagate} 
along the boundary or surface of the lattice media; these are called edge modes. Wave propagation in the interior of the media (well away from any boundaries) is still prohibited at these frequencies; these are known as bulk modes. To induce this behavior, one type of system referred to as {\it Floquet systems}, are generated by driving the lattice potential and creating equations with coefficients that are periodic in $z$. {The} associated linear eigenmodes possess so-called {\it topological invariants} which through the a principle known as the bulk-edge correspondence, indicate the presence of {\it topologically protected} edge states which propagate unidirectionally. These modes are localized along the domain boundary and are exceptionally robust to defects in the lattice system; they  do not {suffer from} backscatter and only move forward {(See Fig. \ref{exptlatticeA}).}
Furthermore,  weak nonlinearity induces {\it edge solitons} {see \cite{abccyp2014,abjc2017,abjc2019}. These solitons} inherit the topological properties of their linear counterpart, yet also manage to  balance dispersion and nonlinearity, like a typical soliton. Several linear and nonlinear results are discussed in this review. {Another notable} TI system { is} 
 the Su-Schrieffer-Heeger (SSH) model{. It} occurs in {non-driven} waveguide lattices where the coupling strength among adjacent sites alternates. 
 }




{Before outlining the content of this article, we note {this review does not include} {a thorough discussion of}
parity-time (PT) symmetric systems. Since their theoretical proposal in the photonic systems in 2008 \cite{Makris2008,Musslimani2008}, this class of systems {has} 
{been heavily studied. 
Indeed,  stable} PT-symmetric modes were experimentally realized in a waveguide array  \cite{Ruter2010,Guo2009}. We omit details {of this subfield, which is extensive, in order} to focus on the basics of optical waveguides {and}  topological insulators systems.}

{Another important realization of waveguide arrays that we do not consider in detail is that of planar lattices governed by  the linearly polarized 2D Maxwell's equations. 
For non-magnetized systems, the governing PDE for time-harmonic solutions is the variable-coefficient Helmholtz equation \cite{LeeThorp2019}. The {variable} coefficient is {due to} the permittivity function that models the dielectric of the waveguides. These systems can exhibit similar properties to those found in Schr\"odinger operators {(mentioned below) which} possess Dirac points in honeycomb lattices \cite{Cassier2021,Ammari2020} and localized edge states \cite{Hu20}. Moreover, these systems  also extend to topological insulators. Indeed, the seminal works of \cite{Haldane2008} and \cite{Wang2008,Wang2009} {showed the existence of} in-plane TE and TM topologically protected modes, respectively.}

{Now we outline the topics covered in this review. {The} general methodology of the tight-binding approximation {is discussed} in Sec.~\ref{fundamental}.  
{In Sec.~\ref{simple_lattice} the equations governing the tight-binding equations and envelope dynamics of a class of simple lattices are derived.}
As an example, a simple square lattice is considered in Sec.~\ref{square_sec}. The two-dimensional harmonic oscillator and its relationship to the orbital approximation are shown in Sec.~\ref{harmonic_oscillate}.} Next, tight-binding models for non-simple lattices is described in Sec.~\ref{honeycomb_sec}. 

{From here, the realization of topological insulators in optical waveguides is {explored in Sec.~\ref{TI_sec}}. The well-known} 
{one dimensional} 
{SSH model is relatively simple to realize in an optical waveguide setting; it is described in {Sec.~\ref{ssh_model_sec}}. In {Sec.~\ref{long_drive_lattice}} a {class of} longitudinally driven{, Floquet-type}} {2D} 
{{lattices are} shown to support unidirectional edge mode propagation with associated Chern invariants. {We conclude in Sec.~\ref{conclude_sec}.}}

\section{Fundamentals}
\label{fundamental}

The analysis here follows closely that in {\cite{Ablowitz2009a,AZ10,AZ13}}. We will
consider the 2D  lattice nonlinear Schr\"odinger (NLS) equation with cubic
nonlinearity, written in dimensionless form:
\begin{equation}
\label{LNLS}
i\psi_z + \nabla^2 \psi - V(\mathbf{r})\psi + \sigma|\psi|^2 \psi= 0,
\end{equation}
where $\mathbf{r}=(x,y)$, {$z$ is a temporal variable}, $V(\mathbf{r})$ is the periodic potential and
$\sigma$ is a constant that is positive for focusing nonlinearity and negative
for defocusing nonlinearity. This model can be used to describe {paraxial} light propagation in a
periodic Kerr nonlinear medium {\cite{Boyd2008,Yang2010}} 
and 
Bose--Einstein condensates trapped in a
2D optical lattice. 
{\cite{Pethick2008}.}

The potential $V(\mathbf{r})$ is a 2D periodic{,} bounded{, and} real-valued function
with two primitive lattice vectors{,} 
$\mathbf{v}_1$ and $\mathbf{v}_2$. {The potential has the translational symmetry $V({\bf r } + m\mathbf{v}_1 + n \mathbf{v}_2) = V({\bf r })$, for any $m,n \in \mathbb{Z}$.} We
denote $\mathbb{P}=\{m\mathbf{v}_1+n\mathbf{v}_2: m,n \in \mathbb{Z}\}$ as
the set of lattice vectors and $\mathbf{k}_1$ and $\mathbf{k}_2$ as the
primitive reciprocal lattice vectors and
$\mathbb{G}=\{m\mathbf{k}_1+n\mathbf{k}_2: m,n \in \mathbb{Z}\}$ as the set of
reciprocal lattice vectors. The unit cell of the physical lattice{, denoted by $\Omega$,} is the
parallelogram with $\mathbf{v}_1$ and $\mathbf{v}_2$ as its two sides and the
unit cell of the reciprocal lattice, {$\Omega'$},
 is the
parallelogram determined by $\mathbf{k}_1$ and $\mathbf{k}_2$. The relationship
between lattice and reciprocal lattice is
$\mathbf{v}_m\cdot\mathbf{k}_n=2\pi\delta_{mn}$. 

\begin{figure}
\centerline{\includegraphics[width=0.25\textwidth]{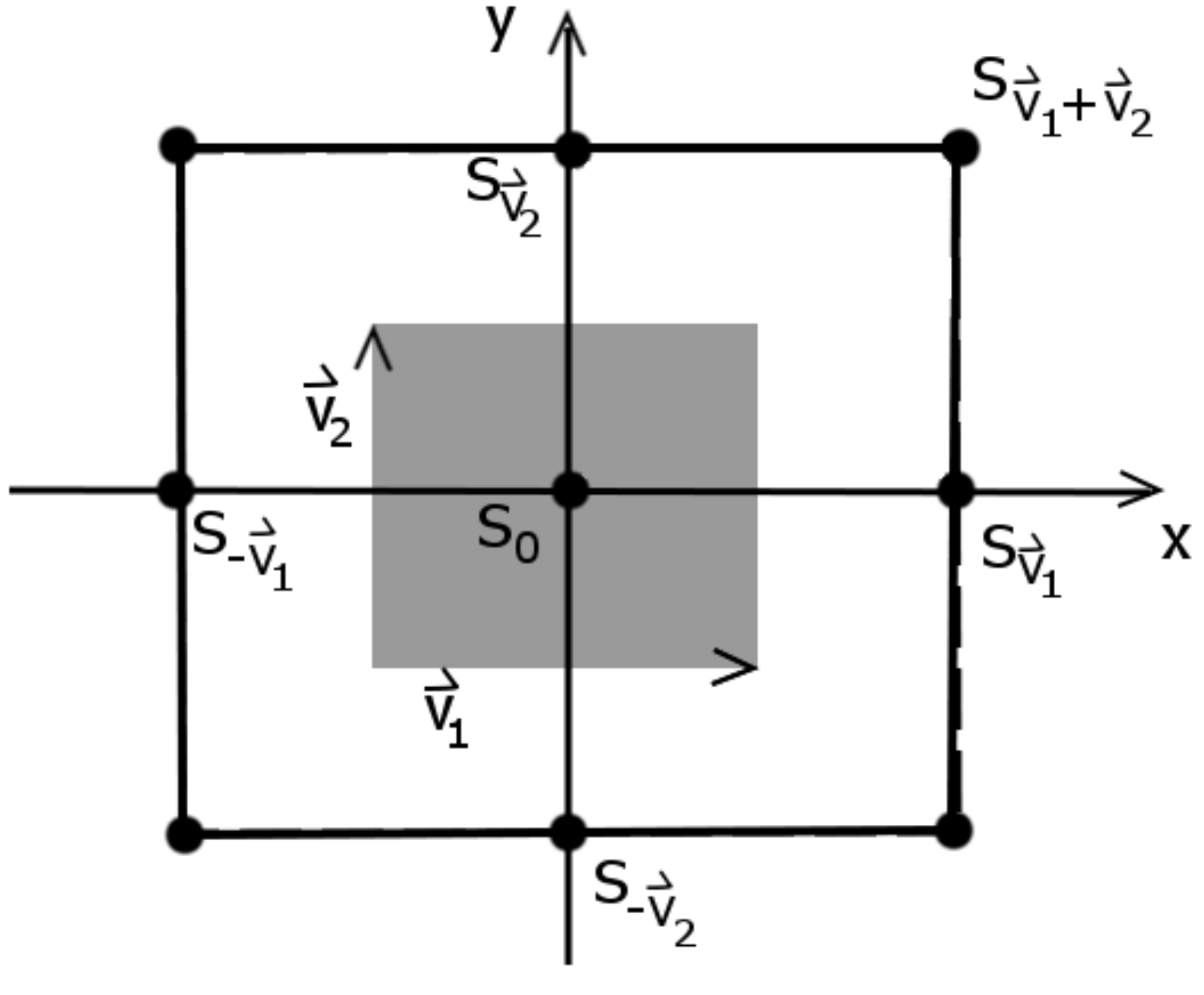}
  \includegraphics[width=0.25\textwidth]{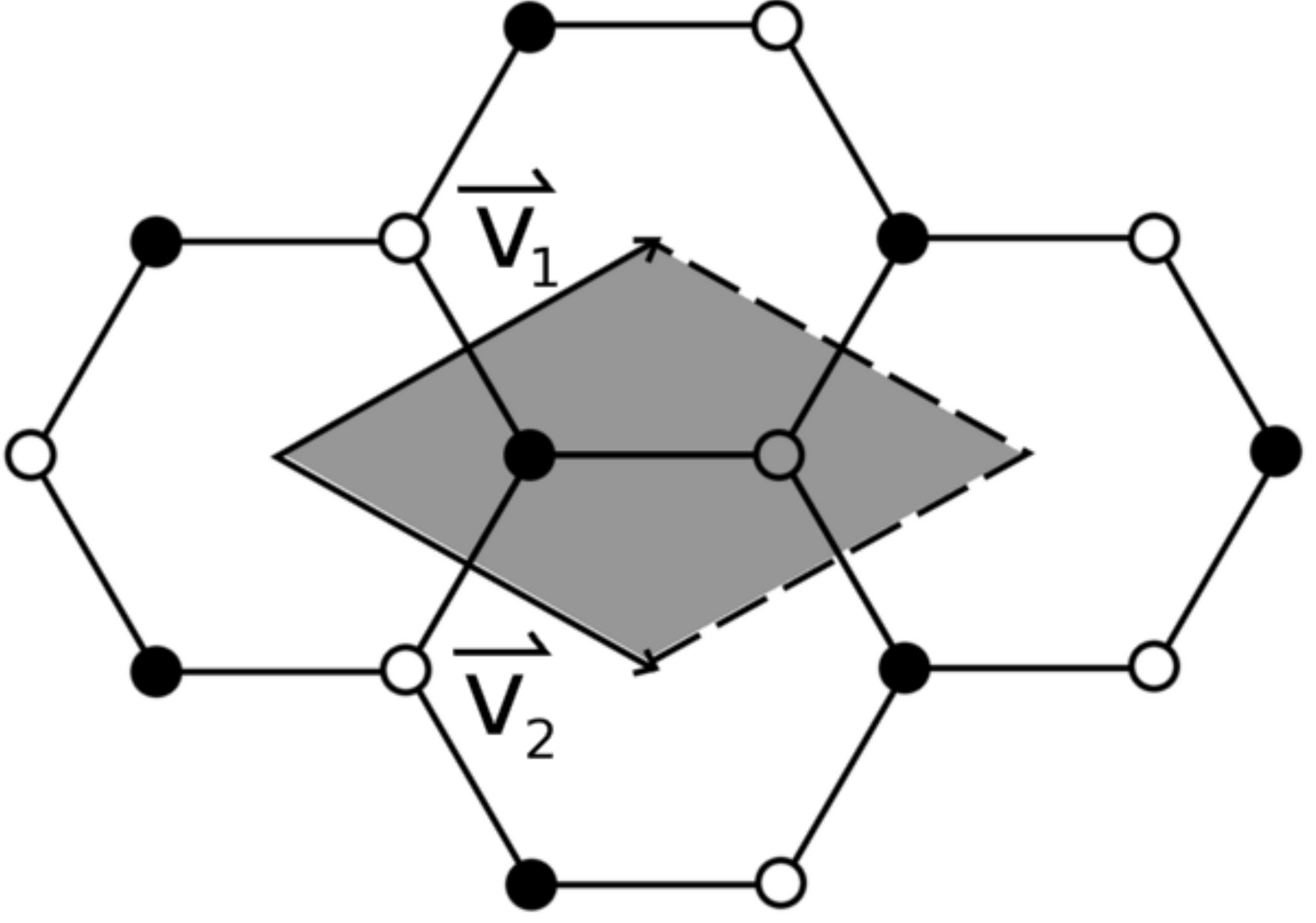}}
\caption{(left) A typical simple lattice. (right) A non-simple honeycomb lattice. {In each case the lattice sites (corresponding to {minima} 
of potential $V({\bf r})$) are denoted by dots. 
{The shaded} {region is a fundamental cell.}
}}
\label{lattices}
\end{figure}

We {first} consider  a simple periodic lattice {that has} 
one local {minimum site per unit} 
cell.   With a starting point and the lattice vectors, all the positions of the
sites can be constructed. All sites form a discrete lattice in the $\mathbf{r}$
plane{, namely $\mathbb{P}$}. We use $S_\mathbf{v}$ to denote the position of the site with index
$\mathbf{v}$ and $S_\mathbf{v}=S_\mathbf{0}+\mathbf{v}$ where $S_\mathbf{0}$ is
the starting point of the site lattice, i.e, $S_\mathbf{0}\in \Omega$.  Due to
translational symmetry, one unit cell has all the information of periodic
functions. For simplicity, we  place $S_\mathbf{0}=\mathbf{0}$ and choose the
parallelogram determined by $\mathbf{v}_1$ and $\mathbf{v}_2$ whose center is
$S_\mathbf{0}$ as the primitive unit cell $\Omega$. We also choose the
parallelogram determined by $\mathbf{k}_1$ and $\mathbf{k}_2$ whose center is
$\mathbf{k}=\mathbf{0}$ as the primitive reciprocal unit cell $\Omega'$. On the
other hand{,} a non-simple lattice may have more than one site in one unit cell.
One may need more than one starting point to construct the 
lattice. An
example of a non-simple lattice is a honeycomb lattice. These two situations
are {illustrated} 
in  Fig.~\ref{lattices}{.} 
{For the square lattice,} 
all sites are `black' and they can be constructed by {integer translations of} the two
primitive vectors. {On the other hand, the honeycomb lattice} 
consists of `black' and `white' sites. The black and white sites are
separately constructed from the underlying primitive vectors.

Let us first consider solutions of  Eq.~\eqref{LNLS} when the nonlinear coefficient $\sigma$ is {negligibly} 
small, or
equivalently 
$|\psi|^2 \ll1$, {so that}
\begin{equation}\label{linear}
i\psi_z+\nabla^2 \psi-V(\mathbf{r})\psi=0.
\end{equation}
Special {separable} 
solutions, which form a complete set, take the form
$\psi(\mathbf{r},z)=\varphi(\mathbf{r})e^{-i \mu z}$ and then Eq.~\eqref{linear} transforms to the following eigenvalue problem
\begin{equation}\label{Leigenproblem}
\nabla^2\varphi-V(\mathbf{r})\varphi= - \mu\varphi.
\end{equation}
According to Bloch theory (cf. \cite{Odeh_64}), the eigenfunction, also called {a}
Bloch mode or Bloch wave, has the $\mathbf{k}$-dependent form
\[
\varphi(\mathbf{r};\mathbf{k})=e^{i\mathbf{k}\cdot\mathbf{r}}u(\mathbf{r};\mathbf{k}),
\]
where $u(\mathbf{r};\mathbf{k})$ has the same periodicity as the potential
$V(\mathbf{r})$ for any $\mathbf{k}$. {Physically, ${\bf k}$ is known as quasi{-}momentum.} It is convenient to introduce the
following two operators:
$$
\mathcal{H} \equiv \nabla^2-V(\mathbf{r}),\quad \quad
\mathcal{H}_\mathbf{k} \equiv \nabla^2+2i\mathbf{k}\cdot\nabla-|\mathbf{k}|^2-V(\mathbf{r}) ,
$$
where $\mathcal{H}$ is the Schr\"odinger operator with a periodic potential and $\mathcal{H}_\mathbf{k}$ is a
$\mathbf{k}$-dependent operator, defined on $L^2(\Omega)$; hence $u(\mathbf{r};\mathbf{k})$ satisfies the following
eigenvalue problem,
\[
 \mathcal{H}_\mathbf{k}u(\mathbf{r};\mathbf{k})=-\mu u(\mathbf{r};\mathbf{k}); \quad
 u(\mathbf{r}+\mathbf{v}_s;\mathbf{k})=u(\mathbf{r};\mathbf{k}); \;\;s=1,2,
 \]
where $ \mu=\mu(\mathbf{k})$  is called the dispersion relation. {On the other hand,} Bloch mode $\varphi(\mathbf{r};\mathbf{k})$
satisfies the eigenproblem {with quasi-periodic boundary condition}
\begin{equation}\label{Leigenproblem2}
 \mathcal{H}\varphi(\mathbf{r};\mathbf{k})=-\mu \varphi(\mathbf{r};\mathbf{k}); 
 \varphi(\mathbf{r}+\mathbf{v}_s;\mathbf{k})=e^{i\mathbf{k}\cdot\mathbf{v}_s}\varphi(\mathbf{r};\mathbf{k})
 \end{equation}
 for $s=1,2$.
 {Note that after one period the Bloch mode comes back to its original value, up to a phase factor.}

{Assume that $\mu(\mathbf{k}) = \mu(\mathbf{k} + \mathbf{g})$ for any $\mathbf{g} \in \mathbb{G}$.} We also note that $\varphi(\mathbf{r};\mathbf{k})$ and
$\varphi(\mathbf{r};\mathbf{k}+\mathbf{g})$ satisfy the same eigenvalue problem
and boundary condition Eq.~\eqref{Leigenproblem2} due to
$\mathbf{v}\cdot\mathbf{g}=2\pi m$ {for any} 
$\mathbf{v}\in\mathbb{P}$ and
$\mathbf{g}\in\mathbb{G}$.  So for any $\mathbf{r}$,
$\varphi(\mathbf{r};\mathbf{k})$ is periodic with respect to $\mathbf{k}$, thus
$\mathbf{k}$ is restricted in the parallelogram determined by {$\mathbf{k}_1$}
and {$\mathbf{k}_2${.}}

It is noted that for an arbitrary value of $\mathbf{k}$, the Bloch mode
$\varphi(\mathbf{r};\mathbf{k})$ is usually not a periodic function of
$\mathbf{r}$. However, there are some special values of $\mathbf{k}$, where
$\varphi(\mathbf{r};\mathbf{k})$ is periodic or anti-periodic with two periods
$\mathbf{v}_1$ and $\mathbf{v}_2$. For example, At the $\Gamma$ point
($\mathbf{k}=\mathbf{0}$) that is located at the center of the Brillouin zone,
$\varphi(\mathbf{r};\mathbf{0})$ is periodic {(see boundary condition in Eq.~(\ref{Leigenproblem2}))}. At the $\text{X}_s,s=1,2$ points
($\mathbf{k}=\frac{1}{2}\mathbf{k}_s, s=1,2$), which are located at the center
of one side of the Brillouin parallelogram, we have:
$\varphi(\mathbf{r}+\mathbf{v}_s;\frac{1}{2}\mathbf{k}_s)=-\varphi(\mathbf{r};\frac{1}{2}\mathbf{k}_s)$
and
$\varphi(\mathbf{r}+\mathbf{v}_{3-s};\frac{1}{2}\mathbf{k}_s)=\varphi(\mathbf{r};\frac{1}{2}\mathbf{k}_s)$.
At the M point ($\mathbf{k}=\frac{1}{2}\mathbf{k}_1+\frac{1}{2}\mathbf{k}_2$)
that is located at one vertex of the Brillouin parallelogram, we have:
$\varphi(\mathbf{r}+\mathbf{v}_1;\frac{1}{2}\mathbf{k}_1+\frac{1}{2}\mathbf{k}_2)=-\varphi(\mathbf{r};\frac{1}{2}\mathbf{k}_1+\frac{1}{2}\mathbf{k}_2)$
and
$\varphi(\mathbf{r}+\mathbf{v}_2;\frac{1}{2}\mathbf{k}_1+\frac{1}{2}\mathbf{k}_2)=-\varphi(\mathbf{r};\frac{1}{2}\mathbf{k}_1+\frac{1}{2}\mathbf{k}_2)${; See Fig. \ref{Rec_construction} below.}
Thus  at the $\Gamma$, $\text{X}_s$ and M points, which are special locations in
the {reciprocal lattice fundamental cell,} 
the corresponding Bloch modes are {either} periodic or
anti-periodic. In addition, at these special points the Bloch modes can be made
real since {the multiplicative factor in Eq.~(\ref{Leigenproblem2}) is real and consequently, the eigenfunctions can also be taken to be real.}


For each $\mathbf{k}$, the operator $\mathcal{H}_\mathbf{k}$ 
has an
infinite set of discrete eigenvalues $\mu(\mathbf{k})=\mu_j(\mathbf{k}),
j=0,1,2,\dots $. Thus, the dispersion relation $\mu$, its corresponding
eigenfunctions and the associated Bloch modes could  have an additional
subscript $j$ to indicate different eigenvalues. Here, for simplicity, we will
usually omit the subscript $j$. {Hence} 
the  spectrum of the Schr\"odinger
operator $\mathcal{H}$ has multiple band structures and {therefore} may {exhibit} 
band gaps between two dispersion surfaces where bounded Bloch modes are not allowed.
As $\mathbf{k}$ varies, the discrete eigenvalue $\mu(\mathbf{k})$ and the
corresponding eigenfunctions $u(\mathbf{r};\mathbf{k})$ as functions of
$\mathbf{k}$ are assumed to be smooth  over $\mathbf{k}$. {Later on, it will no longer {necessarily} be the case that the eigenfunctions are smooth in ${\bf k}$. Chern insulators contain eigenmodes with  discontinuous phase topology {\cite{Brouder2007}.}}

Since $\varphi(\mathbf{r};\mathbf{k})$ is periodic in $\mathbf{k}$, we can represent {it}
as a Fourier series {
\begin{equation}
\label{WannDec}
\varphi(\mathbf{r};\mathbf{k})=\displaystyle{\sum_{\mathbf{v} \in \mathbb{P}}\phi_{\mathbf{v}}(\mathbf{r})e^{i\mathbf{k}\cdot\mathbf{v}}}.
\end{equation}
} where {the Fourier coefficient} $\phi_{\mathbf{v}}(\mathbf{r})$ {is} defined as
\begin{equation}\label{Wannier}
\phi_{\mathbf{v}}(\mathbf{r})=\frac{1}{|\Omega'|}\int_{\Omega'}
\varphi(\mathbf{r};\mathbf{k})e^{-i\mathbf{k}\cdot\mathbf{v}}d\mathbf{k}
\end{equation}
{and} is the so-called Wannier function \cite{Wannier_37}{. From here on,} 
the sum over $\mathbf{v}$ means $\mathbf{v}$ takes all values in $\mathbb{P}$,
i.e., $\mathbf{v}=m\mathbf{v}_1+n\mathbf{v}_2$, for all $m,n\in\mathbb{Z}$.

From 
definition Eq.~\eqref{Wannier}, we can see that
\begin{align*}
\phi_{\mathbf{v}}(\mathbf{r})&=\frac{1}{|\Omega'|}\int_{\Omega'}
\varphi(\mathbf{r};\mathbf{k})e^{-i\mathbf{k}\cdot\mathbf{v}}d\mathbf{k} \\ &=\frac{1}{|\Omega'|}\int_{\Omega'}
u(\mathbf{r};\mathbf{k})e^{i\mathbf{k}\cdot\mathbf{(r-v)}}d\mathbf{k}\\&=\frac{1}{|\Omega'|}\int_{\Omega'}
\varphi(\mathbf{r-v};\mathbf{k})d\mathbf{k}=\phi_{\mathbf{0}}(\mathbf{r-v})
\end{align*}
{due to} 
the periodic nature of $u({\bf r};{\bf k})$. This equation shows that all Wannier modes are merely translations of the primitive Wannier mode, $\phi_{\mathbf{0}}(\mathbf{r})$.
Usually, the subscript $\mathbf{0}$ is omitted and the Wannier function is
referred to as $\phi(\mathbf{r}-\mathbf{v})$. Wannier functions have all the
information of the Bloch modes{, yet they do not depend on ${\bf k}$}. If one has {all} 
Wannier function {coefficients, then} the exact Bloch mode can be constructed through
\eqref{WannDec}{,} or vice versa {via Eq.~(\ref{Wannier})}. In general, is not possible to compute either
Bloch modes or Wannier functions explicitly. However, under some limits{,} such as
tight-binding/deep lattice limit (i.e., {$|V(\bf r)| \gg 1$}), they can be constructed by asymptotic
analysis that in turn provides {crucial} {analytical} understanding. Details will be discussed
below.

For a periodic potential{,} the local minima 
are  called the sites. Physically,
local minima 
are the positions of 
potential wells and in optics {correspond to} 
increased refractive index {to which} 
the electric field is attracted{.} 
In the tight-binding limit{,} the potential well at each site is very deep, {hence it often turns out that} the
Wannier function defined in Eq.~\eqref{Wannier} is localized at the site
$S_\mathbf{v}$, {becoming more localized as the depth increases. Physically speaking,}
Bloch modes {tend to concentrate} 
most of their energy in the
neighborhood of these sites. {The lattice sites are waveguides that effectively trap the wave function with some weak coupling among nearby lattice sites}.

The potential {function describing} 
the periodic lattice can be written in the form
\begin{equation}
\label{lattice_sum}
V(\mathbf{r})=\displaystyle{\sum_\mathbf{v}}V_s(\mathbf{r}-\mathbf{v}).
\end{equation}
where $V_s(\mathbf{r})$ denotes the potential {well} at the site $S_\mathbf{0}$.
It originally is defined only in the cell
$\Omega$ (i.e., its support is only the primitive unit cell $\Omega$).
We also define
$$
\Delta V(\mathbf{r})=V(\mathbf{r})-V_s(\mathbf{r}).
$$
Moreover, we will extend the
domain of $V_s(\mathbf{r})$ to the whole plane with fast decaying tails.
Since the overall value of
the potential is not important, {here} we take the potential to satisfy
$\displaystyle{\max_\mathbf{r}}\{V(\mathbf{r})\}=0$. For an arbitrary potential
that does not satisfy this requirement, we can just simply subtract its
maximum value {through a phase transformation of $\psi({\bf r},z)$}. Mathematically, a way to construct a periodic function is to let
$V_s(\mathbf{r})$ be a rapidly decaying function and then repeat this function
under translational shifts {of the} 
lattice vectors. A periodic function is now  a
sum of rapidly decaying functions that are the same up to a {spatial} 
shift.  In
the tight-binding limit:  $V_s(\mathbf{r})$ of a simple periodic potential can
be approximated by {$V(\mathbf{r})\approx  - V_0  e^{-k_0^2(x^2+y^2)} $ with $k_0^2 \gg 1$}.   We note that
if the potential has more than one local minima in a unit cell, i.e., a
non-simple lattice,  then we {apply} 
this approximation near each {distinct} site {type}.

\section{Simple lattices, nonlinear envelope dynamics}
\label{simple_lattice}

\subsection{Dispersion relations}

In order to understand the envelope dynamics in weakly nonlinear periodic
media, we need a good understanding of the associated linear problem. The
linear problem is governed by a linear Schr\"odinger equation with a periodic
potential and the dispersion relation,  $\mu(\mathbf{k})${,}  
plays a key role.

Since $\mu(\mathbf{k})$ is a periodic function of $\mathbf{k}$, it can be
represented in a Fourier series
\begin{equation}\label{dis_Fourier_1}
\mu(\mathbf{k})=\hat{\mu}_\mathbf{0}+\displaystyle{\sum_{\mathbf{v} \not= {\bf 0}}}\hat{\mu}_\mathbf{v}
e^{i\mathbf{k}\cdot\mathbf{v}}.
\end{equation}
{where {$\hat{\mu}_\mathbf{v}=\hat{\mu}^*_\mathbf{-v}$} since $\mu(\mathbf{k})$ is real}.
For a simple 2D periodic potential we will estimate the order of
$\hat{\mu}_\mathbf{v}$ and find the leading order {contributions.} 
The 1D lattice is a
special case. For a 1D lattice, {it turns out that}
$\mu(k)\approx\hat{\mu}_0+2\hat{\mu}_1\cos(kl)$ 
{with} $|\hat{\mu}_1|\gg |\hat{\mu}_n|, n>1$ where $l$ is the 1D period.

In the tight-binding limit, we assume the Wannier functions Eq.~\eqref{Wannier} are
localized at the {lattice} sites and decay exponentially. This allows us to use WKB
expansions where the harmonic oscillator is a good approximation. 
To leading order, the Wannier function can be approximated by ``orbitals'', defined as
\begin{equation}
\label{Orbital}
\left[ \nabla^2-V_s(\mathbf{r})\right] \phi(\mathbf{r})=-E \phi(\mathbf{r}),
\end{equation}
where $E$ is the real discrete eigenvalue of the operator
$\nabla^2-V_s(\mathbf{r})${,} 
also called orbital energy. In
other words, Wannier functions defined in Eq.~\eqref{Wannier} satisfy the
eigenvalue problem Eq.~\eqref{Orbital} to leading order. We do not distinguish
between orbitals and Wannier functions here. For convenience, we require that
the orbitals are real and  have norm 1, i.e., {$\int
\phi^2(\mathbf{r})d\mathbf{r}=1$}. We define
$$
\mathcal{H}^{\mathbf{v}} \equiv \nabla^2-V_s(\mathbf{r}-\mathbf{v})
$$
where $\mathcal{H}^{\mathbf{v}}$ is {a self-adjoint operator} defined in $L^2(\mathbb{R}^2)$. So, $E$ and
{$\phi(\mathbf{r} - {\bf v})$} are the eigenvalue and corresponding eigenfunction of
$\mathcal{H}^{\mathbf{v}}${. Moreover,} 
$\mathcal{H}^{\mathbf{v}}$ usually has a infinite
number of discrete eigenvalues if $V_s(\mathbf{r})$ is bounded. In this
chapter, we will discuss  the lowest  band, {where} 
using orbitals to construct the
Bloch mode is reasonable if $V_s(\mathbf{r})$ is deep: $|V_s|\gg 1$.

Next, we use a discrete approach to compute the dispersion relation.
Substituting the Bloch mode Eq.~\eqref{WannDec} into the eigenvalue problem
\eqref{Leigenproblem}, we get
\[
[\mathcal{H}^{\mathbf{0}}+E] \varphi(\mathbf{r})=[E-\mu+\Delta V(\mathbf{r})] \varphi(\mathbf{r}).
\]
For the ground state, {we assume} the nullspace of the  operator
$\mathcal{H}^{\mathbf{0}}+E$ is one dimensional. 
Then the Fredholm condition
associated with $\mathcal{H}^{\mathbf{0}}$ gives
\[
\int \phi(\mathbf{r})  \left[ (E-\mu)+\Delta V(\mathbf{r}))\right] \varphi(\mathbf{r})  d\mathbf{r}=0.
\]
Substituting the decomposition of the Bloch mode Eq.~\eqref{WannDec} into the above
condition yields the dispersion relation
\begin{equation}
\label{Dispersion_nondegenerate0}
\mu=E+\frac{\displaystyle{\sum_\mathbf{v}}\lambda_\mathbf{v}e^{i\mathbf{k}\cdot\mathbf{v}}}{\displaystyle{\sum_\mathbf{v}}\kappa_\mathbf{v}e^{i\mathbf{k}\cdot\mathbf{v}}},
\end{equation}
where
\begin{align*}
& \lambda_\mathbf{v}=\int \phi(\mathbf{r})\Delta V(\mathbf{r})\phi(\mathbf{r}-\mathbf{v})d\mathbf{r} \\ &
\kappa_\mathbf{v}=\int \phi(\mathbf{r})\phi(\mathbf{r}-\mathbf{v})d\mathbf{r}=\kappa_{-\mathbf{v}}.
\end{align*}

The dispersion relation in Eq.~\eqref{Dispersion_nondegenerate0} can be simplified.
Note that $\kappa_\mathbf{0}=1$ and $\lambda_\mathbf{v}\ll1$ and
$\kappa_\mathbf{v}\ll1$ when $\mathbf{v}\neq\mathbf{0}$ because
$\phi(\mathbf{r})$ is localized. To  leading order, the dispersion relation is:
$\mu \sim E+\lambda_\mathbf{0} \sim \hat{\mu}_\mathbf{0}$; i.e., the mean value
of $\mu$. Since
$\displaystyle{\sum_{\mathbf{v}\neq\mathbf{0}}}\kappa_\mathbf{v}e^{i\mathbf{k}\cdot\mathbf{v}}\ll1$,
we have
{\begin{equation}
\label{dis_Fourier_2}
\mu \approx E+\lambda_\mathbf{0}+\displaystyle{\sum_{\mathbf{v} \not= {\bf 0}}}C_\mathbf{v}e^{i\mathbf{k}\cdot\mathbf{v}}
\end{equation}}
where
$$
C_\mathbf{v}=\lambda_\mathbf{v}-\lambda_\mathbf{0}\kappa_\mathbf{v}.
$$
Comparing Eqs.~\eqref{dis_Fourier_1} and~\eqref{dis_Fourier_2}, we see
$\hat{\mu}_\mathbf{v}\approx C_\mathbf{v},$ {for} $  \mathbf{v}\neq\mathbf{0}${. Hence,} 
we have calculated the {first few} Fourier coefficients of the dispersion relation
$\mu=\mu(\mathbf{k})$.

Furthermore, we need only take the leading order terms of
$\displaystyle{\sum_{\mathbf{v}}}C_\mathbf{v}e^{i\mathbf{k}\cdot\mathbf{v}}$.
Note that both $\lambda_\mathbf{v}$ and $\kappa_\mathbf{v}$ decay fast as
$|\mathbf{v}|\to \infty$. So we only need to consider the nearest neighbor and
on-site interactions for the dominant contributions. This is the tight-binding
approximation that has been widely used in solid state physics to calculate
electronic band structure (cf. \cite{Callaway_1991}). Then we get the
dispersion relation
\begin{equation}
\label{Dispersion_nondegenerate}
\mu ({\bf k})=E+\lambda_\mathbf{0}+\displaystyle{\sum_{\langle \mathbf{v} \rangle}}C_\mathbf{v}e^{i\mathbf{k}\cdot\mathbf{v}} .
\end{equation}
Here and afterwards $\langle \mathbf{v} \rangle$ {indicates} 
the sum over $\mathbf{v}$ only takes
nearest (nonzero) neighbor shift vectors. For convenience, we also define
$$
\omega(\mathbf{k})=\displaystyle{\sum_{ \langle \mathbf{v} \rangle}}C_\mathbf{v}e^{i\mathbf{k}\cdot\mathbf{v}}.
$$
It is noted that $E+\lambda_\mathbf{0}$ has no $\mathbf{k}$ dependence and only
determines the {mean} 
value of the {frequency.} 
The $\mathbf{k}$
dependence of the dispersion relation is determined by $\omega(\mathbf{k})$.
When {considering} 
nearest neighbor interactions, we {typically} 
assume that
$C_\mathbf{v}$ for all nearest neighbor shift vectors have the same order and
denote
\begin{equation*}
C=C_{\mathbf{v}_1}.
\end{equation*}
So, for any nearest neighbor shift vector $\mathbf{v}$, $C_\mathbf{v}\sim O(C)$.
In the tight-binding limit, $C$ is very small. So $\omega(\mathbf{k})$ is order
$O(C)$. It is also seen that as $V_0\to \infty, ~C\to 0$ and consequently
$\omega(\mathbf{k})\to 0$. Hence the dispersion surface {in Eq.~(\ref{Dispersion_nondegenerate})} becomes flatter and
flatter. On the other hand, the two nearest orbital energy difference
$E_{j+1}-E_{j}\sim O(\sqrt{V}_0)$, so $E_{j+1}-E_j \to \infty$. Consequently,
there may exist a gap between $\mu_{j+1}(\mathbf{k})$ and
$\mu_{j}(\mathbf{k})$.

Note that the ground state (lowest eigenfunction) of the operator
{$\mathcal{H}^{\bf v}$} {is} 
taken to be simple; however, the eigenvalues associated
with the higher excited states can be degenerate; i.e., there can be multiple
eigenfunctions corresponding to one eigenvalue. The interested reader can find
a discussion of the higher states in \cite{AZ13}.

\subsection{Envelope dynamics}

Similar to Fourier modes, the Bloch modes form a complete set in the space of
$L^2$ functions  {\cite{East73,Kuch93}.}
{As a result,} 
an $L^2$ function can be decomposed {into} Bloch
mode components {\cite{Ilan2010,Pelinovsky2011}}.

In the linear limit, the dynamics of Bloch modes are
determined by the dispersion relation. Due to the superposition principle of
linear problems, different Bloch modes have different dynamics and they do not
{mix} 
with each other.  However, when nonlinearity is present, the dynamics
is more subtle. {Although the derivation of the equation for a continuous envelope in space-time is well-known,
it is not obvious how one can derive the equations for a discrete, in space, envelope.}

When $\mu(\mathbf{k})$ has a single dispersion relation branch, we assume to leading order
\begin{equation}
\label{envelope_nondegenearte}
\psi(\mathbf{r},z,Z) \sim
\displaystyle{\sum_{\mathbf{v}}a_\mathbf{v}(Z)\phi(\mathbf{r}-\mathbf{v})e^{i \left[ \mathbf{k}\cdot\mathbf{v} -  \mu({\bf k}) z \right]}} .
\end{equation}
Here $a_\mathbf{v}$ represents the Bloch wave {mode} envelope at the site
$S_{\mathbf{v}}$. We 
assume the envelope $ a_\mathbf{v}(Z)$ varies
slowly under evolution, where $Z=\varepsilon z$ {for a} 
small parameter
$\varepsilon$ {that} will be determined later.

Substituting the envelope representation Eq.~\eqref{envelope_nondegenearte} into
the lattice NLS Eq.~\eqref{LNLS}, one obtains
\begin{align}
\label{Dy_non}
&&\left(\mathcal{H}^{\mathbf{p}}+E\right)\left(\displaystyle{\sum_{\mathbf{v}}}a_\mathbf{v}(Z)\phi(\mathbf{r}-\mathbf{v})e^{i\mathbf{k}\cdot\mathbf{v}}\right)
\nonumber\\&&=-\displaystyle{\sum_{\mathbf{v}}}\left(\varepsilon
i\frac{da_\mathbf{v}}{dZ}+a_\mathbf{v}\left[ \mu-E-\Delta
V(\mathbf{r}-\mathbf{p})\right] \right)\phi(\mathbf{r}-\mathbf{v})e^{i\mathbf{k}\cdot\mathbf{v}}\nonumber\\&&-
\sigma\left(\displaystyle{\sum_{\mathbf{v}}a_\mathbf{v}\phi(\mathbf{r}-\mathbf{v})e^{i\mathbf{k}\cdot\mathbf{v}}}\right)^2\left(\displaystyle{\sum_{\mathbf{v}}a_\mathbf{v}\phi(\mathbf{r}-\mathbf{v})e^{i\mathbf{k}\cdot\mathbf{v}}}\right)^*,
\end{align}
where $\mathbf{p}\in \mathbb{P}$. The Fredholm condition associated with $ \mathcal{H}^{\mathbf{p}}$, i.e., $\int F
\phi(\mathbf{r}-\mathbf{p})=0$ where $F$ represents the right hand side (RHS) of Eq.~\eqref{Dy_non}, yields
\begin{align*}
& \sum_{\bf v'} i\varepsilon \kappa_{{\bf v}'} \frac{da_{\mathbf{p} + {\bf v}'}}{dZ}e^{i\mathbf{k}\cdot\mathbf{v}'}  +\displaystyle{\sum_{\mathbf{v}'}}a_{\mathbf{p}+\mathbf{v}'} \left[ (\mu-E)\kappa_{\mathbf{v}'}-\lambda_{\mathbf{v}'} \right]e^{i\mathbf{k}\cdot\mathbf{v}'}  \\ & +\sigma\displaystyle{\sum_{\mathbf{v}_1}\sum_{\mathbf{v}_2}\sum_{\mathbf{v}_3}\gamma_{\mathbf{v}_1\mathbf{v}_2\mathbf{v}_3}a_{\mathbf{v}_1}a_{\mathbf{v}_2}a_{\mathbf{v}_3}^*}=0,
\end{align*}
where {$ {\bf v} = {\bf p} + {\bf v}'$} and only leading order terms are considered and
\begin{align*}
& \gamma_{\mathbf{v}_1\mathbf{v}_2\mathbf{v}_3}= e^{i\mathbf{k}\cdot(\mathbf{v}_1+\mathbf{v}_2-\mathbf{v}_3)} \times  \\ & \int\phi(\mathbf{r}-{\bf p} - \mathbf{v}_1)\phi(\mathbf{r}-{\bf p} - \mathbf{v}_2)\phi(\mathbf{r}- {\bf p} - \mathbf{v}_3)\phi(\mathbf{r}-\mathbf{p})d\mathbf{r}.
\end{align*}

When only on-site and nearest neighbor interactions are taken into account,
the governing equation{, after dropping the prime notation,} is
\begin{equation}\label{D_govern_Non}
i\varepsilon\frac{da_\mathbf{p}}{dZ}+\omega(\mathbf{k})a_\mathbf{p}-\displaystyle{\sum_{\langle \mathbf{v} \rangle}}a_{\mathbf{p}+\mathbf{v}}C_\mathbf{v}e^{i\mathbf{k}\cdot\mathbf{v}}+g\sigma|a_\mathbf{p}|^2a_\mathbf{p}=0,
\end{equation}
where {$g=\gamma_{\mathbf{000}}= \int \phi(\mathbf{r})^4d\mathbf{r}$} is the only
on-site interaction term taken for the nonlinear term. Here we assume that
$\varepsilon$, $\sigma$ and $C$ all have the same order to ensure maximal
balance.

After rescaling,  we obtain the nonlinear discrete evolution equation
\begin{equation}\label{D_govern_Non_final}
i\frac{da_\mathbf{p}}{dZ}+\widetilde{\omega}(\mathbf{k})a_\mathbf{p}-\displaystyle{\sum_{ \langle \mathbf{v} \rangle}}a_{\mathbf{p}+\mathbf{v}}\widetilde{C}_\mathbf{v}e^{i\mathbf{k}\cdot\mathbf{v}}+gs(\sigma)|a_\mathbf{p}|^2a_\mathbf{p}=0,
\end{equation}
where for convenience we have taken $\varepsilon=|C|=|\sigma|$;
$\widetilde{\omega}=\frac{\omega}{|C|}$;
$\widetilde{C}_\mathbf{v}=\frac{C_\mathbf{v}}{|C|}$ and $s(\sigma)$ is the sign of
$\sigma$. Eq.~\eqref{D_govern_Non_final} is the unified discrete nonlinear wave
system that describes the dynamics of a single envelope in any simple
nonlinear periodic lattice. Note that the {linear} coefficients 
of the 
equation are directly related to the coefficients of the linear
dispersion relation in the tight-binding limit{, defined in Eq.~(\ref{Dispersion_nondegenerate})}. We also note that the 1D {reduction} 
is obtained as a special case; i.e., either the vector $\mathbf{v}$ is one
dimensional or the 2D lattice is well-approximated by a 1D lattice. So if 
$\mathbf{v_1}=l \ihat$ and we omit  $\mathbf{v_2}${,} the 1D lattice {equation }
is given
in 1D notation by
\begin{align}
\label{one_d_lattice}
\nonumber
i\frac{da_p}{dZ}+ &\widetilde{\omega}(k)a_p-(a_{p+l}\widetilde{C}_{p+l}e^{ikl}+a_{p-l}\widetilde{C}_{p-l}e^{-ikl}) \\ & +gs(\sigma)|a_p|^2a_p=0.
\end{align}

{The derivation of the tight-binding models above can be made rigorous. In particular, it is possible to show that the Wannier expansion (\ref{WannDec}) approaches the solution of the lattice NLS equation (\ref{LNLS}) in the deep-lattice limit, i.e. $V_0, |V_s|  \rightarrow \infty$, using an appropriate Sobolev norm. The works of \cite{Ablowitz2012,Fefferman2018,Pelinovsky2010,Pelinovsky2008} have proven this for various  lattice Schr\"odinger equations with different types of deep, but periodic potentials.}

\subsection{Continuum Reduction}
{Next, we} 
consider the continuous limit.
{Assume} 
that the envelope $a_{\mathbf{v}}$ varies slowly over $\mathbf{v}$.
In other words, the envelope takes the form
\begin{align*}
 \psi(\mathbf{r},Z) & \sim
\displaystyle{\sum_{\mathbf{v}}a_\mathbf{v}(Z)\phi(\mathbf{r}-\mathbf{v})e^{i\mathbf{k}\cdot\mathbf{v}}}e^{-i\mu\, z}
\\ & \approx
\displaystyle{\sum_{\mathbf{v}}a(\mathbf{R},Z)\phi(\mathbf{r}-\mathbf{v})e^{i\mathbf{k}\cdot\mathbf{v}}}e^{-i\mu\,z}
\end{align*}
where $\mathbf{R}=(X,Y)=\nu \mathbf{r}$ now denotes the coordinate of the
envelope and $\nu\ll1$.  To leading order, {$a_\mathbf{v}\approx\int
a(\mathbf{R})\phi^2(\mathbf{r}-\mathbf{v})d\mathbf{r}\approx  a(\nu ({\bf R} - 
S_\mathbf{v}))$ where $a_\mathbf{v}$} is defined at site points.

Before proceeding, we recall our assumption that the dispersion relation is
sufficiently smooth at the $\mathbf{k}$ value we are studying. We also
introduce some further notation{:  $\partial_m \equiv \frac{\partial}{\partial
\mathbf{r}_m}$}  and {$\nabla \equiv (\partial_1, \partial_2)^{T}$;
 $\widetilde{\partial}_m \equiv \frac{\partial}{\partial \mathbf{R}_m}$ and $\widetilde{\nabla} \equiv (\widetilde{\partial}_1,
 \widetilde{\partial}_2)^{T}$;
 $\overline{\partial}_m \equiv \frac{\partial}{\partial \mathbf{k}_m}$ and $\overline{\nabla} \equiv (\overline{\partial}_1,
 \overline{\partial}_2)^{T}$;
 $\partial_{m,n} \equiv \partial_m\partial_n$. Here $m = 1$ denotes the $x$-direction and $m = 2$ is the $y$-direction.}

Using Taylor expansion, we get
\[
a_{\mathbf{p}+\mathbf{v}}\approx a_\mathbf{p}+\nu\mathbf{v}\cdot\widetilde\nabla
a_{\bf p}+\frac{\nu^2}{2}\mathbf{v} \mathbf{H}\mathbf{v}^T  a_{\bf p},
\] 
where $\mathbf{H}=\left(\begin{array}{ll} \widetilde{\partial}_{11} \quad
&\widetilde{\partial}_{12}\\ \widetilde{\partial}_{21} \quad &\widetilde{\partial}_{22}
\end{array}\right)$ is the Hessian matrix operator.

Then
\begin{align}
\nonumber
& \displaystyle{\sum_{\langle \mathbf{v} \rangle}}a_{\mathbf{p}+\mathbf{v}}C_\mathbf{v}e^{i\mathbf{k}\cdot\mathbf{v}} \approx \\ & a_\mathbf{p}\displaystyle{\sum_{\langle \mathbf{v}\rangle}}C_\mathbf{v}e^{i\mathbf{k}\cdot\mathbf{v}}+\nu\widetilde\nabla
a_{\bf p}\cdot
\displaystyle{\sum_{\langle \mathbf{v}\rangle}}\mathbf{v}C_\mathbf{v}e^{i\mathbf{k}\cdot\mathbf{v}}+\frac{\nu^2}{2}\displaystyle{\sum_{ \langle\mathbf{v}\rangle}}C_\mathbf{v}e^{i\mathbf{k}\cdot\mathbf{v}}\mathbf{v} \mathbf{H}\mathbf{v}^T a_{\bf p} \nonumber\\
&=a_\mathbf{p}\omega(\mathbf{k})- i\nu\overline{\nabla}\mu\cdot\widetilde{\nabla}a_{\bf p}-\frac{\nu^2}{2}\displaystyle{\sum^2_{m,n=1}}\overline{\partial}_{m,n}\mu\widetilde{\partial}_{m,n}a_{\bf p}.
\label{Taylor_Expan}
\end{align}
Substituting Eq.~\eqref{Taylor_Expan} into Eq.~\eqref{D_govern_Non}
yields{, to leading order,} the equation
\begin{align*}
i\varepsilon\frac{\partial a}{ \partial Z} + i\nu\overline{\nabla}\mu\cdot\widetilde{\nabla}a+
\frac{\nu^2}{2}\displaystyle{\sum^2_{m,n=1}}\overline{\partial}_{m,n}\mu\widetilde{\partial}_{m,n}a+g \sigma
|a|^2a=0 , 
\end{align*}
{where $a \equiv a_{p}$ is a continuous function now.}
The above equation{, whose coefficients depend on $\mu({\bf k})$,} governs the dynamics of a single  Bloch mode envelope in
nonlinear simple periodic media. It is valid for any value of $\mathbf{k}$. 
In analogy to homogeneous media, $\overline{\nabla}\mu$ plays the role of the group
velocity; it is the velocity of the envelope. In special cases,
$\overline{\nabla}\mu=\mathbf{0}$. This condition gives the extrema of the
dispersion surface, and at these points, the group velocity is zero and the
envelope will remain at 
{its initial position.} 
The envelope has a
spatial shift in the cross-section when propagating along $z$ direction if
$\overline{\nabla}\mu\neq\mathbf{0}$. However, by defining a moving frame variable
{$ {\bf \xi} = {\bf R} - \overline{\nabla} \mu  Z $} 
we find the equation
\begin{align}
\label{C_govern_non}
i\frac{da}{dZ}+\frac{1}{2}\displaystyle{\sum^2_{m,n=1}}\overline{\partial}_{m,n}\widetilde{\mu}~\widetilde{\partial}_{m,n}a+ s(\sigma)
|a|^2a=0,
\end{align}
where $\widetilde{\mu}=\frac{\mu}{|C|}$ and $s(\sigma)$ is the sign of $\sigma${ {and} the matrix $\overline{\partial}_{m,n}\widetilde{\mu}$ is typically called the Hessian}. Here we have taken the maximal balance
condition
$\varepsilon\sim O(\nu^2 |C|)\sim O(g |\sigma| )$. The above Eq.~\eqref{C_govern_non} is a 2D nonlinear Schr\"odinger
equation.  At different values of $\mathbf{k}$, the {linear} dispersive terms may be elliptic, hyperbolic or even parabolic.

\section{A typical simple lattice--square lattice}
\label{square_sec}

{In the previous section,} we  derived the dispersion relation for arbitrary simple lattices and
the dynamics of Bloch mode envelopes. In this section, we will use a typical
square lattice to apply the above general analysis. For convenience, 
we assume the nonlinearity is focusing, 
i.e., $\sigma>0$.
Square 2D periodic structures are  common in nature and can be readily
engineered in optics (cf. \cite{Fleischer2003Nature} {and Fig.~\ref{2drectlattice}}).

A typical square lattice is
 \begin{equation}
 \label{square_lattice}
 V(x,y)=\frac{V_0}{2}(\sin^2(k_0x)+\sin^2(k_0y)-2),
 \end{equation}
where {$0 \ge V({\bf r}) \ge - V_0, $} $V_0>0$ is the lattice intensity {and} $k_0$ is the scaled wavelength of the
interfering plane waves. {The} 
characteristic vectors are
 \begin{gather*}
\mathbf{v}_1=l\left(1,0\right),\quad\quad
\mathbf{v}_2=l\left(0,1\right),\nonumber\\ 
\mathbf{k}_1=\frac{2\pi}{l}\left(1,0\right),\quad\quad
\mathbf{k}_2=\frac{2\pi}{l}\left(0,1\right),
\end{gather*}
where $l=\frac{\pi}{k_0}$ is the lattice constant. {Clearly, ${\bf v}_i \cdot {\bf k}_j = 2 \pi \delta_{ij}$ and the potential has periodicity $V(x + m l , y + n l) = V(x,y)$ for $m,n \in \mathbb{Z}$.}

\begin{figure}
\centering
\includegraphics*[width=0.35\textwidth]{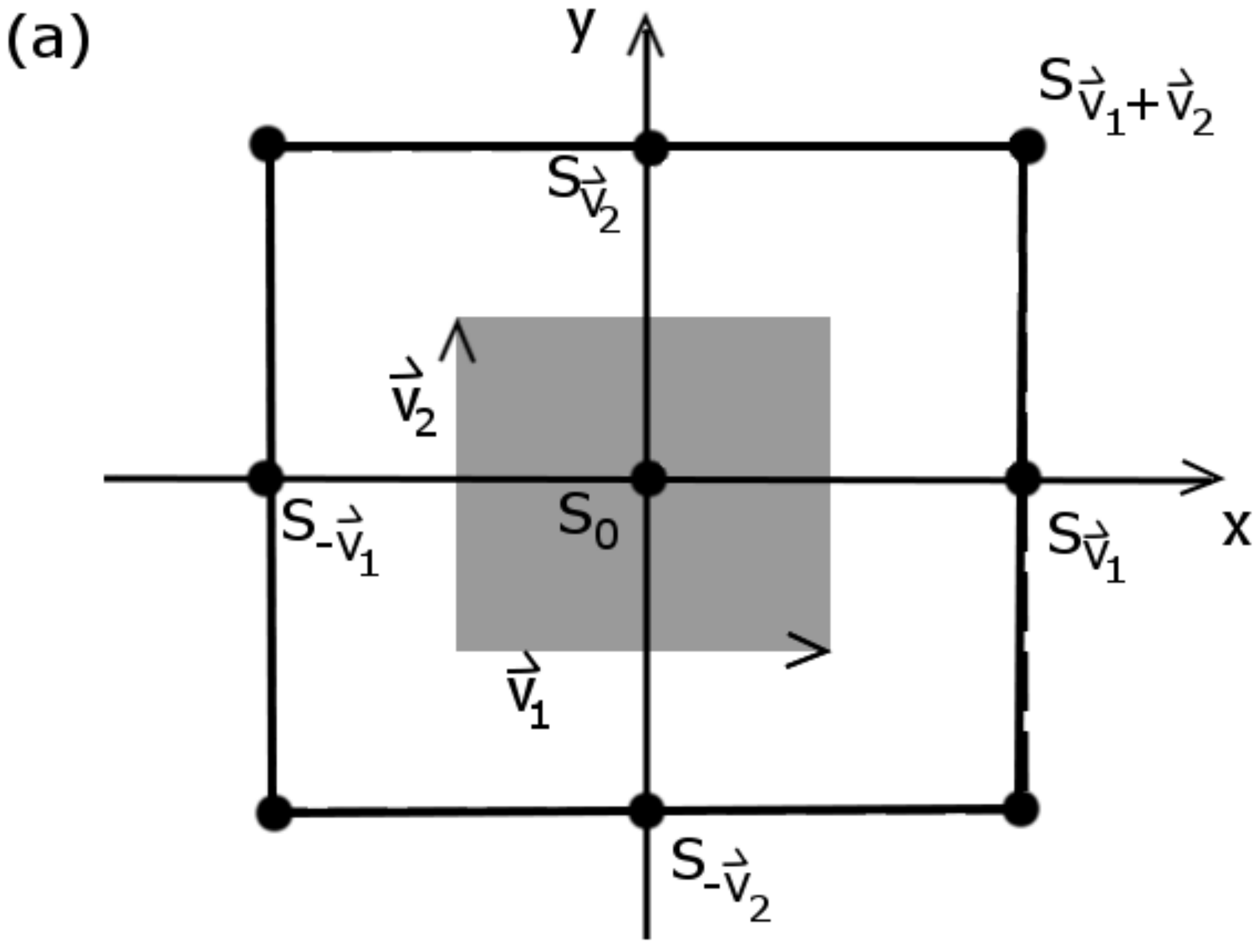}
\includegraphics*[width=0.35\textwidth]{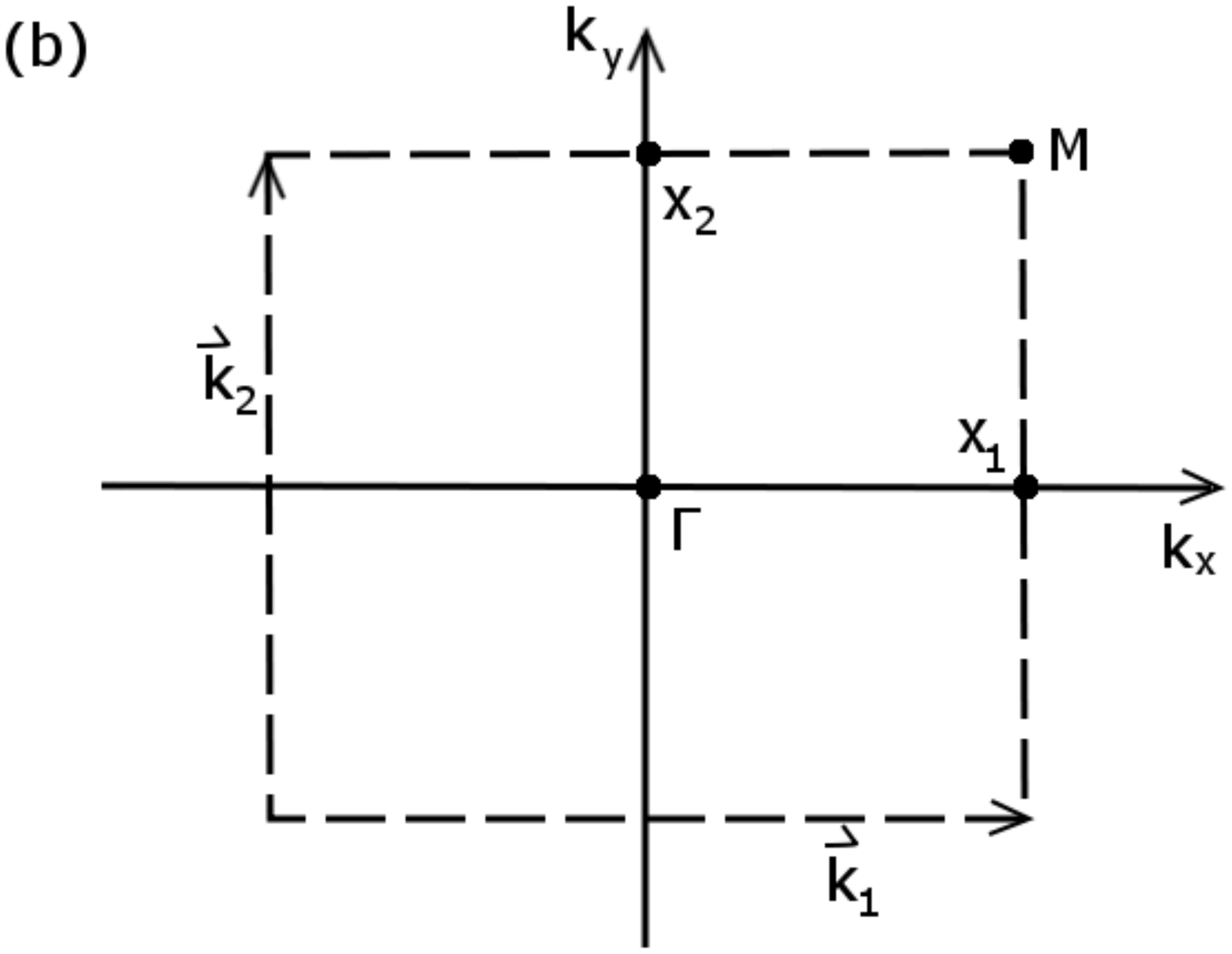}
\includegraphics*[width=0.35\textwidth]{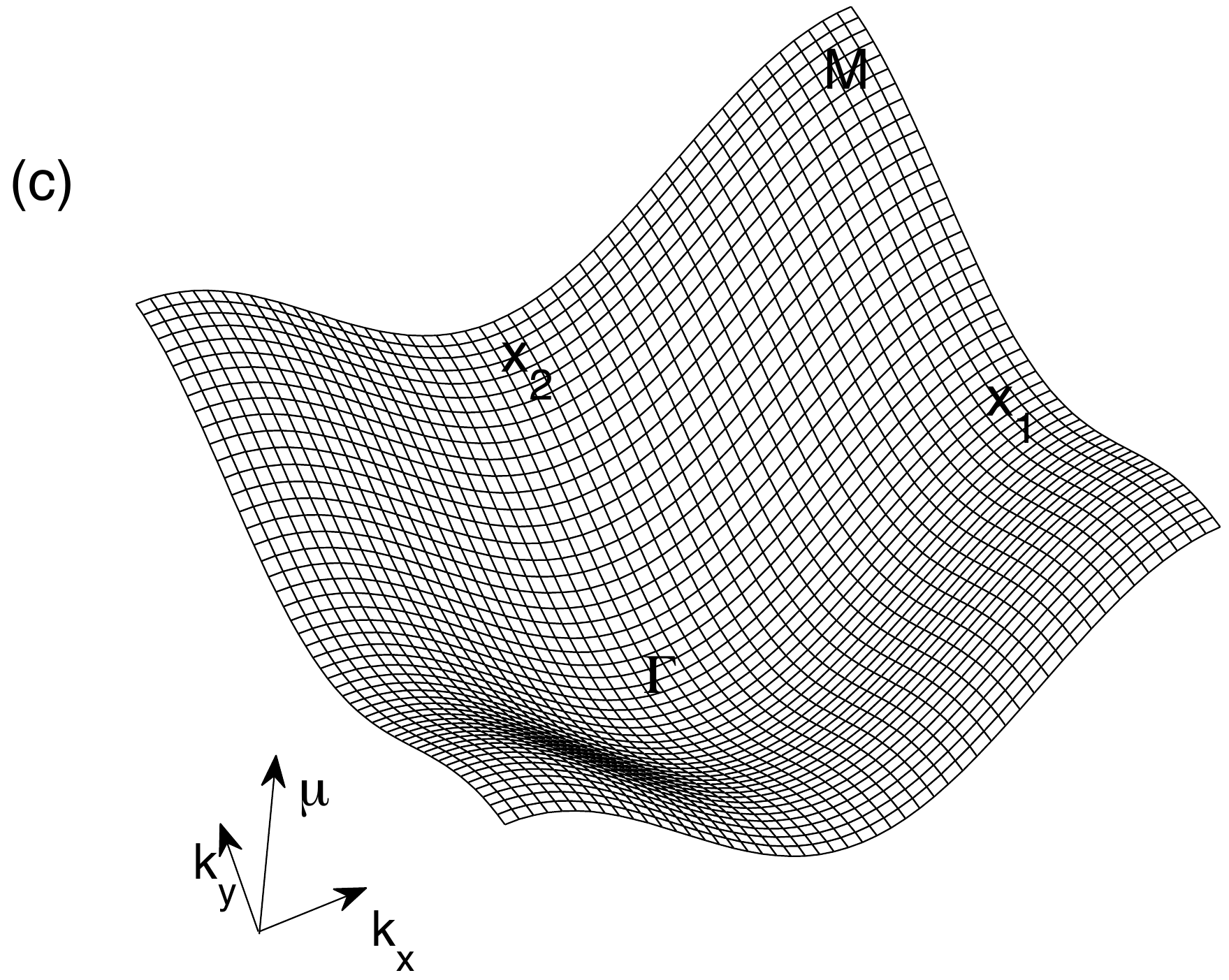}
\includegraphics*[width=0.35\textwidth]{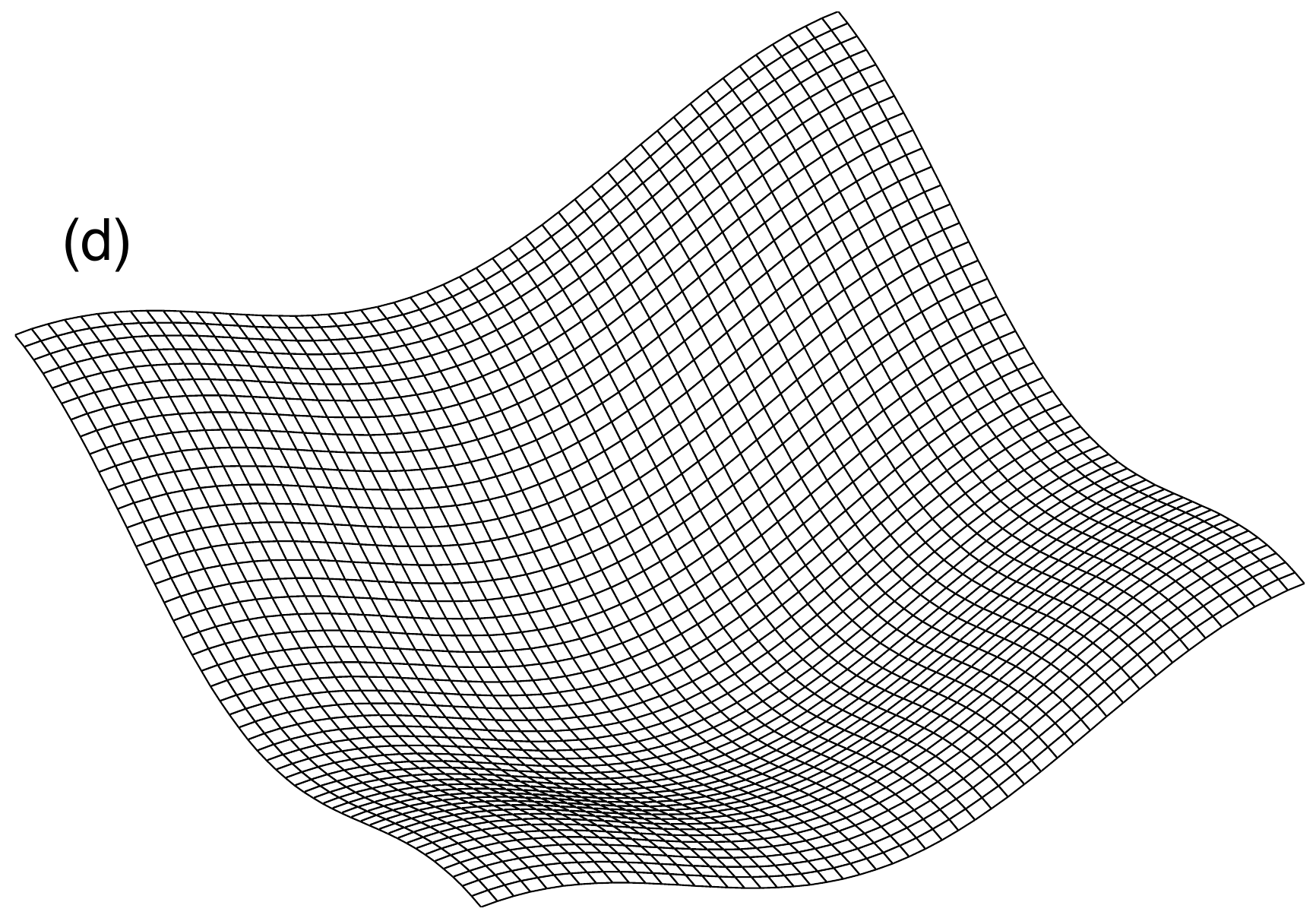}
\caption{\label{Rec_construction}(a) The site distributions of the square lattice Eq.~\eqref{square_lattice}. The shadow
region is the primitive unit cell. (b) The Brillouin zone of the square lattice--the square surrounded by the dashed
lines. The first dispersion relation band (c) {obtained} from direction simulation and (d) from the {approximate} formula
\eqref{Dispersion_rectangular_0}.}
\end{figure}

The site distribution is displayed in Fig.~\ref{Rec_construction}(a). For this
potential, {each} 
site has {four} 
nearest neighbors. Note that
$S_{\mathbf{v}_1+\mathbf{v}_2}$ is not one {of the nearest neighbors} of
$S_{\mathbf{0}}$. The nearest shift vectors are $\mathbf{v}_1$,
$-\mathbf{v}_1$,$\mathbf{v}_2$,$-\mathbf{v}_2$.

For the first band, we {find} that
{$$
C_{\mathbf{v}_1}=C_{\mathbf{v}_2}=  C \approx -0.056 V_0 \exp\left(-\frac{\sqrt{V_0}\;\pi^2}{4\sqrt{2}k_0}\right)<0
$$}
 (see below for further details). So the dispersion relation of the first band
is
{\begin{equation}\label{Dispersion_rectangular_0}
  \begin{aligned}
\mu(\mathbf{k}) &=E+\lambda_\mathbf{0} + 2C \left[ \cos(\mathbf{k}\cdot\mathbf{v}_1)+\cos(\mathbf{k}\cdot\mathbf{v}_2) \right] ,\\ 
&=E+\lambda_\mathbf{0} + 2C \left[ \cos(lk_x)+\cos(lk_y) \right].
\end{aligned}
\end{equation}}

The Brillouin zone that is also a square is displayed in Fig.~\ref{Rec_construction}(b), as well as special points. The dispersion relation
obtained by direct numerical simulation of the eigenproblem
\eqref{Leigenproblem2} is displayed in Fig.~\ref{Rec_construction}(c) and it
agrees {both qualitatively and quantitatively} very well with the dispersion relation obtained by the formula
\eqref{Dispersion_rectangular_0}{, shown in Fig.~\ref{Rec_construction}(d)}.

From the analytical formula Eq.~\eqref{Dispersion_rectangular_0}, we {readily obtain} 
\begin{align}
\overline{\mathbf{H}}\mu= - 2l^2C\left(\begin{array}{cc}\cos(lk_x)&0\\ 0&\cos(lk_y)\end{array}\right) .
\end{align}
Here $\overline{\mathbf{H}}=\left(\begin{array}{ll} \overline{\partial}_{11} \quad
&\overline{\partial}_{12}\\ \overline{\partial}_{21} \quad &\overline{\partial}_{22}
\end{array}\right)$ is the Hessian matrix operator with respect to
$\mathbf{k}$.

{{Next} we describe the dispersive nature of the system at special symmetry points.}
At the $\Gamma$ point, the Hessian matrix is
\begin{align*}
\overline{\mathbf{H}}\mu = - 2l^2C\left(\begin{array}{ll} 1 \quad &0\\ 0 \quad &1 \end{array}\right).
\end{align*}
So $\Gamma$ is a minimum point. {Furthermore,} since $C<0$ the governing equation of the
envelope is a focusing NLS equation. It is expected that band gap solitons will
bifurcate from this point, see \cite{Yang2007}.

At the $\text{M}$ point, the Hessian matrix is
\begin{align*}
\overline{\mathbf{H}}\mu= - 2l^2C\left(\begin{array}{ll} -1 \quad &0\\ 0 \quad &-1 \end{array}\right).
\end{align*}
So $M$ is a maximum point {since $- C > 0.$} The governing equation of the envelope is a
defocusing NLS equation. There may exist dark solitons.

At the $\text{X}_1$ point, the Hessian matrix is
\begin{align*}
\overline{\mathbf{H}}\mu= - 2l^2C\left(\begin{array}{ll} -1 \quad &0\\ 0 \quad &1 \end{array}\right).
\end{align*}
So $\text{X}_1$ is a saddle point. The governing equation is a {hyperbolic} focusing NLS equation{. We note that  this version of NLS describes deep water waves \cite{AbSe81}.}

At the $\text{X}_2$ point, the Hessian matrix is
\begin{align*}
\overline{\mathbf{H}}\mu= - 2l^2C\left(\begin{array}{ll} 1 \quad &0\\ 0 \quad &-1 \end{array}\right).
\end{align*}
So $\text{X}_2$ is also a saddle point  but with opposite negative and positive
eigen-directions to $\text{X}_1$.

\section{2D {quantum} harmonic oscillator}
\label{harmonic_oscillate}

In this section,  we discuss the case when $V(\mathbf{r})$ is large  and is
locally harmonic at each site, i.e.,
{for $V_0 \gg 1$  we approximate $V(\mathbf{r}) = - V_0  e^{-\hat{k}^2(x^2+y^2)} $ by
$$
V(\mathbf{r})\approx
V_0( \hat{k}^2(x^2+y^2)-1)
$$
as $(x,y)\to (0,0)$, or} $(x,y)\to (x_0,y_0)$ where $(x_0,y_0)$ is the coordinate of an arbitrary
site. We call this the two {quantum} dimensional harmonic oscillator, which
has been well studied. One can find the results in many books on quantum
mechanics (cf. \cite{Griffiths2004}). Below we list some results for
formulae that we have used in the above sections.

The 2D harmonic oscillator is the eigenvalue problem
\[
\left[ \nabla^2
-d_0(x^2+\,y^2)\right] \eta(\mathbf{r})=-\epsilon\eta(\mathbf{r}),
\]
where
$d_0>0$ is called the intensity and $\epsilon$ the energy. This problem can be
solved by separation of variables into  two 1D oscillators, by assuming
$\eta(x,y)=f(x)g(y)$
\[
\left(\frac{d^2}{dx^2} -d_0x^2\right)f(x)=-\epsilon_x f(x),
\]
and
\[
\left(\frac{d^2}{dy^2} -d_0y^2\right)g(y)=-\epsilon_y g(y),
\]
where $\epsilon=\epsilon_x+\epsilon_y$.

Each of the 1D oscillators are solved in terms of Hermite functions; it
follows that $\epsilon_x=\left(1+2 m\right) \sqrt{d_0}, \quad m=0,1,2,\dots $ and
the associated normalized eigenfunctions are
\[
f_m(x)=\sqrt{\frac{1}{2^mm!}}\frac{d_0^{1/8}}{ \pi ^{1/4}}e^{-\frac{\sqrt{d_0}}{2}x^2}H_m(d_0^{1/4}x),
\]
where $H_m(x)$ is the $m$th Hermite polynomial.
Similarly, $\epsilon_y=\left(1+2 n\right) \sqrt{d_0}, \quad n=0,1,2,\dots $ and
the associated normalized eigenfunctions are
\[
g_n(x)=\sqrt{\frac{1}{2^nn!}}\frac{d_0^{1/8}}{ \pi ^{1/4}}e^{-\frac{\sqrt{d_0}}{2}y^2}H_n(d_0^{1/4}y).
\]
So the total eigenvalue is {$\epsilon = \epsilon_{m,n}=\sqrt{d_0}[ (1+2m)+(1+2n)]$} and the associated normalized eigenfunctions are
\begin{align*}
&\eta_{m,n}(x,y)=f_m(x)g_n(y), \\
  & =\sqrt{\frac{1}{2^{m+n}m!n!}}\frac{d_0^{1/4}}{ \pi
^{1/2}}e^{-\frac{\sqrt{d_0}}{2}(x^2+y^2)}H_n(d_0^{1/4}x)H_m(d_0^{1/4}y).
\end{align*}
We note that the above calculations show that the ground state, or lowest
eigenvalue, is simple but the higher ones, e.g the first excited state, {can have} 
eigenvalues that are multiple (note that eigenvalue
$\epsilon_{1,0}=\epsilon_{0,1}$).

Next, we use the above functions to estimate the parameters for the lowest
eigenvalue; i.e., the ground state.  As mentioned above, when $V_0$ is very
large, {an approximation of $V_s(\mathbf{r})$ is
{$V_s(\mathbf{r})=V_0(\hat{k}^2(x^2+y^2)-1) \approx -V_0 e^{-\hat{k}^2(x^2+y^2)} $}}.  Thus the associated orbitals can be approximated
by the wave functions of the harmonic oscillator and the corresponding orbital
energy $E$ is approximated by $E=\epsilon-V_0$. There are two parameters:
$V_0$,  the depth of the potential and $k_0$,  the width of the potential.
{The} 
validity of this approximation is  due to WKB theory. With the above
approximation, $\Delta
V(\mathbf{r})=-V_0\displaystyle{\sum_{\mathbf{v}\neq\mathbf{0}}}e^{-\hat{k}^2\parallel\mathbf{r}-\mathbf{v}\parallel^2}$,
where $\parallel \mathbf{r}\parallel=\sqrt{x^2+y^2}$ is the standard {Euclidean} norm.
Recall 
we have assumed that the position of the first site
$S_{\mathbf{0}}=\mathbf{0}$, so $S_{\mathbf{v}}=\mathbf{v}$. So the square
lattice Eq.~\eqref{square_lattice} has the asymptotic behavior near the first site
$V(\mathbf{r})\approx V_0 \left[ \frac{1}{2}k_0^2(x^2+y^2)-1 \right]$ {and} 
$V_s(\mathbf{r})=-V_0 e^{-\frac{k_0^2}{2}(x^2+y^2)}$. {The behavior near all other sites is merely a translation of this argument.}

As mentioned above, we consider only the lowest band. With the above
approximation, the orbital energy and the orbital are
\begin{align*}
&&E=\epsilon_{0,0}-V_0=2\hat{k}\sqrt{V_0}-V_0,\\
&&\phi(\mathbf{r})=\eta_{0,0}(\mathbf{r})=\frac{(\hat{k}^2V_0)^{1/4}}{\pi
^{1/2}}e^{-\frac{\sqrt{\hat{k}^2V_0}}{2}(x^2+y^2)}.
\end{align*}
{After some further} 
calculations, the parameters in the dispersion relation Eq.~\eqref{Dispersion_nondegenerate} are {found to be}
\begin{align*}
\kappa_\mathbf{v}&=\int
\phi(\mathbf{r})\phi(\mathbf{r}-\mathbf{v})d\mathbf{r}\approx\text{exp}\left(-\frac{1}{4}\hat{k}
\sqrt{V_0}\,\parallel\mathbf{v}\parallel^2\right),\\
\lambda_\mathbf{v}&=\int \phi(\mathbf{r})\Delta
V(\mathbf{r})\phi(\mathbf{r}-\mathbf{v})d\mathbf{r},\\
&\approx-\frac{V_0^{3/2}}{\hat{k}+\sqrt{V_0}}\text{exp}\left(-\frac{\hat{k}\sqrt{V_0}(2\hat{k}+\sqrt{V}_0)}{4(\hat{k}+\sqrt{V}_0)}\parallel\mathbf{v}\parallel^2\right),\\
\lambda_\mathbf{0}&=\int \phi(\mathbf{r})\Delta
V(\mathbf{r})\phi(\mathbf{r})d\mathbf{r}\\
&\approx-\frac{V_0^{3/2}}{\hat{k}+\sqrt{V_0}}\displaystyle{\sum_{ \langle \mathbf{v} \rangle}}\text{exp}\left(-\frac{\hat{k}^2
\sqrt{V_0}}{\hat{k}+\sqrt{V_0}}\;\parallel\mathbf{v}\parallel^2\right).
\end{align*}
Since $C_\mathbf{v}=\lambda_\mathbf{v}-\lambda_\mathbf{0}\kappa_\mathbf{v}$,
\begin{align*}
&C_\mathbf{v}
\approx\frac{V_0^{3/2}}{\hat{k}+\sqrt{V_0}}\exp\left(-\frac{\hat{k}
\sqrt{V_0}}{4}\parallel\mathbf{v}\parallel^2\right)
\\  & \times \Bigg[ \displaystyle{\sum_{ \langle \mathbf{v} \rangle}}\text{exp}\left(-\frac{\hat{k}^2
\sqrt{V_0}}{\hat{k}+\sqrt{V_0}}\parallel\mathbf{v}\parallel^2\right) \\ &   -\text{exp}\left(-\frac{\hat{k}^2
\sqrt{V_0}}{2 (\hat{k}+\sqrt{V_0})}\parallel\mathbf{v}\parallel^2\right)\Bigg].
\end{align*}
For simplicity, we only take the leading order of $C_{\mathbf{v}}$ under the
limit $V_0\gg1$, {and} 
get
\begin{align*}
C_\mathbf{v}\approx V_0~\text{exp}\left(-\frac{\hat{k}
\sqrt{V_0}}{4}\parallel\mathbf{v}\parallel^2\right) & \Bigg[ \displaystyle{\sum_{ \langle \mathbf{v} \rangle}}\text{exp}\left(-\hat{k}^2
\parallel\mathbf{v}\parallel^2\right) \\ & -\text{exp}\left(-\frac{\hat{k}^2
}{2}\parallel\mathbf{v}\parallel^2\right)\Bigg].
\end{align*}
As $V_0$ goes to infinity, $C_\mathbf{v}$ goes to zero exponentially with
respect to $V_0${,} while $\lambda_\mathbf{0} \sim O(V_0)$ goes to negative
infinity,  $E+\lambda_{\mathbf{0}}\to -\infty$. Since we consider a  square
lattice Eq.~\eqref{square_lattice} {with lattice period $\ell$ and} $\hat{k}^2=\frac{1}{2}k_0^2${, it follows that} 
{\begin{align*}
C_{\mathbf{v}_1}=C_{\mathbf{v}_2} \approx V_0\text{exp}\left(-\frac{k_0\sqrt{V_0}\;l^2}{4\sqrt{2}}\right) & \Bigg[ 4~\text{exp}\left(-\frac{k_0^2l^2}{2}\right) \\ & -\text{exp}\left(-\frac{k_0^2l^2}{4}\right)\Bigg].
\end{align*}}
Note that $k_0l=\pi$, so
{\begin{equation*}
C_{\mathbf{v}_1}=C_{\mathbf{v}_2}\approx -0.056V_0\text{exp}\left(-\frac{\sqrt{V_0}\;\pi^2}{4\sqrt{2}k_0}\right).
\end{equation*}}

\section{{A typical non-simple lattice--honeycomb lattice}}
\label{honeycomb_sec}

As mentioned earlier,  a non-simple lattice may have more than one site, i.e.,
one minima, in a unit cell. In this case one may need more than one initial
site to describe the lattice. An example of a non-simple lattice is {the} 
honeycomb
lattice. The right-hand lattice in Fig.~\ref{lattices} is a non-simple
honeycomb lattice. {It's} sites {(potential minima)} consist of `black' and `white' sites. The black
and white sites are separately  constructed from the underlying primitive
vectors. Hence, we need two initial sites to describe the honeycomb lattice.

A perfect hexagonal lattice is composed of two standard triangular sublattices:
A and B sublattices. The lattice vectors $\mathbb{P}$ should form a triangular
lattice. To generate the other sublattice, extra information is needed to
determine the shift from the B site to the A site in the same unit cell. We
denote this shift as a vector $\mathbf{d}_1$. 
{The lattice vectors are given by ${\bf v}_1$ and ${\bf v}_2$,  and a displacement between adjacent A and B sites is}
$\mathbf{d}_1=-\frac{1}{3}(\mathbf{v}_1+\mathbf{v}_2)$. We also introduce two
other vectors. $\mathbf{d}_2=\mathbf{v}_2+\mathbf{d}_1$
$\mathbf{d}_3=\mathbf{v}_1+\mathbf{d}_1$. The vectors and their relations are
shown in Fig.~\ref{lattice_construction}. 
By connecting all the nearest
neighbors, a perfect hexagonal lattice is obtained. It is noted that all A
(filled-black) form a triangular sublattice and all 
B (open-white) sites form the other triangular sublattices. The distance between two nearest A sites
or two nearest B sites {(next-nearest neighbors)} is $l$. However, the nearest neighbors of A sites are
three B sites {are a distance $l / \sqrt{3}$ apart} and the shifts are determined by $\mathbf{d}_1, \mathbf{d}_2$ and
$\mathbf{d}_3$.

\begin{figure}
\includegraphics*[width=0.4\textwidth]{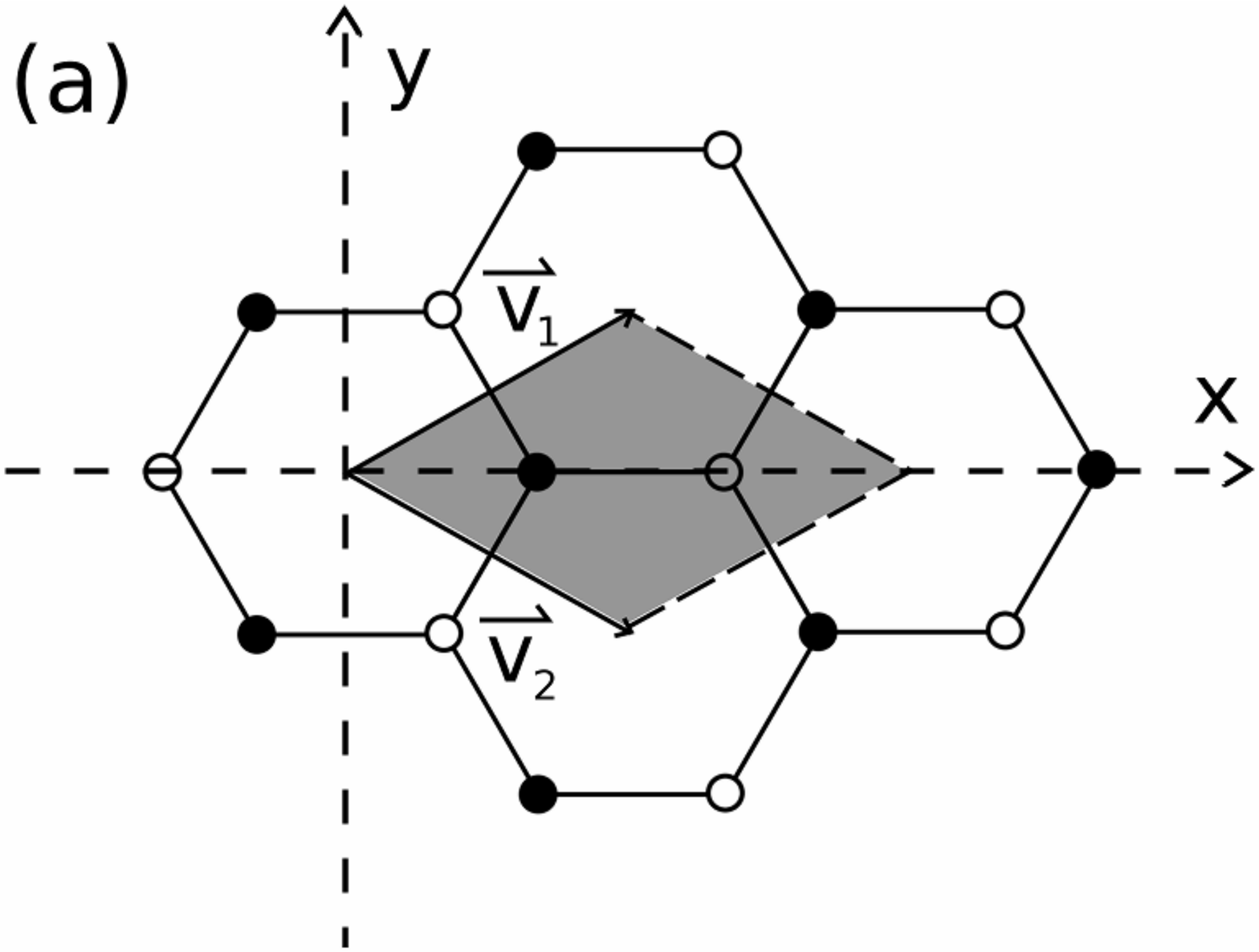}
\includegraphics*[width=0.4\textwidth]{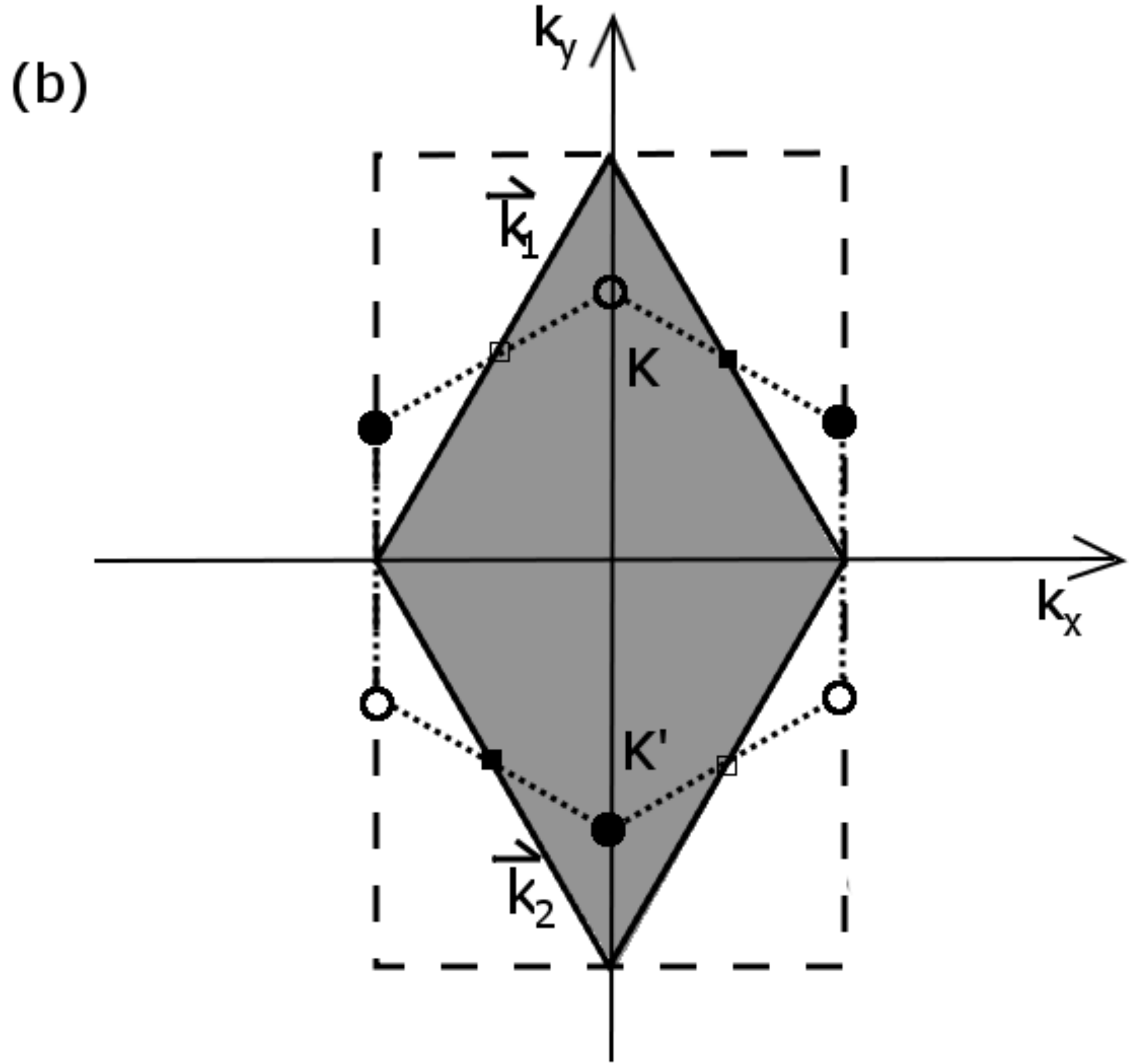}
\includegraphics*[width=0.4\textwidth]{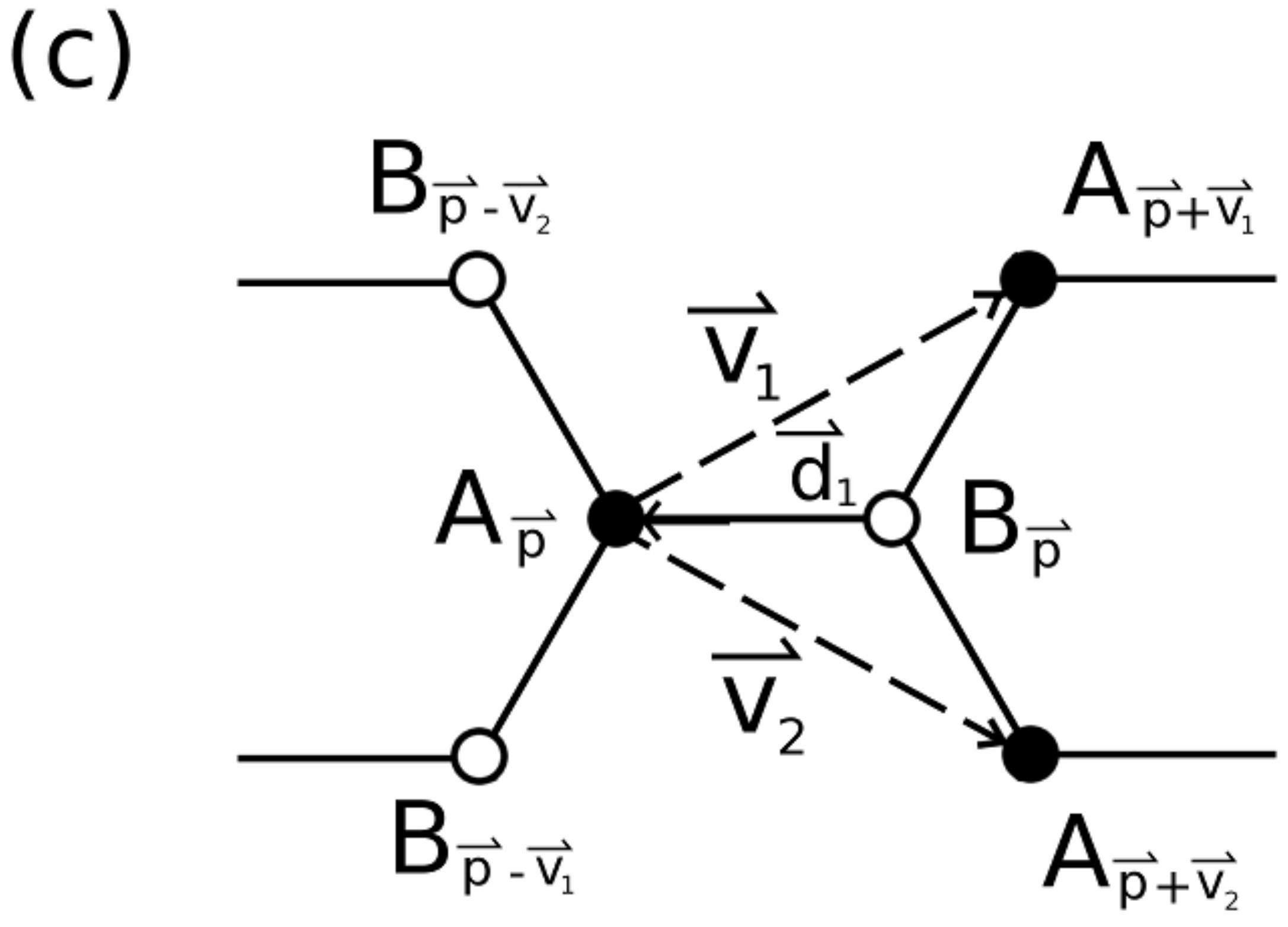}
\caption{\label{lattice_construction} The hexagonal lattice (a) and the extended Brillouin zone (b). {Shaded} 
regions {in (a) and (b)} are the unit cell $\Omega$ and {reciprocal unit cell} 
$\Omega'${, respectively}. The construction of the A (filled-black) and B (open-white)
sublattices is also depicted (c).} 
\end{figure}

A honeycomb lattice can be constructed by three interfering plane waves
\begin{equation}\label{potential}
V(\mathbf{r})=\frac{V_0}{9}\left(\left|e^{ik_0\mathbf{b}_1\cdot\mathbf{r}}+e^{ik_0\mathbf{b}_2\cdot\mathbf{r}}+e^{ik_0\mathbf{b}_3\cdot\mathbf{r}}\right|^2-9\right),
\end{equation}
where {$0 \ge V({\bf r}) \ge - V_0$ and} $\mathbf{b}_1=(0,1)$,
$\mathbf{b}_2=(-\frac{\sqrt{3}}{2},-\frac{1}{2})$ and
$\mathbf{b}_3=(\frac{\sqrt{3}}{2},-\frac{1}{2})$; $V_0>0$ is the
lattice intensity; $k_0$ is the scaled wave length of the interfering plane waves.  The characteristic vectors for
this
potential are
\begin{align*}
\mathbf{v}_1=l\left(\frac{\sqrt{3}}{2},\frac{1}{2}\right),\quad\quad
\mathbf{v}_2=l\left(\frac{\sqrt{3}}{2},-\frac{1}{2}\right);\\
\mathbf{k}_1=\frac{4\pi}{\sqrt{3}l}\left(\frac{1}{2},\frac{\sqrt{3}}{2}\right),\quad\quad
\mathbf{k}_2=\frac{4\pi}{\sqrt{3}l}\left(\frac{1}{2},-\frac{\sqrt{3}}{2}\right),
\end{align*}
where $l=\frac{4\pi}{3k_0}$ {and ${\bf v}_i \cdot {\bf k}_j = 2 \pi  \delta_{ij}$ The lattice in Eq.~(\ref{potential}) has the periodicity $V({\bf r} + m {\bf v}_1 + n {\bf v}_2) = V({\bf r})$ for any $m,n \in \mathbb{Z}$}.

As earlier, the dispersion relation is determined from Eq.~\eqref{Leigenproblem}. For a honeycomb lattice it is convenient to write the
potential  in the form
\begin{equation}
\label{HC_potential}
V(\mathbf{r})=\sum_\mathbf{v}[ V_A(\mathbf{r}-\mathbf{v})+V_B(\mathbf{r}-\mathbf{v})]
\end{equation}
where $V_A$, $V_B$ denote the potentials generated from the two sites in the primitive unit cell. 
In the tight-binding approximation they have sharp minima near the A and B sites, respectively;  the sum over $\mathbf{v}$ means
$\mathbf{v}$ takes all values in $\mathbb{P}$, i.e., $\mathbf{v}=m\mathbf{v}_1+n\mathbf{v}_2$, for all
$m,n\in\mathbb{Z}$.  The Bloch mode {is assumed to take} 
the form
\[
\varphi(\mathbf{r};\mathbf{k})=\displaystyle{a\sum_{\mathbf{v}}\phi_A(\mathbf{r}-\mathbf{v})e^{i\mathbf{k}\cdot\mathbf{v}}+b\sum_{\mathbf{v}}\phi_B(\mathbf{r}-\mathbf{v})e^{i\mathbf{k}\cdot\mathbf{v}}}
\]
where  $\phi_A(\mathbf{r})$ and $\phi_B(\mathbf{r})$ represent an orbital (i.e., Wannier function) of a single $V_A$ or
$V_B$ potential{,} respectively; they have the same eigenvalue denoted as $E$. That is to say,
\begin{equation}
\label{HC_orbital}
\left[ \nabla^2 -V_j(\mathbf{r})\right] \phi_j(\mathbf{r})=-E\phi_j(\mathbf{r}),
\end{equation}
where $j$ is A or B. Here we only consider the lowest band {energy level}, so there is no
subindex to {denote} 
different bands. We also assume $\phi_A$ and $\phi_B$ are
real and normalize them with norm 1, i.e.,
{$\int\phi_A^2 
d\mathbf{r}=\int\phi_B^2 
d\mathbf{r}=1$.} It is convenient
to introduce the notation $$\Delta
V_j(\mathbf{r} )=V(\mathbf{r})-V_j(\mathbf{r} )=\displaystyle{\sum_{\mathbf{v}\neq\mathbf{0}}\left[ V_j(\mathbf{r}-\mathbf{v})+V_l(\mathbf{r}-\mathbf{v})\right]+V_l(\mathbf{r})},
 $$ for $l\neq j.$ Again, we consider the tight-binding limit, i.e., $V_0\gg1$ which
means the potential well at each site is very deep, and only on-site and
nearest neighbor interactions will need to be considered \cite{Fefferman2018}.

\subsection{{Dispersion Relation}}

As in the simple lattice case, we first determine the dispersion relation. We
can use Fredholm alternative conditions or equivalently the following method. 
Substituting the above Bloch mode $\varphi(\mathbf{r};\mathbf{k})$
into the eigenproblem Eq.~\eqref{Leigenproblem}{, and applying the orbital relation in Eq.~(\ref{HC_orbital}),} we get
\begin{align}\label{dis_mu}
\displaystyle{\sum_\mathbf{v}}\left[ (\mu-E)[a\phi_A(\mathbf{r}-\mathbf{v})\right.&+b\phi_B(\mathbf{r}-\mathbf{v}) ]-a\Delta
V_A(\mathbf{r}-\mathbf{v}) \phi_A(\mathbf{r}-\mathbf{v})\nonumber\\ &- \left. b\Delta V_B(\mathbf{r}-\mathbf{v})
\phi_B(\mathbf{r}-\mathbf{v})\right]e^{i\mathbf{k}\cdot\mathbf{v}}=0.
\end{align}
Multiplying $\phi_j(\mathbf{r}), j=A,B$ to 
Eq.~\eqref{dis_mu} and integrating over the whole plane, we get {the} 
matrix eigenvalue problem,
\begin{align}\label{Matrixpro}
& \left(\begin{array}{cc}\mu-E-c_0& [(\mu-E)c_1-c_2] \gamma(\mathbf{k})\\ \left[ (\mu-E)c_1-c_2 \right] \gamma^*(\mathbf{k})&\mu-E-c_0\end{array}\right)\left(\begin{array}{c}a\\b\end{array}\right) \\ \nonumber
&=\left(\begin{array}{c}0\\0\end{array}\right).
\end{align}
Here only on-site and nearest neighbor interactions are considered because of the tight-binding limit; and
\begin{align*}
&\gamma(\mathbf{k})=(1+e^{-i\mathbf{k}\cdot\mathbf{v}_1}+e^{-i\mathbf{k}\cdot\mathbf{v}_2})\\
&c_0=\int\phi_A({\bf r}) \Delta
V_A({\bf r} ) \phi_A({\bf r} ) d\mathbf{r}=\int\phi_B({\bf r}) \Delta
V_B({\bf r} ) \phi_B({\bf r} ) d\mathbf{r};\\
&c_1 \left(|| {\bf v}|| \right)=\int\phi_A({\bf r}) \phi_B({\bf r} - {\bf v}) d\mathbf{r}=\int\phi_B({\bf r}) \phi_A({\bf r} - {\bf v}) d\mathbf{r};\\
&c_2 \left(|| {\bf v}|| \right) =\int\phi_A({\bf r})\Delta
V_B({\bf r} - {\bf v}) \phi_B({\bf r} - {\bf v}) d\mathbf{r} \\ & =\int\phi_B({\bf r}) \Delta
V_A({\bf r} - {\bf v}) \phi_A({\bf r} - {\bf v}) d\mathbf{r}.
\end{align*}
It is noted that $c_0,c_1,c_2$ are all real and {$c_1$ and $c_2$} 
are very small{. Moreover, the coefficients $c_1$ and $c_2$ are functions of the distance $|| {\bf v}||$, and as a result, each nearest neighbor term is identical, up to a phase.}
Note we
can use $V_B(\mathbf{r})=V_A(\mathbf{r}-\mathbf{d}_1)$ and
$\phi_B(\mathbf{r})=\phi_A(\mathbf{r}-\mathbf{d}_1)$ in these calculations.

The 
system {in} Eqs.~\eqref{Matrixpro} has non-trivial solutions if and only
if the determinant is zero{. The dispersion relation that follows is} 
\[
\mu(\mathbf{k})-E-c_0=\pm | ( \mu(\mathbf{k})-E)c_1+c_2 | \cdot   |\gamma(\mathbf{k})|.
\]
Since {$c_1, c_2 \ll1$}, 
the above dispersion relation becomes (higher order terms are omitted)
\[
\mu(\mathbf{k}) \approx E+c_0 \pm  C|\gamma(\mathbf{k})|,
\]
where $C= c_0c_1 - c_2$. Since the {asymptotic} behavior of the honeycomb
potential Eq.~\eqref{potential} near the site is
$V(\mathbf{r})\approx-V_0(\frac{1}{4}k_0^2((x-x_0)^2+(y-y_0)^2)-1)$, we can
find that {$C=-0.297V_0e^{-\frac{2\sqrt{V_0}\pi^2}{9k_0}} < 0$} with the same
approximations we used in the simple lattice case. 

A typical dispersion surface containing the {two} lowest {spectral bands} 
is depicted in the
left hand Fig. \ref{HCband}; an intensity plot of a hexagonal lattice is given
in the right side. {The touching points, also referred to as Dirac points, correspond to the zeros of $\gamma({\bf k})$.} 

It is also known that 
material graphene has honeycomb lattice structure. In
the graphene literature, it has been shown that two different energy bands can
touch each other at certain isolated points that are called Dirac points; such
Dirac points are {sometimes} termed diabolical points
\cite{Berry_Jeffrey06,Berry_Jeffrey}.  Thus {Dirac} points also exist in
the band structure of two-dimensional honeycomb lattices. The tight-binding
approximation is often used in the study of graphene and it is found that
structure of the dispersion relation near these Dirac points is conical in
nature \cite{Wallace,Novoselov}; the regions in the neighborhood of Dirac
points are called Dirac cones. 

\begin{figure}
\includegraphics*[width=0.2\textwidth]{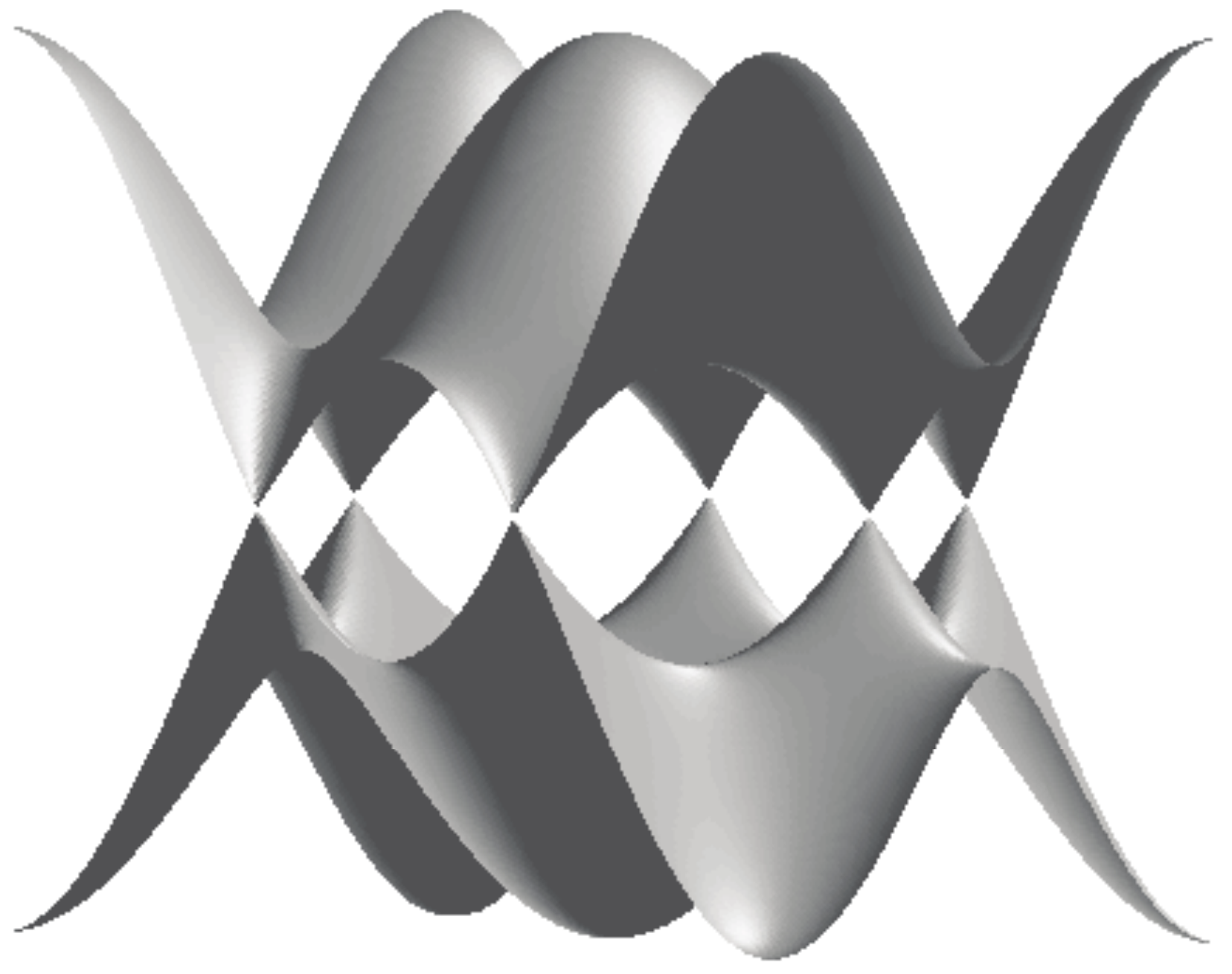}
\includegraphics*[width=0.25\textwidth]{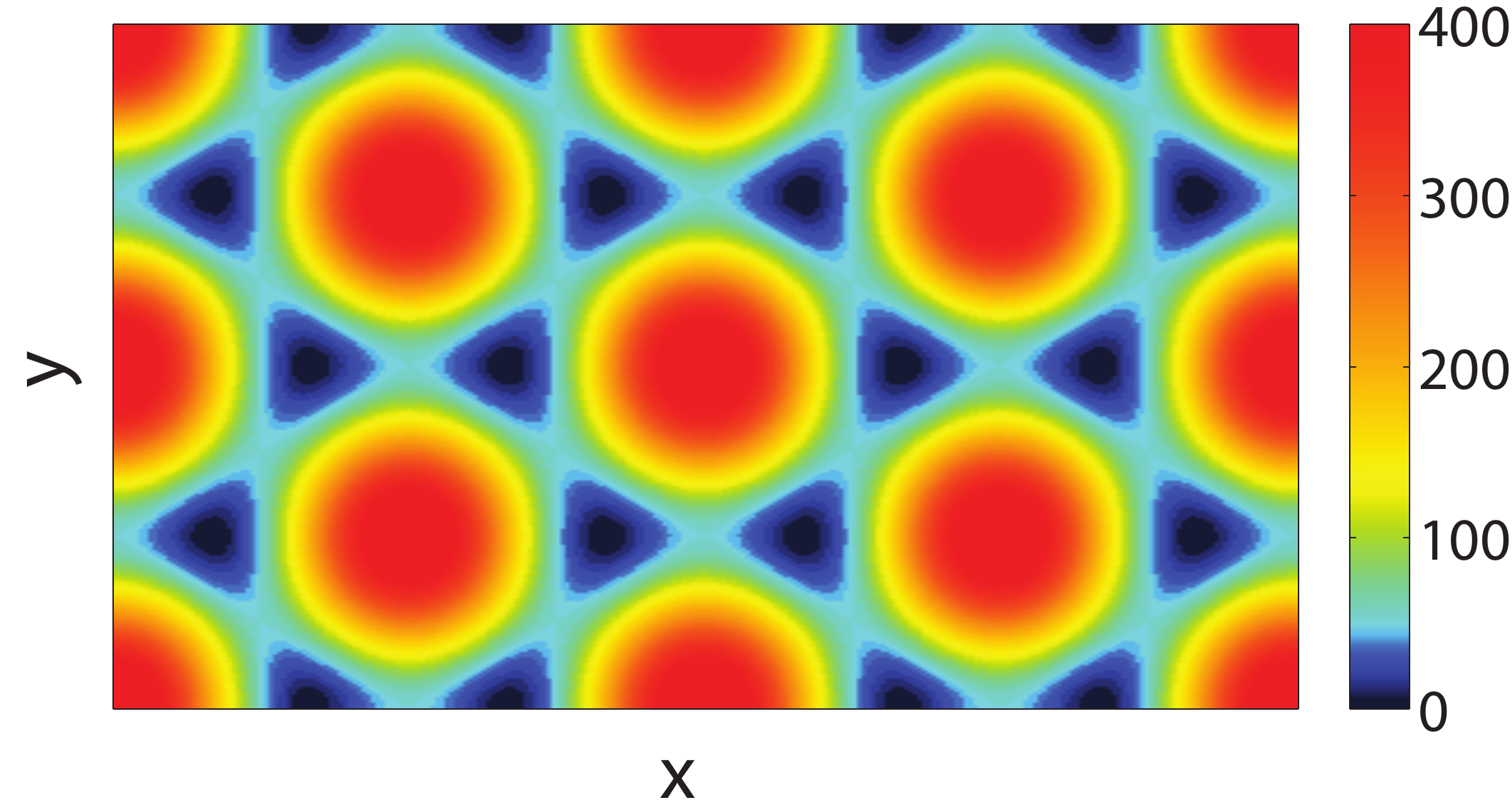}
\caption{Left: the {two} lowest {spectral} bands 
of a typical honeycomb lattice. {The touching points are the Dirac points} {\cite{FW2012}}. Right: an intensity plot {of potential (\ref{potential}).} The local minima (blue)
are identified as sites. The local maxima are at the centers of a triangular lattice of  hexagons.}
\label{HCband}
\end{figure}

Note that $\gamma(\mathbf{k})$ is periodic {in} 
$\mathbf{k}$. In one reciprocal
unit cell, there are two zeros {known as Dirac points}, which we denote $K$ and $K'$. For the above
special potential Eq.~\eqref{potential}, the location of the Dirac points are 
\begin{equation}
\label{dirac_pts}
K=\left( 0,\frac{4\pi}{3l}\right) ~~\text{and} 
~~~~~ K'=\left(0,-\frac{4\pi}{3l}\right).
\end{equation} All the zeros of $\gamma(\mathbf{k})$ form the
reciprocal hexagonal lattice{, which also happens to be the Brillouin zone}. At these points, $\mu-E-c_0=0$ and so the matrix
in Eq.~\eqref{Matrixpro} is identically equal zero. Thus, $a$ and $b$ are free.
Thus the eigenspace is two dimensional. The associated original linear
Schr\"odinger eigenproblem has degeneracy. In other words, when $\mu=E+c_0$,
the eigenproblem Eq.~\eqref{Leigenproblem} has two independent Bloch modes.

\subsection{{Envelope Dynamics}}
Suppose we input a Bloch wave envelope into the crystal. To 
leading order{,} the envelope {is taken to vary} 
slowly along $z$,
\begin{align}\label{envelope}
\psi
\sim \Big( \sum_{\mathbf{v}}a_\mathbf{v}(Z)\phi_A(\mathbf{r}-\mathbf{v})e^{i\mathbf{k}\cdot\mathbf{v}}  + \sum_{\mathbf{v}}b_\mathbf{v}(Z)\phi_B(\mathbf{r}-\mathbf{v})e^{i\mathbf{k}\cdot\mathbf{v}}\Big)e^{-i\mu
z}.
\end{align}
Since $\psi$ is not a Bloch mode anymore, the intensities are different at
different sites, i.e., $a$ and $b$ have subindex $\mathbf{v}$ that are sites
on the A,B lattices, respectively, and $Z=\varepsilon z$; the small parameter
$\varepsilon$ will be determined later.

Substituting the envelope solution Eq.~\eqref{envelope} into the lattice NLS Eq.~\eqref{LNLS}, one obtains
\begin{align}\label{envelope2}
&\sum_{\mathbf{v}}\Big(\varepsilon i\frac{da_\mathbf{v}}{dZ} + a_{\bf v} \Big[\nabla^2
-V(\mathbf{r}-\mathbf{v})\Big] \phi_A(\mathbf{r}-\mathbf{v}) \nonumber  \\ &   +\mu a_{\bf v}
\phi_A(\mathbf{r}-\mathbf{v})\Big)e^{i\mathbf{k}\cdot\mathbf{v}}  \nonumber
\\&+ \sum_{\mathbf{v}}\Big(\varepsilon
i\frac{db_\mathbf{v}}{dZ}+ b_{\bf v}\Big[\nabla^2 -V(\mathbf{r}-\mathbf{v})\Big]\phi_B(\mathbf{r}-\mathbf{v})\nonumber \\ & +\mu b_{\bf v}
\phi_B(\mathbf{r}-\mathbf{v})\Big)e^{i\mathbf{k}\cdot\mathbf{v}} \nonumber
\\&+
\sigma \left({\sum_{\mathbf{v}}\left(a_\mathbf{v}(Z)\phi_A(\mathbf{r}-\mathbf{v})+b_\mathbf{v}(Z)\phi_B(\mathbf{r}-\mathbf{v})\right)e^{i\mathbf{k}\cdot\mathbf{v}}}\right)^2\nonumber 
\\&\times\left({\sum_{\mathbf{v}}\left(a_\mathbf{v}(Z)\phi_A(\mathbf{r}-\mathbf{v})+b_\mathbf{v}(Z)\phi_B(\mathbf{r}-\mathbf{v})\right)e^{i\mathbf{k}\cdot\mathbf{v}}}\right)^*=0.
\end{align}
To simplify the steps, rather than employing Fredholm conditions, we can do the
following. Multiply
$\phi_j(\mathbf{r}-\mathbf{p})e^{-i\mathbf{k}\cdot\mathbf{p}}, j=A,B,$ where
$\mathbf{p}\in \mathbb{P},$  to 
Eq.~\eqref{envelope2} and integrate
over the whole plane to get
\begin{align*}
&\varepsilon i\frac{da_\mathbf{p}}{dZ}+(\mu-E-c_0)a_\mathbf{p}+[(\mu-E)c_1-c_2] \mathcal{L}_1b_\mathbf{p}+\sigma
g|a_\mathbf{p}|^2a_\mathbf{p} \\ \nonumber & =0;\\
&\varepsilon i\frac{db_\mathbf{p}}{dZ}+(\mu-E-c_0)b_\mathbf{p}+[(\mu-E)c_1-c_2] \mathcal{L}_2a_\mathbf{p}+\sigma
g|b_\mathbf{p}|^2b_\mathbf{p} \\ \nonumber & =0,
\end{align*}
where
\begin{gather*}
\mathcal{L}_1b_\mathbf{p}=b_\mathbf{p}+b_{\mathbf{p}-\mathbf{v}_1}e^{-i\mathbf{k}\cdot\mathbf{v}_1}+
b_{\mathbf{p}-\mathbf{v}_2}e^{-i\mathbf{k}\cdot\mathbf{v}_2},\\
\mathcal{L}_2a_\mathbf{p}=a_\mathbf{p}+a_{\mathbf{p}+\mathbf{v}_1}e^{i\mathbf{k}\cdot\mathbf{v}_1}
+a_{\mathbf{p}+\mathbf{v}_2}e^{i\mathbf{k}\cdot\mathbf{v}_2},
\end{gather*}
and $g=\int\phi_A^4d\mathbf{r}=\int\phi_B^4d\mathbf{r}$. {Recall that $c_1 = c_1(||{\bf v}||)$ and $c_2 = c_2(||{\bf v}||)$, that is the coefficients are functions of distance with respect to a displacement vector ${\bf v}$.}

Away from the Dirac points{,} the situation is essentially the same as in the
simple lattice. Here the determinant  of the system  Eqs.~\eqref{Matrixpro} is
nonzero and $a_\mathbf{p}$ is proportional to $b_\mathbf{p}$ and the equations
reduce to those discussed earlier in the simple lattice case. So, next we only
consider  the case when we are near Dirac points, so {that} $\mathbf{k}$ takes the
value near $K$, for example. At that point, considering $\mu-E-c_0=0$, the
envelope equation is,  after rescaling (recall $C<0$),
\begin{align}
&&i\frac{da_\mathbf{p}}{dZ}+\mathcal{L}_1b_\mathbf{p}+s(\sigma)|a_\mathbf{p}|^2a_\mathbf{p}=0;\label{Dis_Dirac1}\\
&&i\frac{db_\mathbf{p}}{dZ}+\mathcal{L}_2a_\mathbf{p}+s(\sigma)|b_\mathbf{p}|^2b_\mathbf{p}=0,\label{Dis_Dirac2}
\end{align}
where we have taken $\varepsilon\sim O(|C|)\sim O(|\sigma|)$ to ensure maximal
balance and again $s(\sigma)$ is the sign of $\sigma${,} or zero if there is no
nonlinearity. The 
system \eqref{Dis_Dirac1}--\eqref{Dis_Dirac2} is what we
refer to as the discrete Dirac system.

\subsection{{Continuum Reduction}}
Next we consider the continuous limit; {i.e.,} 
we assume the lattice constant $l$
is much smaller than the characteristic scale of the envelope.  
Denote
$a(\mathbf{R})$ and $b(\mathbf{R})$ as the continuous envelopes where
$\mathbf{R}=(X,Y)=\nu \mathbf{r}$, $\nu\ll1$. Then after some  
{expansions at the Dirac point ${\bf k} = K$, similar to Eq.~(\ref{Taylor_Expan}), we obtain} 
$\mathcal{L}_1\approx \frac{\nu\sqrt{3}l}{2}(\partial_X+i\partial_Y) $ and
$\mathcal{L}_2\approx \frac{\nu\sqrt{3}l}{2}(-\partial_X+i\partial_Y) $. {Note that expanding around the other Dirac point, ${\bf k} = K'$, results in the {conjugate} 
system with $\mathcal{L}_1 \approx \frac{\nu\sqrt{3}l}{2}(\partial_X- i\partial_Y)$ and $\mathcal{L}_2\approx \frac{\nu\sqrt{3}l}{2}(-\partial_X- i\partial_Y).$} Thus 
the discrete system {near ${\bf k} = K$} becomes the following continuous Dirac  system (after
rescaling)
\begin{align}
&&i\frac{da}{dZ}+(\partial_X-i\partial_Y)b+s(\sigma)|a|^2a=0;\label{C_Dirac1}\\
&&i\frac{db}{dZ}+(-\partial_X-i\partial_Y)a+s(\sigma)|b|^2b=0,\label{C_Dirac2}
\end{align}
where we have taken $\varepsilon\sim O(|C|\nu)\sim O(|\sigma|)$ to ensure the
maximal balance. 
The continuous Dirac system governs
broad envelopes of Bloch modes {with quasimomentum ${\bf k} = K$} propagating in the honeycomb lattice. If the
envelope is not wide, i.e., not slowly varying in the transverse direction, the
discrete system is more appropriate than the continuous system to describe the
envelope evolution. If the envelope is very wide, both discrete and continuous
systems are satisfactory, but the continuous system is simpler to use. {Finally, {observe that} combining the linearized version of system (\ref{C_Dirac1})-(\ref{C_Dirac2}) yields the 2D wave equation
$$ \frac{\partial^2 a}{ \partial {Z}^2}  = \frac{\partial^2 a}{ \partial {X}^2} + \frac{\partial^2 a}{ \partial {Y}^2} ,$$
with wave speed $1$.}

\begin{figure}
\includegraphics*[width=0.45\textwidth]{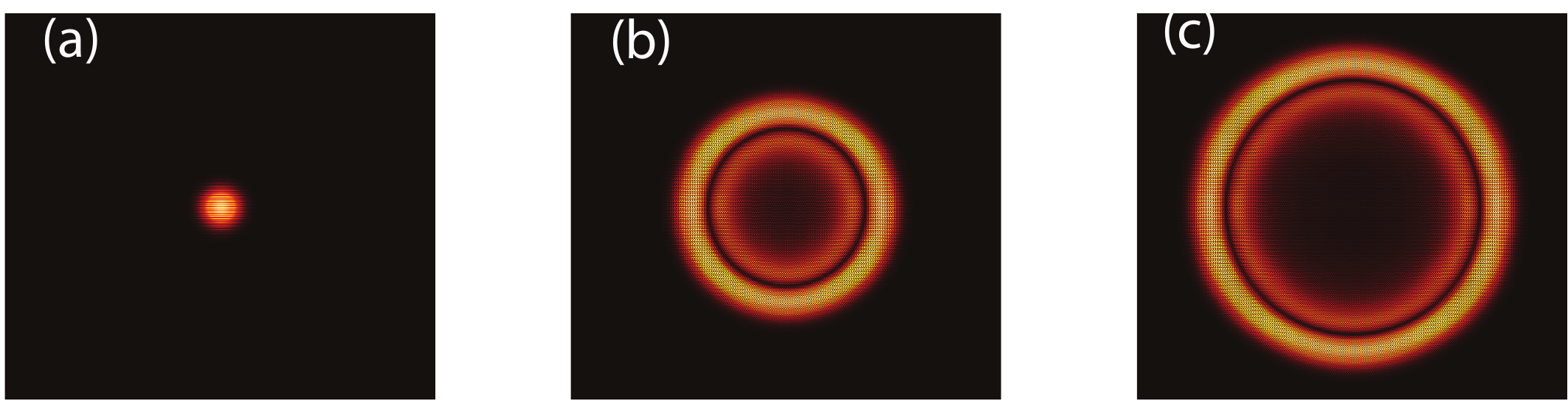}\\
\includegraphics*[width=0.45\textwidth]{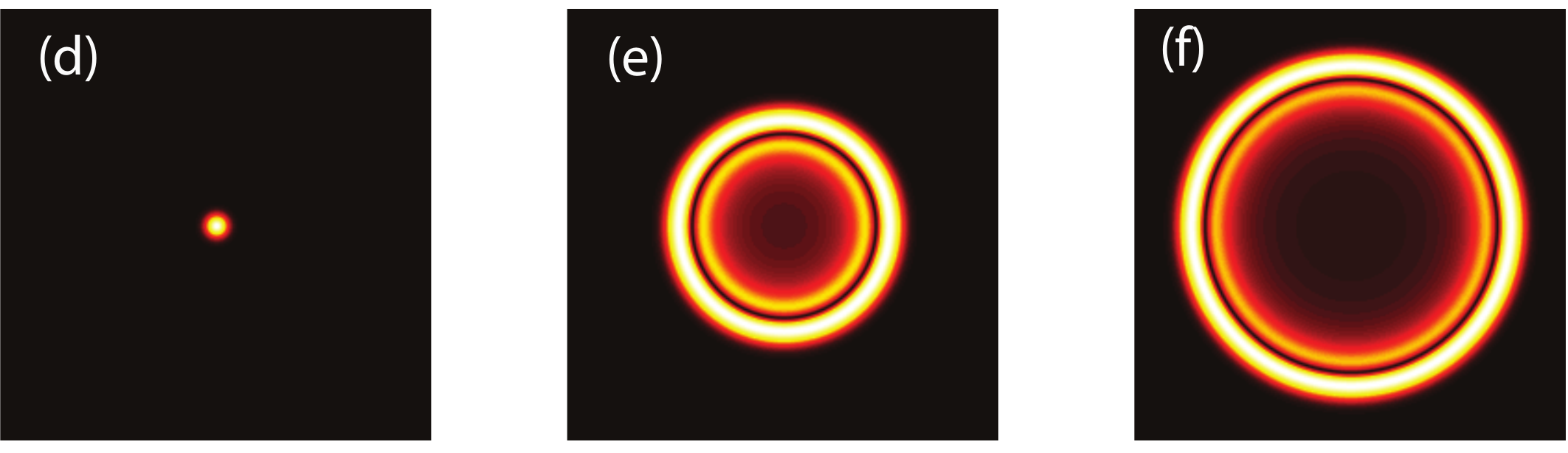}
\caption{\label{conical_figs}The propagation of the {magnitude} 
of a Gaussian Bloch mode envelope associated with a
Dirac
point. Top: simulations of the lattice NLS Eq.~\eqref{LNLS}; Bottom: simulations of the Dirac
Eqs.~\eqref{C_Dirac1}
and~\eqref{C_Dirac2}; here only the $A$-component is shown. Reprinted figure with permission from \cite{AZ10}, copyright (2010) by the American Physical Society.}
\end{figure}

We can compare typical numerical simulations of both lattice NLS equation and
the Dirac system. The comparison between {magnitudes} 
is displayed in Fig.~\ref{conical_figs}. The top panel is from the lattice NLS equation and the
bottom panel is from the Dirac system. From the top panel, we see that a spot
becomes two rings that separated by a  dark ring. The simulation of the Dirac
system gives an excellent match. Thus the Dirac system is a good model to
describe the envelope of Bloch modes near a Dirac point propagating in a
perfect hexagonal lattice. {The system (\ref{Dis_Dirac1}-\ref{Dis_Dirac2}) was
originally  found in \cite{Ablowitz2009a}.}

Thus the existence of Dirac points shows us that certain envelopes associated
with the underlying Bloch modes propagate in an interesting manner: an input
spot becomes two expanding bright rings as the beam propagates in the crystal.
This phenomenon is called conical diffraction
\cite{Berry_Jeffrey06,Berry_Jeffrey} {and} 
is a fundamental feature of crystal
optics and is of interest in mathematics and physics.  It was first predicted
by W. Hamilton \cite{Hamilton} in 1832 and observed by H. Lloyd \cite{Lloyd}
in a biaxial crystal soon afterwards; here a narrow beam entering a crystal
spreads into a hollow cone within the crystal. The existence of the conical
diffraction phenomenon in the light beam propagation in honeycomb lattices was
demonstrated both experimentally and numerically {in} 
\cite{Segev07prl,Segev08OPL}. The theoretical explanation was given {shortly thereafter} \cite{Ablowitz2009a}.

\begin{figure}
\centering
\includegraphics*[width=0.2\textwidth]{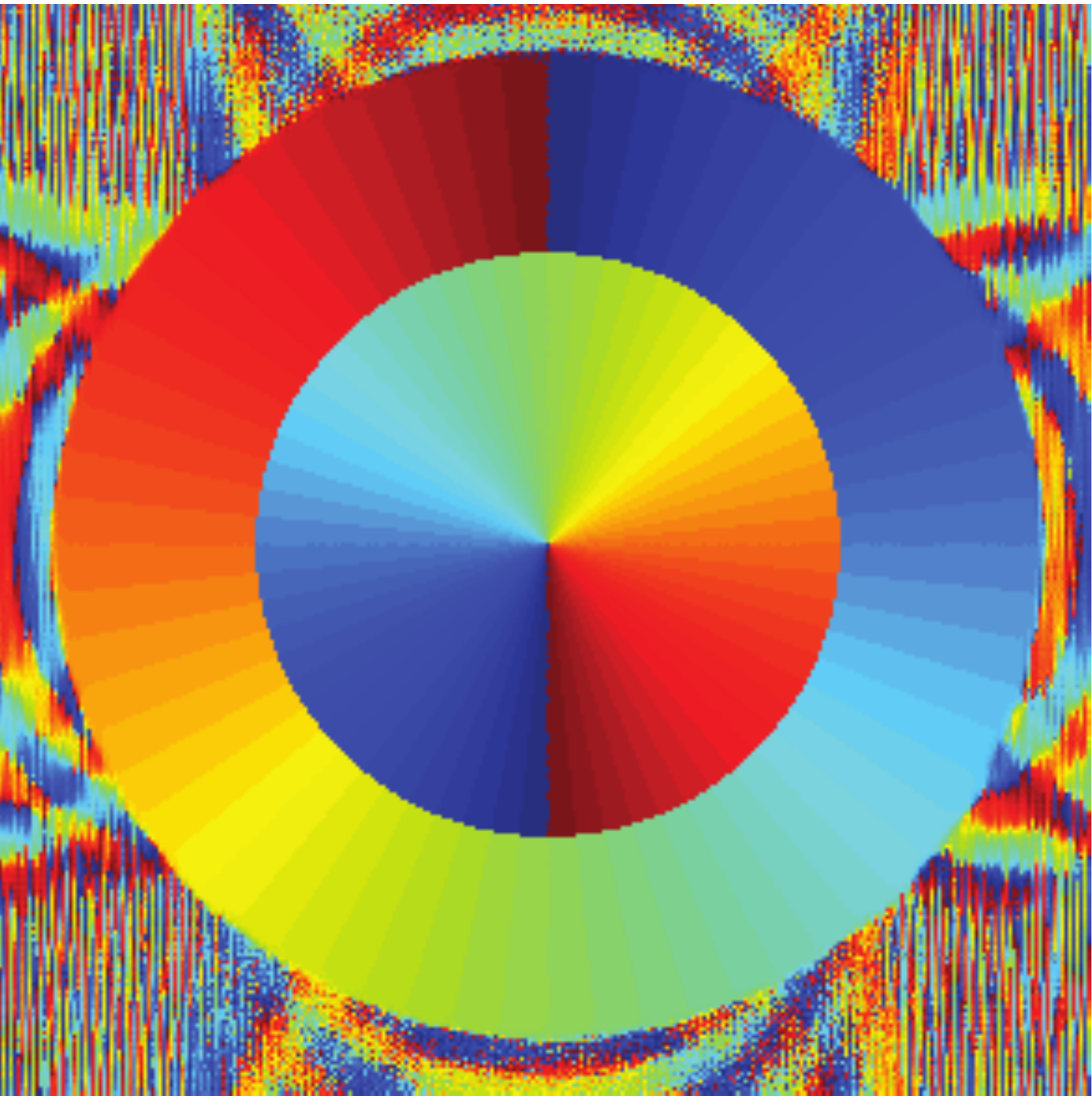}
\includegraphics*[width=0.2\textwidth]{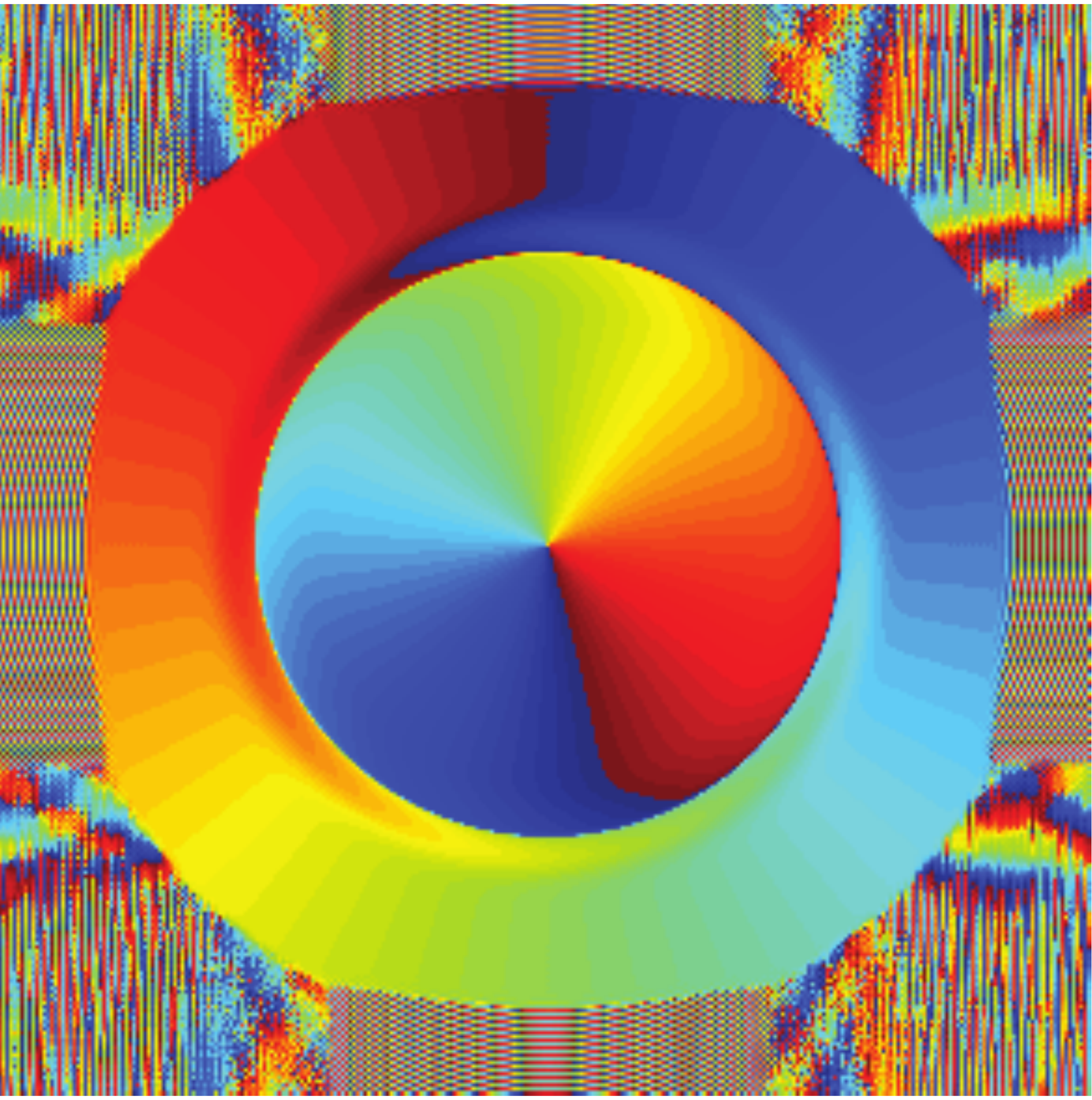}
\caption{Left: phase structure of amplitude A in linear Dirac system; Right: phase structure of the amplitude A in the
nonlinear system. Reprinted figure with permission from \cite{Ablowitz2009a}, copyright (2009) by the American Physical Society.}
\label{diracphase}
\end{figure}


{We} mention that for both the linear and nonlinear lattices the
evolution of the {magnitude (recall Fig.~\ref{conical_figs})} 
is similar and we observe conical refraction.
However, there is some difference in the phase structure--here we {used} amplitude
A. This {is} indicated in Fig.~\ref{diracphase} 
where the left figure is
associated with a linear lattice and the right figure a nonlinear lattice ({see \cite{Ablowitz2009a,AZ10,AZ11}).
When the honeycomb lattice is deformed{,} then we can have elliptical and even straight line diffraction \cite{AZ13A}.
The system of envelope equations changes significantly when one considers shallow lattices \cite{AZ12}.}

\section{{Topological Insulator Systems}}
\label{TI_sec}

{Within the framework of the lattice waveguides described above, it is possible to realize topological insulator systems. Generally speaking, topological insulators behave as insulators (forbid flow of energy) in the bulk or interior of a medium, but act as conductors (allow flow of energy) along the edge or surface. Localized states, called edge modes, decay exponentially fast perpendicular to the medium boundary and propagate parallel to it \cite{Fefferman2016}; {see e.g. Fig. \ref{TI_cartoon}.}
Moreover, these edge states {can be} 
associated with topological invariants. In the case of a nontrivial topological invariant, the bulk-edge correspondence implies the existence of topologically-protected modes.
 These modes tend to be unusually robust and  retain their form, even when {they propagate into/around}
 a material defect.

\begin{figure}
\centering
\includegraphics[scale=.5]{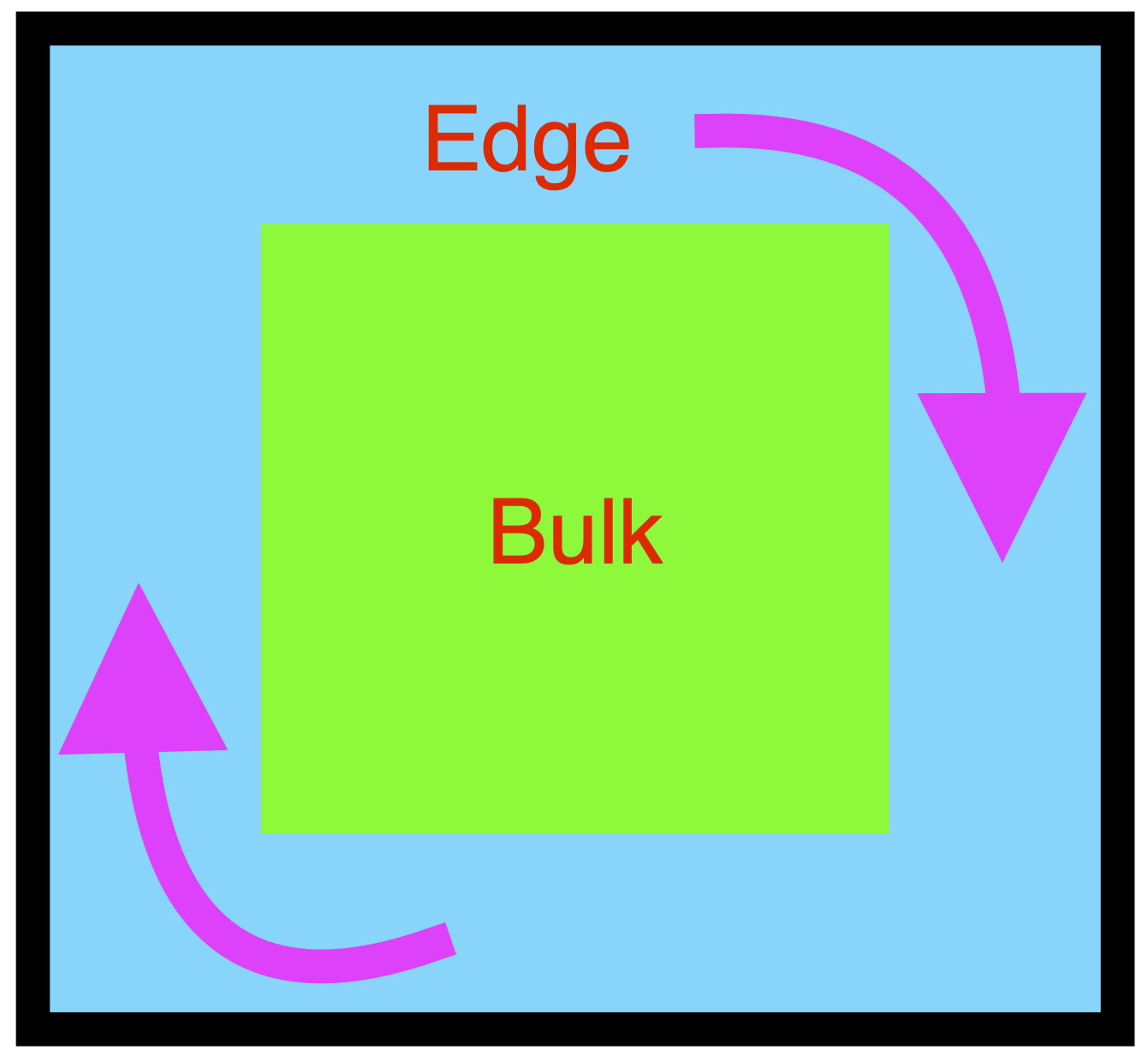}
 \caption{Edge mode propagation along the boundary of a 2D domain. The bulk region is in the middle, separated from the domain walls. \label{TI_cartoon} }
\end{figure}

Two different systems will be presented, each with its own characteristics. The first  is the 1D Su-Schrieffer-Heeger (SSH) model, originally used to {understand the propagation of} solitary waves in hydrocarbon chains \cite{SSH1979}.  The SSH model is similar to the {1D} discrete NLS model in Eq.~(\ref{one_d_lattice}), except the left and right couplings are not equal. The second system is a 2D Floquet topological insulator, which can be realized in photonic lattices by longitudinal modulation of a waveguide array \cite{Rechtsman2013}. In terms of the governing equations, this results in time-dependent coefficients, which can be solved via Floquet theory.

One of the necessary ingredients for inducing modes with nontrivial topological invariants is  the breaking of symmetries. In the case of the SSH model, inversion symmetry is broken by the asymmetric coefficient values. In the case of Floquet photonic insulator, the temporal driving breaks time-reversal symmetry by the time-dependent coefficients. Symmetry breaking  can open spectral band gaps within 
corresponding to topologically-protected modes. 

{It is} possible to find parameter regimes  where the corresponding bulk eigenmodes of these systems acquire nontrivial topological invariants. The topological invariants considered here are defined in terms of  line integrals in their associated spectral planes. These integrals are indirect ways of determining whether or not the modes posses nontrivial phase properties.
For the SSH model, the eigenmodes can possess a nonzero Zak phase \cite{Zak1989}, which {corresponds to a winding number of the phase.} In the Floquet model,  eigenstates can acquire a nonzero Chern number \cite{Thouless1982}, which is {related to} the Berry phase \cite{Berry1984} that indicates a phase discontinuity.

A consequence of a nontrivial topological invariant is remarkably stable modes{, known as topologically protected modes}. The SSH modes are localized at the endpoints of the lattice, and remain {fixed} throughout the evolution. On the other hand, the Floquet edge modes with nonzero Chern number propagate unidirectionally along the boundary and around any defects {they encounter.} Rather than backscatter, as one might expect, here there is 
{unidirectional} mode propagation. This propagation in a preferred orientation is known as chirality.
}

The connection between the bulk topological invariants and topologically protected edges states is the bulk-edge correspondence (see \cite{Asboth2016,Rudner2013,Drouot2019,Drouot2021}). The principle typically consists of the following properties: (1) A chiral edge mode exists for a topological insulators if the corresponding bulk modes have a non-zero topological invariant. (2) The topological number is equal to the net number of chiral edge states. (3) The topological invariant is independent of surface defects or boundary conditions. 

{\section{The SSH Waveguide Lattice}
\label{ssh_model_sec}

\begin{figure}
\centering
\includegraphics[scale=.5]{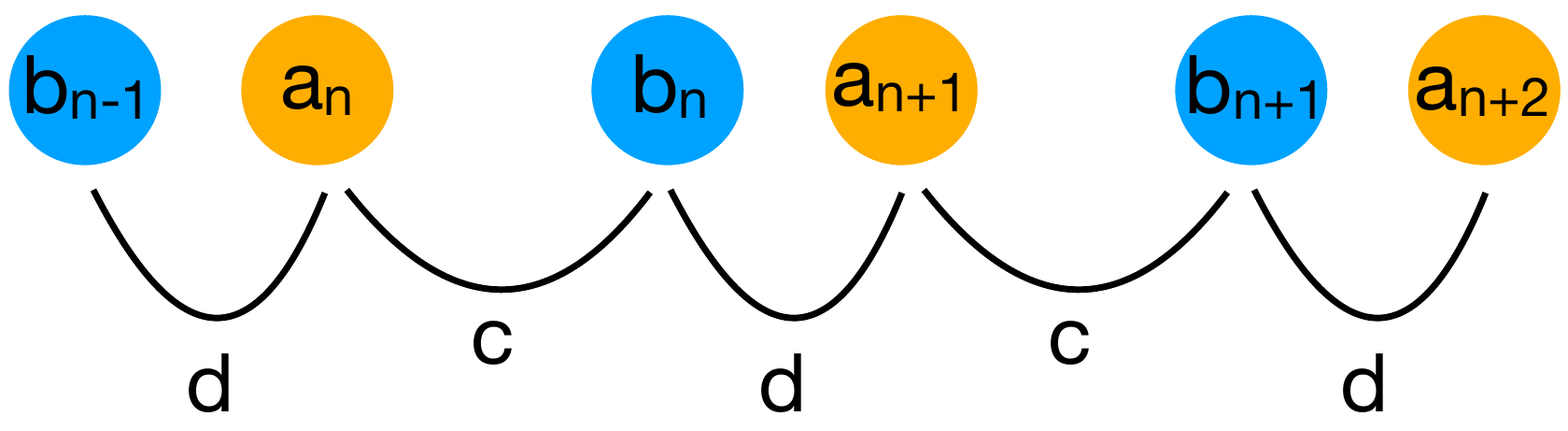}
 \caption{1D SSH lattice waveguide profiles. Nearest neighbor couplings are indicated. Here, $d > c > 0$ due to the spacing. \label{SSH_Lattice_Cartoon} }
\end{figure}

The simplest topological insulator system to realize in a photonic waveguide system is that of the SSH model. The model {can be} formulated by adjusting the waveguide spacings in an alternating manner, like that  in Fig.~\ref{SSH_Lattice_Cartoon}. Experimentally, these types of lattices {have been realized in} laser-etched arrays \cite{Szameit2006} {and}  photorefractive crystals \cite{Malkova2009}.

We assume that the potential minima $V({\bf r})$ and orbital approximation near both $a$ and $b$ lattice sites are of identical form. As a result, the coupling coefficients, which are {inversely proportional to} distance, 
are asymmetric. After transforming and rescaling a set of equations similar to Eq.~(\ref{one_d_lattice}), one obtains the nonlinear SSH system in Kerr media
\begin{align}
\label{ssh_lattice}
& i\frac{da_n}{dz}  + c b_{n} +d b_{n-1}+ \gamma |a_n|^2a_n=0 \\
& i\frac{db_n}{dz}  + c a_{n} +d a_{n+1}+\gamma |b_n|^2b_n=0 
\end{align}
where $n \in \mathbb{Z}$ and $c,d, \gamma $ are {taken to be} non-negative coefficients. If $c > d$, this physically  corresponds to placing the waveguides $a_n$ and $b_n$  closer together than their other neighbors; and vice versa if $c < d$ {see e.g. Fig. \ref{SSH_Lattice_Cartoon}}.
By allowing $c \not= d$, the inversion symmetry of the problem, $V({\bf r}) = V(-{\bf r})$, is broken, unlike the simple square lattice examined in Sec.~\ref{square_sec}.

Some of the main results associated with the SSH model are {presented below}; a more comprehensive treatment can be found in \cite{Asboth2016}.
To highlight the topological nature of this system, consider plane wave solutions on the infinite line domain of the form 
$$\begin{pmatrix} a \\ b \end{pmatrix}_n = \begin{pmatrix} \alpha \\ \beta \end{pmatrix}(k) ~e^{i (k n - \lambda z) } .$$
In the linearized problem ($\gamma =  0$), {this} 
yields  {the} eigenvalue system
\begin{equation}
\mathcal{H}(k) {\bf c} = - \lambda {\bf c} , ~~~~ {\bf c} = \begin{pmatrix}  \alpha \\ \beta \end{pmatrix}
\end{equation}
for {(the spectral Hamiltonian)} 
\begin{equation*}
\mathcal{H}(k) = \begin{pmatrix} 0 & c + d e^{- i k}   \\  c + d e^{ i k}  & 0 \end{pmatrix},
\end{equation*}
which is $2\pi$-periodic in $k$. The two dispersion relations are given by
\begin{equation}
\label{SSH_eigvalue}
\lambda_{\pm}(k) = \pm {\left|c + d e^{- i k}  \right|} , 
\end{equation}
{leading to} a gap width of $2 |c - d|$. The corresponding normalized eigenfunctions are
\begin{equation}
\label{SSH_eigfunc}
{\bf c}_{\pm}(k) = \frac{1}{\sqrt{2}} \begin{pmatrix} \mp e^{i \theta(k)} \\ 1 \end{pmatrix} ,
\end{equation}
where $ \theta(k) = \tan^{-1} \left( - \frac{d \sin k}{ c + d \cos k} \right)$ is the {counterclockwise} angle from the positive real axis. A plot of the dispersion relations for different values of $c$ and $d$ is shown in Fig.~\ref{SSH_bands_plot}. Notice that when  inversion symmetry is broken ($c \not= d$), a gap opens between the bands. When the symmetry is preserved ($c = d$), the gap closes at $k = \pm \pi $.

\begin{figure}
\centering
\includegraphics[scale=.35]{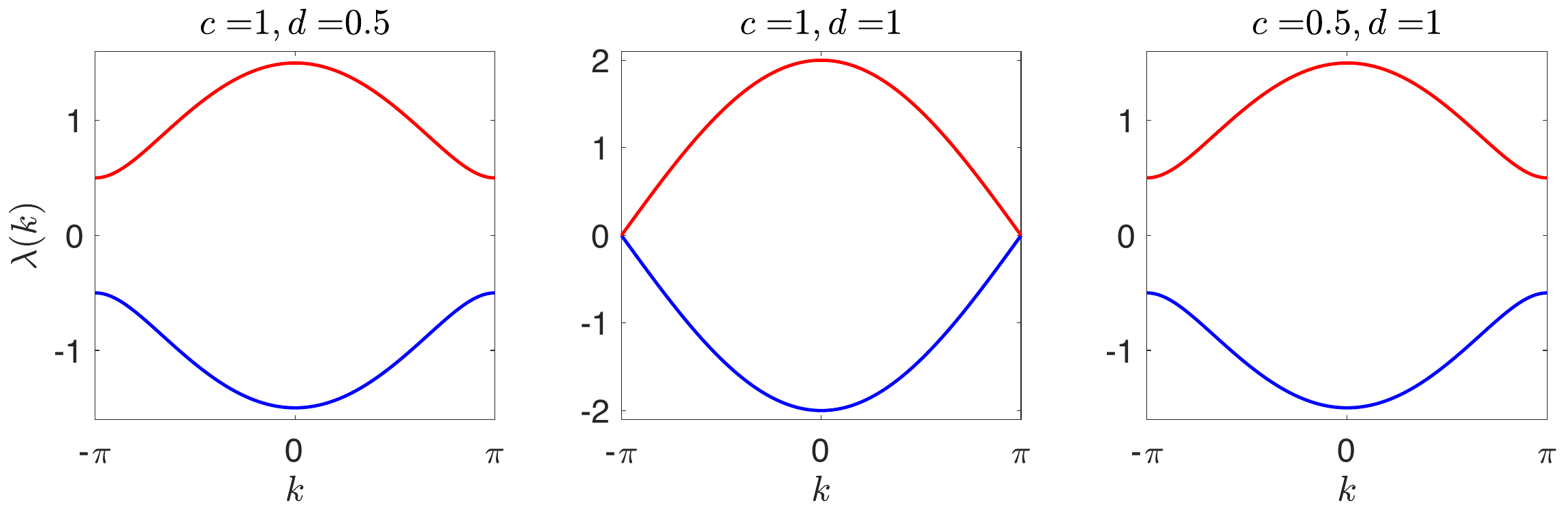}
 \caption{SSH model bulk dispersion relations in Eq.~(\ref{SSH_eigvalue}). \label{SSH_bands_plot} }
\end{figure}

The topological quantity associated with this system is the {Berry/}Zak phase
\begin{equation}
\label{zak_phase}
\mathcal{Z} = i \oint \left\langle {\bf c} ~\bigg| \frac{\partial {\bf c}}{ \partial k} \right\rangle dk ,
\end{equation}
where $\langle {\bf f} | {\bf g} \rangle = {\bf f}^{\dag} {\bf g}$ and $\dag$ denotes the complex conjugate transpose. The Zak phase is an indirect way of measuring the winding number of the eigenfunction phase ${\theta}(k)$ over one period in $k$. To see this, take the eigenfunction in Eq.~(\ref{SSH_eigfunc}) and observe that
\begin{equation*}
\mathcal{Z} =  i \oint \left\langle {\bf c}_{\pm} ~\bigg| \frac{\partial {\bf c}_{\pm}}{ \partial k} \right\rangle dk  = \frac{i}{2} \int_0^{2 \pi} i \frac{d \theta}{d k} dk = - \frac{1}{2} \theta(k) \bigg|_{k = 0}^{k = 2 \pi} . 
\end{equation*}
A graphical depiction  of the path $c + d e^{-ik}$ for topologically distinct parameter sets is shown in Fig.~\ref{SSH_wind_num}. In the topological case ($d > c$), the path encircles the origin and corresponds to a Zak phase of $\mathcal{Z} = - \frac{1}{2} \left[ \theta(2 \pi) - \theta(0) \right] = - \frac{1}{2} \left[ - 2 \pi - 0 \right] = \pi.$ On the other hand, in the non-topological case ($d < c$), the loop does not enclose the origin and $\mathcal{Z} = - \frac{1}{2} \left[ 0 - 0 \right] = 0.$

\begin{figure}
\centering
\includegraphics[scale=.35]{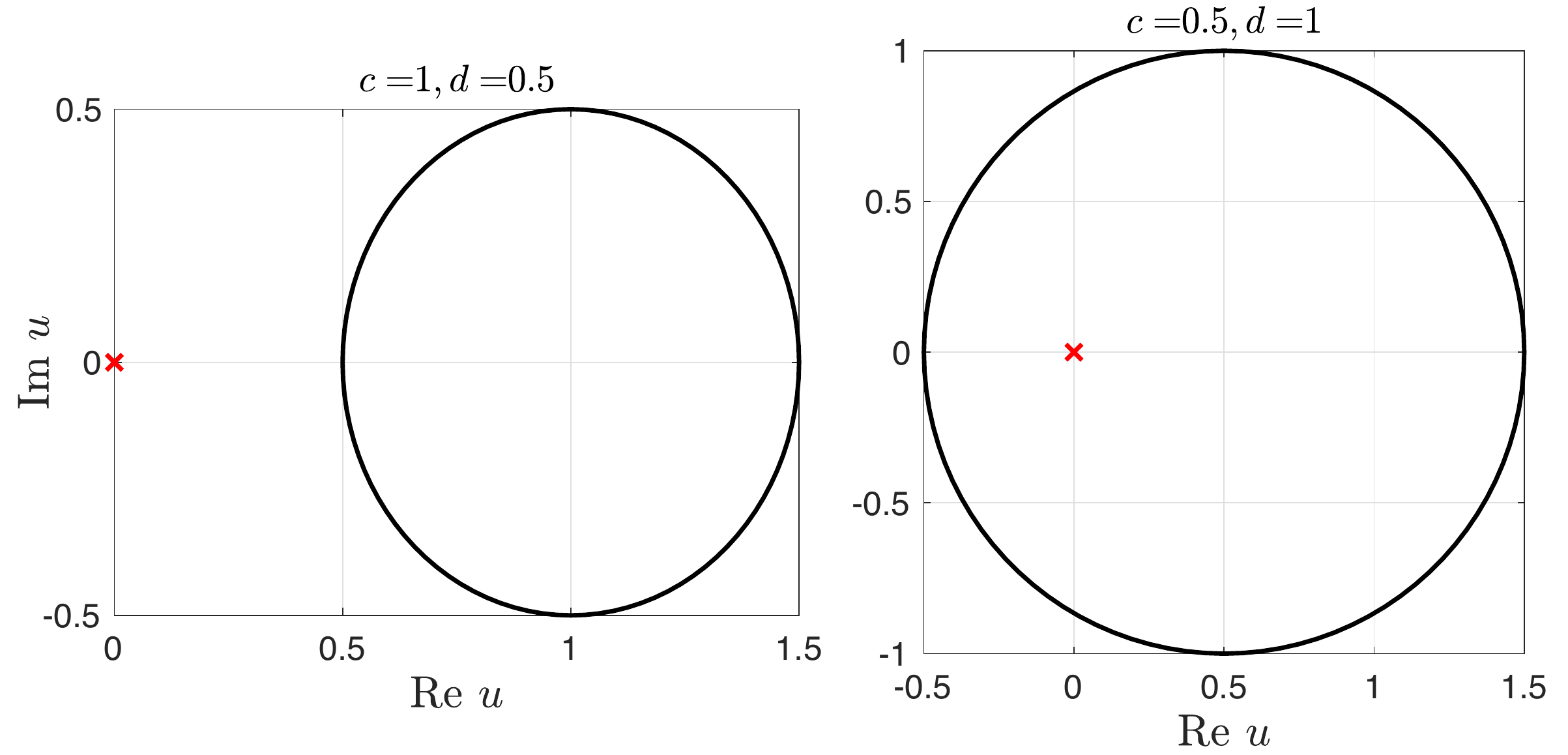}
 \caption{Path of complex function $u(k)= c + d e^{-  i k}$ for $k \in [0, 2\pi]$. Note that the loop is clockwise oriented. \label{SSH_wind_num} }
\end{figure}

Now let us examine the finite domain problem, which may support edge modes.
Here the topological (nonzero Zak phase) case corresponds to localized {chiral} edge modes at the endpoints of the lattice {(recall the bulk-edge correspondence)}. We impose the Dirichlet zero boundary conditions 
\begin{equation*}
a_n, b_n = 0, ~~~ \text{for} ~~~ n < 1~  \text{and} ~ n > N
\end{equation*}
in the linearized version of Eq.~(\ref{ssh_lattice}) with $N \gg 1$. For time-harmonic solutions of the form $a_n(t) = \alpha_n e^{- i \lambda t}$ and $b_n(t) = \beta_n e^{- i \lambda t}$ The corresponding system is given by
\begin{equation}
\label{edge_dispersion_bands}
\mathbb{M} {\bf c}_n = - \lambda {\bf c}_n
\end{equation}
where $\mathbb{M}$ is the $2N \times 2N$ matrix
\begin{equation*}
\mathbb{M} =\left( 
\begin{array}{c|c} 
  O& \mathcal{M} \\ 
  \hline 
  \mathcal{M}^T & O 
\end{array} 
\right) , ~~~ \mathcal{M} = \begin{pmatrix} c & &  \\ d & c &   \\ & d & \ddots  \\ & & \ddots  \\ \\ & & & & d & c \end{pmatrix} ,
\end{equation*}
{$O$ is {a} 
zero matrix} and ${\bf c}_n = (\alpha_1,  \alpha_2,  \dots,  \alpha_N ~|~ \beta_1,  \beta_2,  \dots,  \beta_N)^T.$
The solution of system (\ref{edge_dispersion_bands}) for different parameter values is shown in Fig.~\ref{SSH_edge_bands}. Noticeably, when $c > d$ (non-topological case) there are no localized edge modes. On the other hand, when $d > c$ (topological case) there are two zero energy ($\lambda = 0$) edge states. {Via the bulk-edge correspondence, we infer that these eigenmodes {correspond} 
to a chiral edge state.}

Plots of the  eigenmodes are shown in Fig.~\ref{SSH_top_eig_modes}. There are two zero energy, localized eigenmodes: {symmetric} 
and anti-symmetric; both are real. Analytically, one can show that the zero energy modes along the left edge are of the form
\begin{equation}
a_n(t) = \left( - \frac{c}{d} \right)^n , ~~~~~ b_n (t) = 0,
\end{equation}
and decay as $n \rightarrow \infty$ when $c < d$. A similar form exists on the right edge, except the $b_n$ mode is decaying and $a_n$ is zero. All other nonzero energy modes are bulk modes and they are not localized. 

Physically speaking, the presence of edges modes corresponds to isolated endpoints in Fig.~\ref{SSH_Lattice_Cartoon}, well-separated from the next interior site. In the topological regime, the mode propagation manifests itself as {an} electromagnetic field concentrated at the endpoints. 

\begin{figure}
\centering
\includegraphics[scale=.35]{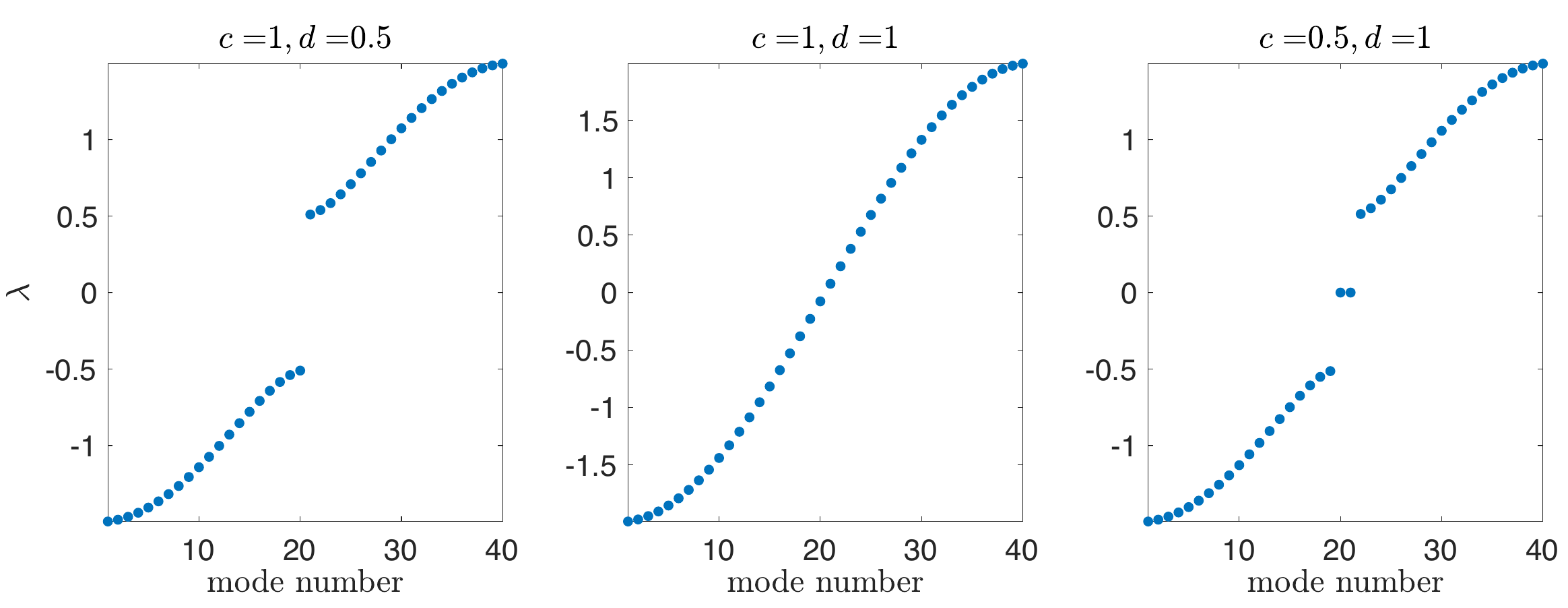}
 \caption{SSH model edge dispersion relations in Eq.~(\ref{edge_dispersion_bands}) using $N = 20$ {sites}. Recall that $c > d$ ($c < d$) corresponds to the non-topological (topological) case in Fig.~\ref{SSH_wind_num}. \label{SSH_edge_bands} }
\end{figure}

Theoretical and experimental research on the nonlinear SSH model is still ongoing. A number of works have established the existence of nonlinear solitons for the system (\ref{ssh_lattice}) in the bulk \cite{Vicencio2009,Solnyshkov2017,Tuloup2020} and at the edge \cite{Smirnova2019,Ma2021}. Topological edge solitons appear  rather stable, as long as the energies {are} well-removed from the balanced limit, $c \approx d$ \cite{Ma2021}.
We point out that {with} {non-Kerr type} nonlinearities, 
fascinating phenomena such as nonlinear-induced topological transition \cite{Hadad2016,Hadad2018} have 
been theorized.

\begin{figure}
\centering
\includegraphics[scale=.35]{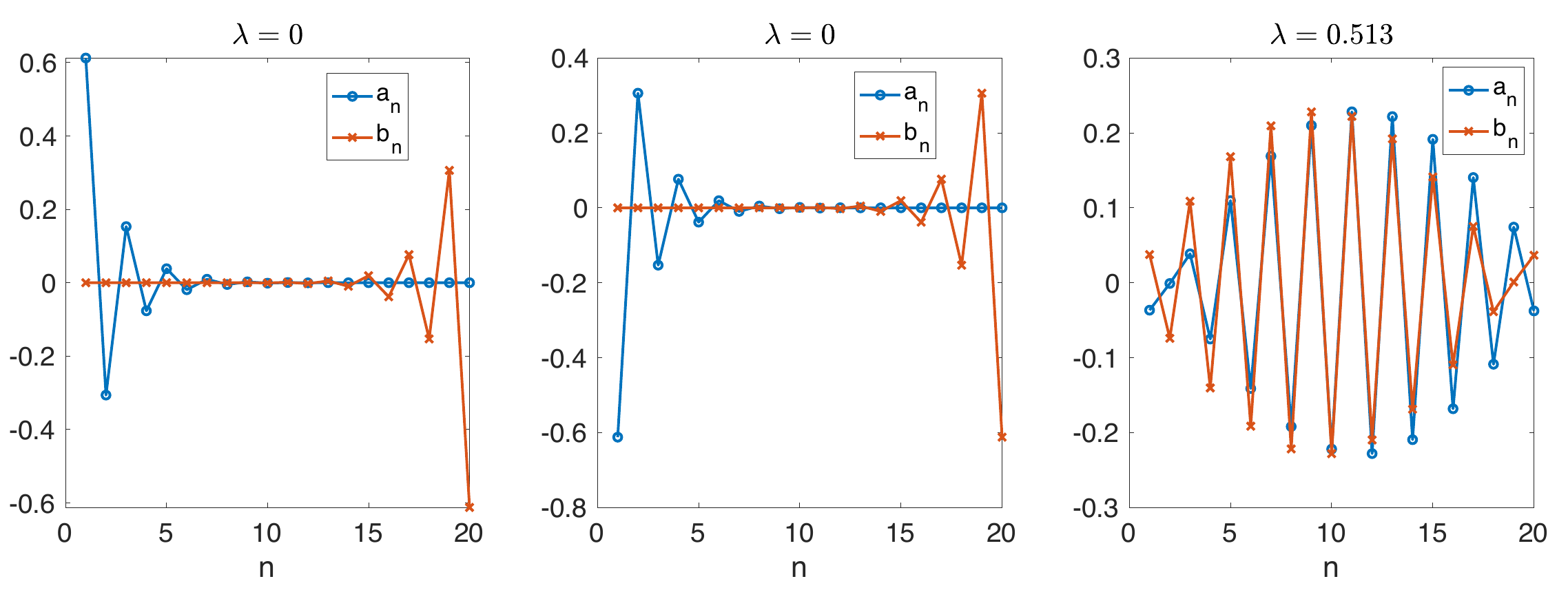}
 \caption{SSH edge eigenmodes  of Eq.~(\ref{edge_dispersion_bands}) using $N = 20$ {sites} and $c = 0.5, d = 1$ (topological case). The  corresponding eigenvalues are shown in Fig.~\ref{SSH_edge_bands}. All eigenmodes  are real. \label{SSH_top_eig_modes} }
\end{figure}

}

{
\section{Longitudinally-driven Photonic Lattices}
\label{long_drive_lattice}

A {photonic} Floquet topological insulator is described in this section. The proposal and experimental realization of this system was originally given in the seminal  work of Rechtsman {et al.} 
\cite{Rechtsman2013}.
Physically, the system is a photonic waveguide array, similar to the one described in Sec.~\ref{intro_sec}. The new {technique introduced} 
is that the waveguides are constructed with a helical-variation in the longitudinal direction (see Fig.~\ref{lattice_cartoon}). As a result, one obtains lattice potentials that are periodic in both the transverse {\it and} longitudinal directions. Lattices that are periodic in the time (or time-like) variable are typically referred to as {\it Floquet lattices} due to the classic mathematical theory {of ODEs} developed by Floquet \cite{East73}. Below, the 
{key ideas and} governing equations {are} described; a more thorough treatment can be found in \cite{abjc2017,abjc2019,abccyp2014}.} 

\begin{figure}[b]
\centering
\includegraphics[scale=.5]{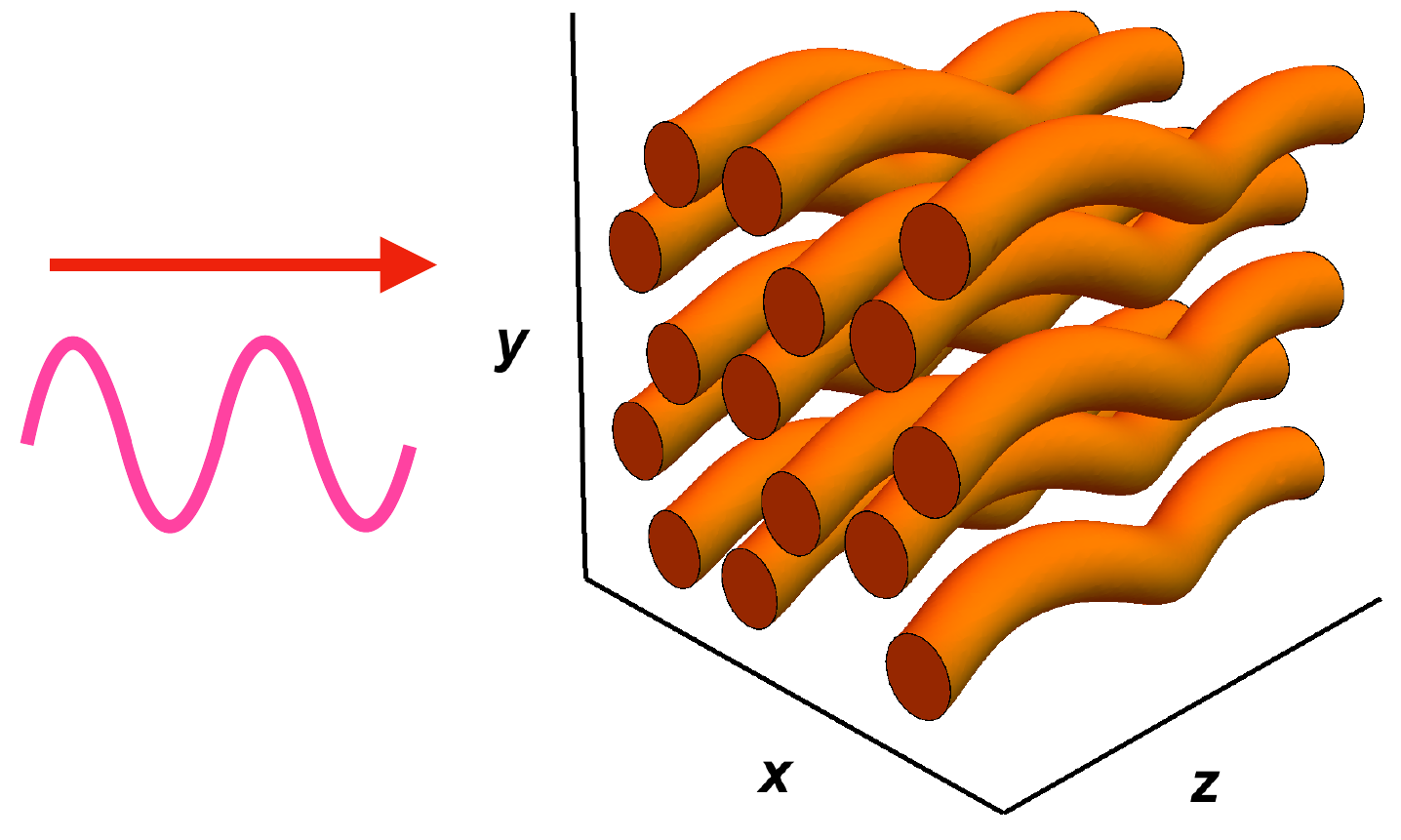}
 \caption{An electromagnetic wave propagating through a honeycomb lattice that is hellically-varying in the longitudinal direction. The lattice rods correspond to regions of higher index of refraction and act as waveguides. \label{lattice_cartoon} }
\end{figure}

{The starting point is a {modification} 
of  Eq.~(\ref{LNLS}), {now with a} longitudinally-varying photonic lattice that is modeled by the potential $V({\bf r} , z)$ that is periodic in $x,y,$ and $z$. The governing equation is
\begin{equation}
\label{FNLS}
i\psi_z + \nabla^2 \psi - V(\mathbf{r},z)\psi + \sigma|\psi|^2 \psi= 0,
\end{equation}
where $V({\bf r}, z) =  V({\bf r} - {\bf h} (z))$ for the  potential given in Eq.~(\ref{lattice_sum}), with driving function ${\bf h}(z)$ that has {period} 
$T$: ${\bf h}(z + T) = {\bf h}(z)$. Physically, this corresponds to waveguides where the lattice sites oscillate with a helical motion in $z$. In \cite{abjc2017,abjc2019} more complicated lattice driving patterns were considered in which each sublattice was allowed to move independently, as long as there was a commensurate period. Here, however, we only consider the 
case where all lattice sites are driven in the same manner. }


{A natural transformation is to the coordinate frame co-moving with helical motion: $\widetilde{\bf r} = {\bf r} - {\bf h}(z)$. Doing so, and introducing the phase
\begin{equation*}
\psi({\bf r},z) ={\widetilde{\psi}}(\widetilde{{\bf r}},{z}) \exp\left(  \frac{ i  \int_0^{{z}} |  {\bf h}'(\zeta) |^2 d \zeta }{4} \right) \; ,
\end{equation*}
{yields}
\begin{equation}
\label{LSE_3}
i \widetilde{\psi}_{z} +  \left[ \widetilde{\nabla} + i  {\bf A}(z)\right]^2 \widetilde{\psi} - V(\widetilde{{\bf r}}) \widetilde{\psi} +  \sigma \big|\widetilde{\psi}\big|^2 \widetilde{\psi} = 0  ,
\end{equation}
where $\widetilde{\nabla} \equiv \partial_{\tilde{x}} \ihat + \partial_{\tilde{y}} \jhat$ {yields the vector potential} 
\[{ \bf A}(z) = - \frac{{\bf h}'(z)}{2} \]
A typical driving function taken is
\begin{equation}
\label{driving_fcn1}
{ \bf A}(z)=\kappa \left( \sin\left( \Lambda z + \chi \right) , - \cos \left( \Lambda z + \chi \right) \right) .
\end{equation}
where $\kappa$ is the relevant helix radius, {$\Lambda = 2 \pi / T$} is the angular frequency, and $\chi$ is an arbitrary phase shift.
There are a few things to note: 
(a) In the helical frame of reference, the potential $V(\widetilde{{\bf r}})$ is stationary. The form of the potential resembles that of Eq.~(\ref{HC_potential}) for a honeycomb lattice. (b) The coordinate transformation has introduced a magnetic vector potential ${\bf A}(z)$. A common feature among Chern insulators is the presence of {a magnetic-type} field. Opposed to other systems, which use actual magnetic fields \cite{Wang2008,Wang2009}, here an effective or pseudo magnetic field is generated by the helically-varying waveguide. (c) The helical driving of the system breaks time reversal symmetry (conjugation + $z \rightarrow -z$) since ${\bf A}(-z) \not= {\bf A}(z)$.
}

{Finally, to simplify the problem, the Peierls phase transformation \cite{Peierls1933,Luttinger1951}
\begin{equation*}
 \widetilde{\psi}(\widetilde{{\bf r}},z) = \varphi(\widetilde{{\bf r}},z) e^{ - i \widetilde{{\bf r}} \cdot {\bf A}(z)}  ,
\end{equation*}
 is applied and {reduces} {Eq.~(\ref{LSE_3}) to}
\begin{equation}
\label{LSE_5}
i  \varphi_z  +  {\nabla}^2 \varphi + {{\bf r}} \cdot {\bf A}'(z) \varphi - V({{\bf r}}) \varphi + \sigma |\varphi|^2 \varphi = 0  ,
\end{equation}
where the tilde notation has been dropped. This is the final form of the PDE, from which the tight-binding model {discussed below} is derived.}

{Next, the field $\phi$ is expanded in terms of an orbital basis. For this system, a direct Wannier expansion is ineffective since a nonzero Chern number eliminates their exponential decay \cite{Brouder2007}; however other {{\it indirect}} 
Wannier approaches may be possible \cite{Ablowitz2020}. To generate a convenient and analytical basis, we examine the weakly driven and linear limit of Eq.~(\ref{LSE_5}) where $|{\bf A}'(z)| \ll 1 $. Physically, a {rapidly varying regime where 
weakly driven regime where $ \Lambda \gg1$ was employed; see \cite{abccyp2014,abjc2017}. Indeed the experiments \cite{Rechtsman2013} were in this rapidly varying helical regime.}
Using these assumptions eliminates all $z$-dependent coefficients in Eq.~(\ref{LSE_5}). Note, however, that the variable ${\bf r}$ here is in the helical frame of reference, so these orbitals are localized 
{at the oscillating lattice sites (in the original frame of reference).}

For a lattice with two  sites per unit cell, we look for solutions of the form
\begin{align}
\label{ansatz_define}
\varphi({\bf r},z) = \sum_{m,n} \left[ \widetilde{a}_{mn}(z) \phi_{A,{mn}}({\bf r} ) + \widetilde{b}_{mn}(z) \phi_{B,{mn}}({\bf r}) \right] e^{- i E z} ,
\end{align}
where $\phi_{j,mn}({\bf r})$ are  orbital functions. In general, the number of distinct orbital terms in the expansion matches the number of lattice sites per unit cell e.g. a lattice with three lattice sites per until cell will have an extra {term} {of the form} $\widetilde{c}_{mn}(z) \phi_{C,mn}({\bf r})$.
In particular, for the honeycomb lattice the orbital functions are defined by $\phi_{A,mn}({\bf r}) = \phi({\bf r} - m {\bf v}_1 - n {\bf v}_2 - {\bf d}_1)$ and $\phi_{B,mn}({\bf r}) = \phi({\bf r} - m {\bf v}_1 - n {\bf v}_2 )$, where $\phi({\bf r})$ satisfies the orbital equations in (\ref{HC_orbital}).


From here, the derivation of a tight-binding model follows similar to that of Secs.~\ref{square_sec} and \ref{honeycomb_sec}. A set of semi-discrete equations are derived by substituting expansion (\ref{ansatz_define}) into Eq.~(\ref{LSE_5}), multiplying by each orbital type, and then integrating over $\mathbb{R}^2$. For typical experimental systems, the potential is deep or has large magnitude at the waveguides, that is $|V({\bf r})| \gg 1$ near the lattice sites. As a result, a tight-binding approximation is applied and only the {on-site and} nearest neighbor interactions are kept. 
Details of the derivation can be found in \cite{abjc2017,abjc2019}.

The {paradigm} 
Floquet tight-binding model {is} 
a honeycomb lattice. Following the procedure described above, the governing tight-binding model is given by
\begin{align}
\label{HC_TB_eqn1}
 &i \frac{d a_{mn}}{dz}  + \sigma g |a_{mn}|^2 a_{mn} \\ \nonumber &+ C \Big[ e^{- i {\bf d}_1 \cdot {\bf A}(z)} b_{mn} \\ \nonumber & +    e^{- i {\bf d}_3 \cdot {\bf A}(z)} b_{m-1,n-1} +e^{- i {\bf d}_2 \cdot {\bf A}(z)} b_{m+1,n-1} \Big] = 0   , \\
\label{HC_TB_eqn2}
& i \frac{d b_{mn}}{dz} + \sigma g |b_{mn}|^2 b_{mn} \\ \nonumber &+  C \Big[ e^{ i {\bf d}_1 \cdot {\bf A}(z)} a_{mn}  \\ \nonumber & +   e^{ i {\bf d}_3 \cdot {\bf A}(z)} a_{m+1,n+1} + e^{ i {\bf d}_2 \cdot {\bf A}(z)}a_{m-1,n+1}  \Big]  = 0 ,
\end{align}
where $g = \int \phi_A^4({\bf r}) d {\bf r} =  \int \phi_B^4({\bf r}) d {\bf r}$.
A couple of notes about this system:
{(a) The above system is essentially the same as the one discussed in section 6 --see Eq.~(\ref{Dis_Dirac1}-\ref{Dis_Dirac2}) only now the coefficients are periodic functions of $z$.}
(b) The coefficient $C = C\left(\parallel {\bf v} \parallel \right)$ is distance-dependent and so it is the same for all nearest neighbor interactions. (c) The indices used here are not in terms of the  lattice vectors ${\bf v}_1$ and ${\bf v}_2$. Instead, the $m$ index is in terms of the vector $ {\bf w}_1 = ({\bf v}_1 - {\bf v}_2)/2 = l (0, 1/2)$ and the $n$-index for ${\bf w}_2 = ({\bf v}_1 + {\bf v}_2)/2 = l (\sqrt{3}/2 , 0)${ (see Fig.~\ref{honey_lattice_fig}); this is useful in edge mode calculations, discussed below.}}

\begin{figure}
\centering
\includegraphics[scale=.7]{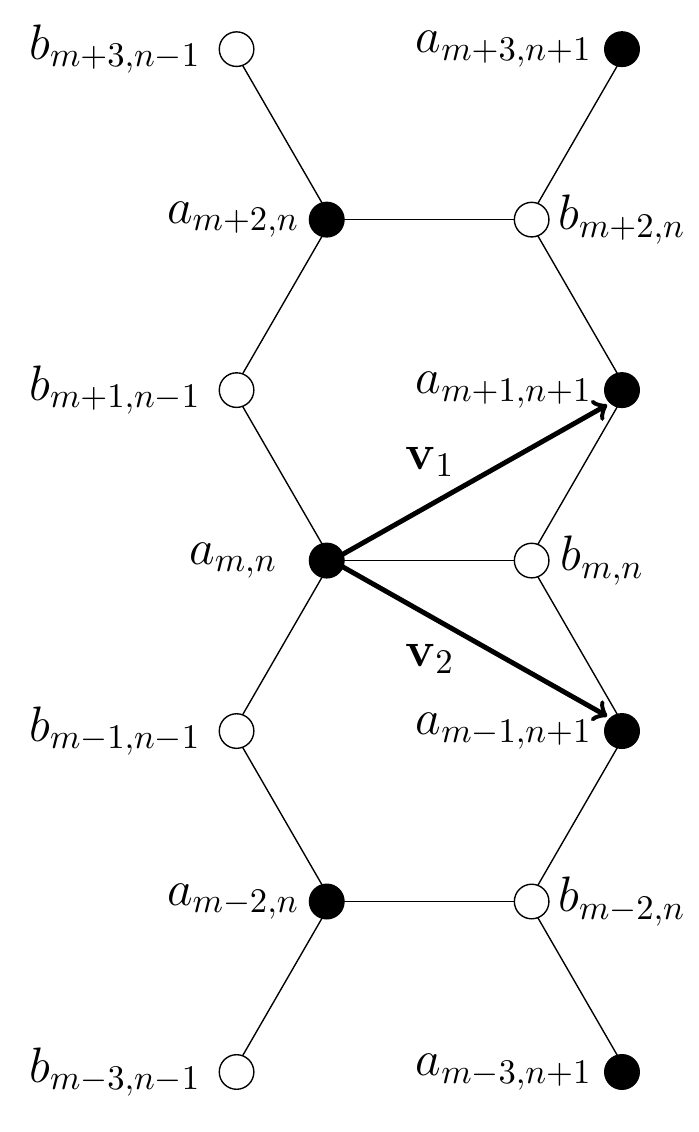}
 \caption{Discrete honeycomb lattice for the tight-binding system (\ref{HC_TB_eqn1})-(\ref{HC_TB_eqn2}). The $a$ lattices sites (black dots) are located at points $\left\{ {\bf v}_a | {\bf v}_a = m {\bf w}_1 + n{\bf w}_2 + {\bf d}_1 \right\}$ and the $b$ sites (white dots) at $\left\{ {\bf v}_b | {\bf v}_b = m {\bf w}_1 + n{\bf w}_2  \right\}$ where $m,n \in \mathbb{Z}$. Zig-zag boundary conditions are those in the vertical direction. \label{honey_lattice_fig} }
\end{figure}

{
\subsection{Floquet Dispersion Bands}
\label{HC_disp_bands_sec}

To begin analyzing the honeycomb Floquet system, 
{we first consider 
the} linearized version of system (\ref{HC_TB_eqn1})-(\ref{HC_TB_eqn2}) on an infinite domain. A linear reduction can be achieved by taking a small intensity field: $| a_{mn} |^2,| b_{mn} |^2 \approx 0$.  The corresponding eigenmodes are known as bulk modes. The spectral dispersion surfaces or bands are computed by looking for Fourier solutions of the form 
\begin{align}
&a_{mn}(z) = A({\bf k},z) e^{ i {\bf k} \cdot (m {\bf w}_1 + n {\bf w}_2 )} \; ,\\ \nonumber
&b_{mn}(z) = B({\bf k},z) e^{ i {\bf k} \cdot (m {\bf w}_1 + n {\bf w}_2 )} \; ,
\end{align}
which yield 
\begin{equation}
\label{HC_spec_sys}
\frac{d {\bf c}}{dz} 
 =  i \mathcal{H}({\bf k},z) {\bf c} \; , ~~~~~ {\bf c} = \begin{pmatrix}
A \\ B
\end{pmatrix}({\bf k},z) \; ,
\end{equation}
{for} 
(the Hamiltonian)
\begin{equation*}
\label{M_matrix_define}
\mathcal{H}({\bf k} ,z) = 
\begin{pmatrix}
0 & \tau({\bf k},z) \\
\tau({\bf k},z)^* & 0
\end{pmatrix} \; ,
\end{equation*}
{and}  $ \tau({\bf k},z) = e^{ - i {\bf d}_1 \cdot {\bf A}(z) } + e^{-  i ( {\bf d}_3  \cdot {\bf A} (z) + {\bf k} \cdot {\bf v}_1)} + e^{-  i ( {\bf d}_2  \cdot {\bf A} (z) + {\bf k} \cdot {\bf v}_2)}$. 
Notice that the matrix $\mathcal{H}$ is $T$-periodic in $z$ and periodic in the spectral plane: $\mathcal{H}({\bf k} + p {\bf k}_1  ,z) = \mathcal{H}({\bf k} ,z ) = \mathcal{H}({\bf k} +q {\bf k}_2 ,z) $ where $p,q \in \mathbb{Z}.$

\begin{figure}
\centering
\includegraphics[scale=.35]{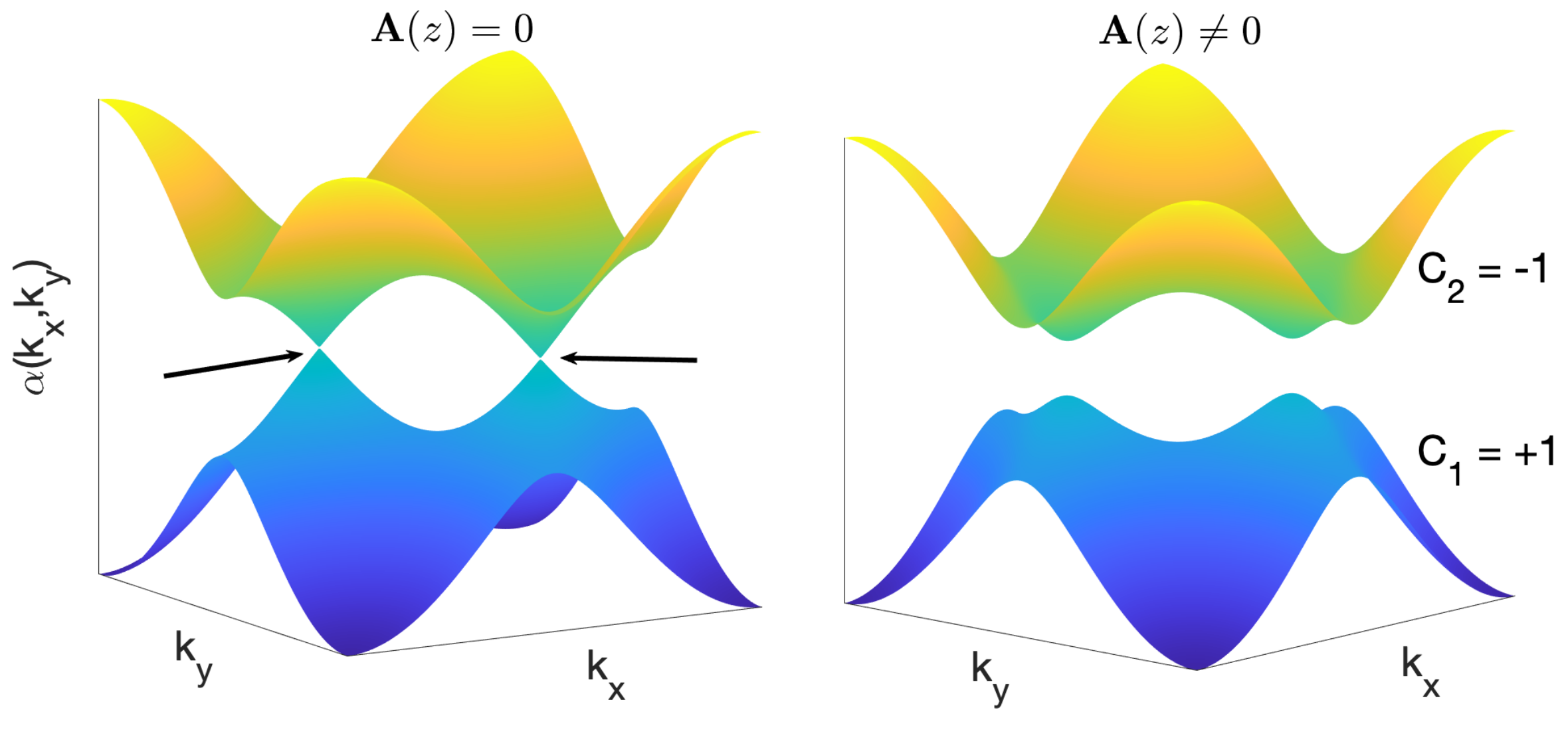}
 \caption{{Typical bulk} dispersion surfaces (\ref{floquet_define}) computed from (\ref{HC_spec_sys}). In the absence of driving {(${\bf A}(z) = 0$)}, the surfaces touch at the Dirac points, $K$ and $K'$. Driving the lattice {(${\bf A}(z) \not=0$)} opens a band gap and the corresponding eigenfunctions acquire nontrivial Chern numbers, as indicated.}
 \label{spectral_surfaces} 
\end{figure}

We look for 
{solutions} of system (\ref{HC_spec_sys}) via Floquet theory \cite{East73}. These solutions are assumed to satisfy the quasi-periodic boundary condition
\begin{equation}
\label{bulk_floqet_BCs}
{\bf c}({\bf k},z+T)  = \rho ~{\bf c}({\bf k},z) , ~~~~~~~ \rho({\bf k}) = e^{- i \alpha({\bf k}) T} .
\end{equation}
The parameter $\rho$ is known as the characteristic or Floquet multiplier and for stable Floquet modes, it lies on the unit circle. 
To find it, the $2 \times 2$ principal fundamental matrix solution of (\ref{HC_spec_sys}) at $z = T$ is  computed numerically. This matrix solution is known as monodromy matrix. Moreover, the eigenvalues of the monodromy matrix are the Floquet multipliers in Eq.~(\ref{bulk_floqet_BCs}). Finally, the so-called Floquet exponents are calculated by
\begin{equation}
\label{floquet_define}
\alpha({\bf k}) = \frac{i \log[\rho({\bf k})] }{T}  .
\end{equation}
The exponential form of the Floquet multiplier in Eq.~(\ref{bulk_floqet_BCs}) implies an infinite number of solutions, due to periodicity in $\alpha({\bf k}); $ i.e. $\rho$ is unchanged by the shift $\alpha \rightarrow \alpha + 2 \pi/T $. For all results shown here, we only present the principal branch $\alpha({\bf k}) \in [-\pi/T , \pi/T. ]$

For a typical set of values, the bulk dispersion surfaces are shown in Fig.~\ref{spectral_surfaces}. In the absence of driving  {(${\bf A}(z) = 0$)}, the bands touch at the Dirac points (\ref{dirac_pts}). Introduction of the helical {driving}  motion  
{(${\bf A}(z) \not=0$)} opens a band gap. Furthermore, as a result of this driving, the corresponding bulk eigenmodes acquire a nontrivial topological number, discussed next.

Through the periodic driving of a waveguide array, it is possible to realize eigenmodes with a nontrivial topological invariant, known as the Chern number. {The relevant topological} Chern number  of eigenfunction ${\bf c}_p$ in Eq.~(\ref{M_matrix_define}), corresponding to the $p^{\rm th}$ spectral band, is {given} by 
\begin{equation}
\label{chern}
C_p = \frac{1}{2 \pi i} \iint_{\rm UC} \left(  \frac{\partial {\bf c}_p^{\dag}}{\partial k_x}  \frac{\partial {\bf c}_p}{\partial k_y}- \frac{\partial {\bf c}_p^{\dag}}{\partial k_y}  \frac{\partial {\bf c}_p}{\partial k_x} \right) ~ d{\bf k} \; , ~~~ p = 1,2
\end{equation}
where UC denotes the reciprocal unit cell defined in terms of the reciprocal lattice vectors ${\bf k}_1$ and ${\bf k}_2$. {We note that} {$C_p$ is $z$-invariant.}
Physically, the Chern number indicates the presence of a nontrivial phase jump inside the reciprocal unit cell. In a rapidly-varying regime ($T \ll 1$), it {is also} possible to derive an averaged version of the bulk system (\ref{M_matrix_define}) that is independent of $z$ \cite{Ablowitz2021b}. {Remarkably,} it turns out that the form of this the averaged-system is {analogous to the} 
well-known Haldane model used to {study} 
the quantum Hall effect \cite{Haldane1988}. On the other hand, to numerically compute  Chern numbers {directly}, the algorithm given in \cite{fukui2005} can be applied.}

{As a remark, nonlinearity can induce localized bulk modes. These nonlinear Floquet modes, predicted in \cite{Lumer2013} and experimentally observed in \cite{Mukherjee2020}, correspond to band gap spectral values and exhibit a cyclotronic motion about a particular lattice site.}

{
{Next, we study the problem on a  finite domain in  the $x$ direction and infinite in the $y$ direction.} 
We look for edge modes that decay exponentially fast perpendicular to the imposed boundary. As a result, we consider modes of the form
\begin{equation}
a_{mn}(z) = a_n({\bf k}, z) e^{ i {\bf k} \cdot m {\bf w}_1}  , ~~ b_{mn}(z) = b_n({\bf k}, z) e^{ i {\bf k} \cdot m {\bf w}_1}  ,
\end{equation}
which reduce system (\ref{HC_TB_eqn1})-(\ref{HC_TB_eqn2}) to
\begin{align}
\label{HC_TB_edge1}
 & i \frac{d a_{n}}{dz} + \sigma g |a_{n}|^2 a_{n}   \\ \nonumber &+ C \Big[ e^{- i {\bf d}_1 \cdot {\bf A}(z)} b_{n} \\ \nonumber & +  \left(  e^{- i \left[ {\bf d}_3 \cdot {\bf A}(z) + {\bf k} \cdot {{\bf w}_1} \right]} +e^{- i \left[ {\bf d}_2 \cdot {\bf A}(z) - {\bf k} \cdot {{\bf w}_1}  \right]} \right) b_{n-1} \Big] = 0  , \\
\label{HC_TB_edge2}
&  i \frac{d b_{n}}{dz} + \sigma g |b_{n}|^2 b_{n} \\ \nonumber &+  C \Big[ e^{ i {\bf d}_1 \cdot {\bf A}(z)} a_{n} \\ \nonumber &  + \left(  e^{ i \left[ {\bf d}_3 \cdot {\bf A}(z) + {\bf k} \cdot {\bf w}_1 \right]}  + e^{ i \left[ {\bf d}_2 \cdot {\bf A}(z) - {\bf k} \cdot {\bf w}_1 \right]} \right)a_{n+1}  \Big] = 0 ,
\end{align}
where ${\bf k} \cdot {\bf w}_1 = \frac{l}{2} k_y$. 
Zero boundary conditions are imposed along a set of zig-zag boundaries:
\begin{align}
\label{Dirichlet_BCs}
&b_n = 0 \; , ~~ {\rm for} ~~ n <0   , ~  n > N-1  ,  \\ \nonumber
&a_n = 0 \; , ~~ {\rm for} ~~ n <1   ,  ~ n > N  ,
\end{align}
(see Fig.~\ref{honey_lattice_fig} for reference). 

\begin{figure}
\centering
\includegraphics[scale=.3]{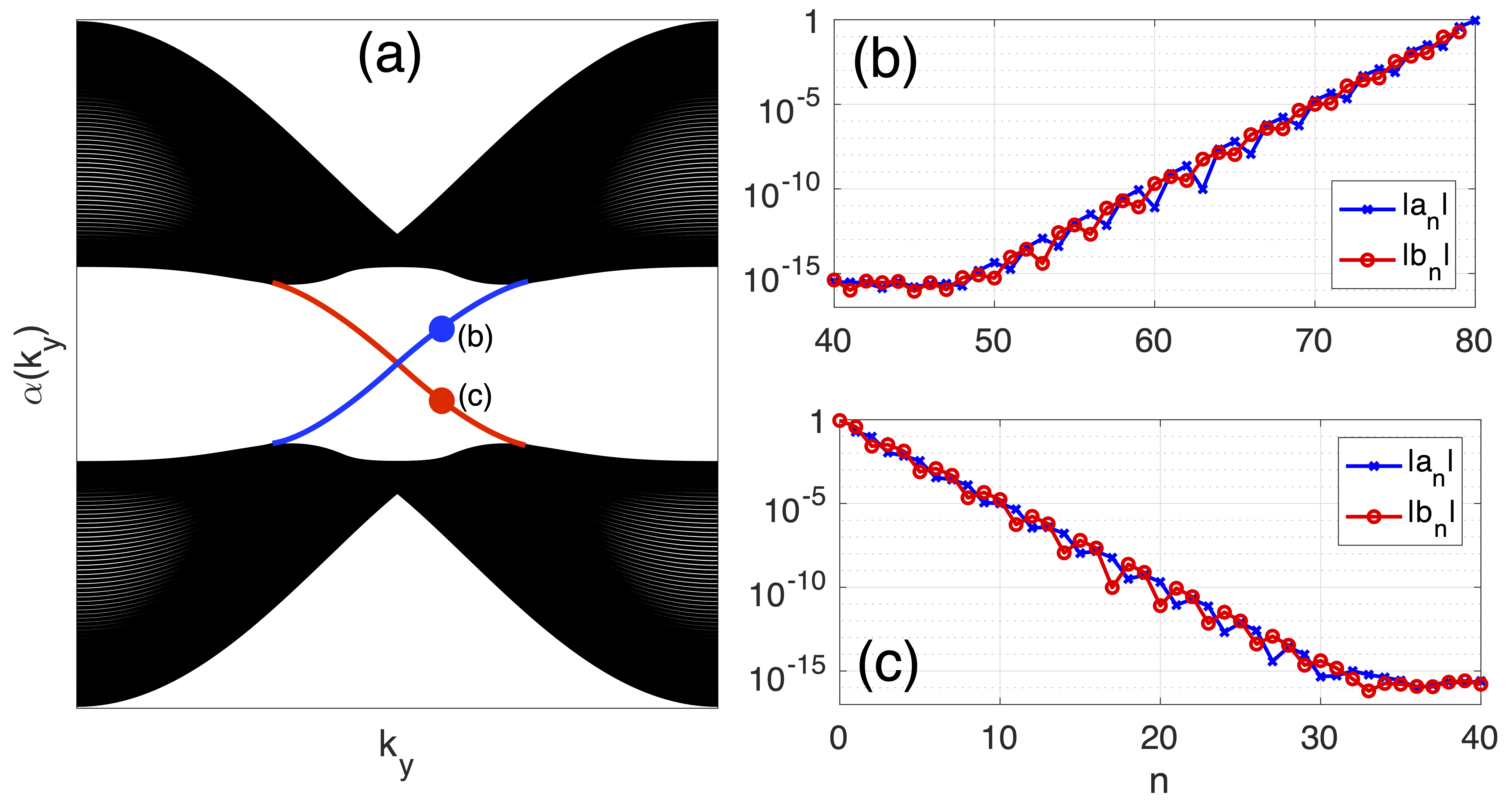}
 \caption{(a) Linear edge dispersion bands given in Eq.~(\ref{edge_floq_mult}). The {red (blue)} 
 gapless bands correspond to edge modes localized along the left (right) domain wall. {Black regions correspond to bulk modes.} Two typical edge modes are shown in panels (b) and (c) on a semi-log plot to highlight their exponential decay.}
  \label{band_diagrams_floq}
\end{figure}

The linear ($\sigma = 0$) edge Floquet modes can be  computed in a manner analogous to that of the bulk problem above. Again, solutions are assumed to satisfy the quasi-periodic boundary condition
\begin{equation}
\label{floquet_mode_form_edge}
\begin{pmatrix}
a(k_y,z+T)  \\ b(k_y,z+T) 
\end{pmatrix}_{n}= \rho(k_y)
\begin{pmatrix}
a(k_y,z) \\ b(k_y,z)
\end{pmatrix}_n , ~~~~~ \rho(k_y) = e^{- i \alpha(k_y) T}.
\end{equation}
For the boundary value problem defined in Eqs.~(\ref{HC_TB_edge1})-(\ref{Dirichlet_BCs}), the corresponding monodromy matrix can be numerically computed at $z = T$ using identity initial conditions. {As before,} the Floquet exponents  are computed from the Floquet multipliers using 
\begin{equation}
\label{edge_floq_mult}
\alpha(k_y) = \frac{i \log[\rho(k_y)] }{T}   ,
\end{equation} 
and  keeping only the fundamental branch.

For a typical driving {function ${\bf A}(z)$} {the}  Floquet band diagram {(corresponding to zero boundary conditions)}  is shown in Fig.~\ref{band_diagrams_floq}(a) using $N = 80$ sites. The  black region corresponds to bulk modes whose corresponding eigenmodes do not decay in $n$. Spanning the bandgap is a chiral edge state, indicated by a family of Floquet quasienergy values (blue and red curves) whose corresponding eigenmodes are localized along the left and right domain walls. In panels Fig.~\ref{band_diagrams_floq}(b) and Fig.~\ref{band_diagrams_floq}(c) the associated edge Floquet modes in Eq.~(\ref{floquet_mode_form_edge}) are shown and their exponential decay is highlighted.

To gain an intuitive understanding of the chirality of this system, consider the group velocity corresponding to {a} 
gapless mode. 
Along both curves the slope and therefore group velocity is sign-definite. Modes with negative (positive) slope correspond to negative (positive) group velocity {localized} along the left (right) boundary. The resulting {(positive)} chiral mode is the combination of these two edge modes; it propagates counterclockwise, as viewed from the waveguide input, along the domain boundary. This is the topological case with nontrivial Chern number. If the Chern number is zero, usually there does not exist chiral edge modes. {But there are counterexamples; e.g. phase} offset sublattice driving patterns \cite{Rudner2013}.

Topologically protected modes are identified through the bulk-edge correspondence. In the topological case,  bulk modes whose band diagrams look like Fig.~\ref{band_diagrams_floq}(a) have a nontrivial Chern number. There is a known algebraic relationship between bulk Chern number and the number of topologically protected edge states. The upper bulk band has Chern number $C_2 = -1$, which {equals} 
the number of topological edge states in the gap above it (zero) minus the number of edge states in the gap below it (one). A similar algebra exists for the lower bulk band.
}

As a final observation, topologically protected  modes can also be created along the boundary of  two topologically distinct  media (e.g. different Chern numbers) fused together. These so-called {\it interface modes} behave similar to the edge modes constructed along a domain wall, i.e. Eq.~(\ref{Dirichlet_BCs}), in that they propagate unidirectionally along the interface. These types of arrangements can allow more precise steering of the electromagnetic waves.
Physically, topologically protected interface modes have been observed in various Chern insulator systems like gyrotropic lattices \cite{Hu20}, Floquet photonic lattices \cite{Shi2021}, and more generally, systems with a sharp transition between the topologically distinct bulk regions \cite{Hu20,Bal2022,Drouot2021}. 

{
\subsection{Edge Mode Dynamics}
\label{floq_edge_mode_dyn}

In this section, the dynamics of the edge modes found in the previous section are {discussed, with particular focus on} the  chiral propagation of the topologically protected modes {mentioned} above. 
Wide spatial envelopes, localized along the domain boundary, {are found to} propagate into and around 
lattice defects, rather than reflecting or disintegrating. When Kerr nonlinearity is relevant, it is possible to realize Floquet edge solitons which also 
propagate unidirectionally {\cite{abccyp2014,abjc2017,abjc2019,Leykam2016}.}

To form an analytical description of edge envelopes, consider waveguides that are rapidly rotating such that the angular frequency in {Eq.~(\ref{driving_fcn1})} is large: $ \Lambda \gg 1$. Furthermore, assume a weakly nonlinear regime where $\sigma g = 1/\Lambda$.
A multiple scales analysis 
{(see \cite{abccyp2014,abjc2017})} reveals{, to leading order,} {edge} states localized along the left boundary  of the form
\begin{equation}
\label{nonlinear_soln_asym}
a_{mn}(Z) \sim 0 \; , ~~~~ b_{mn}(z) \sim B(y_m,Z) r^n e^{i  {k_0} y_m } ,
\end{equation}
where $Z $ is a slow time variable, $y_m = \sqrt{3} m /2$ is the continuous variable $y$ sampled at points on the discrete grid,
and {$|r(k_0)| < 1$} corresponds to an exponentially decaying edge mode as $n \rightarrow \infty$. The edge mode excited corresponds to the mode {$k_y = k_0$} of the edge band diagram. The slowly-varying envelope $B(y,Z)$ satisfies the generalized NLS equation
\begin{align}
\nonumber
i \frac{\partial B}{\partial Z} & +\alpha_*B+ i\alpha_*' \frac{\partial B}{\partial y} + \frac{\alpha_*''}{2} \frac{\partial^2 B}{\partial y^2} -  i \frac{\alpha_*'''}{6} \frac{\partial^3 B}{\partial y^3} + \dots \\ 
\label{gen_NLS_eqn}
& + \alpha_{\rm nl} |B|^2B + \dots = 0 \; ,
\end{align}
such that {$\alpha_{\rm nl} =   || b_n(k_0) ||_4^4/ \left( \Lambda || b_n(k_0) ||_2^2  \right) \ge 0 $ and $\alpha_*^{(p)} = \frac{d^p \alpha}{ dk_y^p} \big|_{k_y =k_0}$} for the {red} 
curve in Fig.~\ref{band_diagrams_floq}(a). A similar calculation  on the right boundary shows that $a_{mn}$ is the nontrivial contribution while $b_{mn}$ is nearly zero. In the linear regime, the governing equation only contains linear contributions, i.e. $\alpha_{\rm nl} = 0$.

\begin{figure} [h]
\centering
\includegraphics[scale=.4]{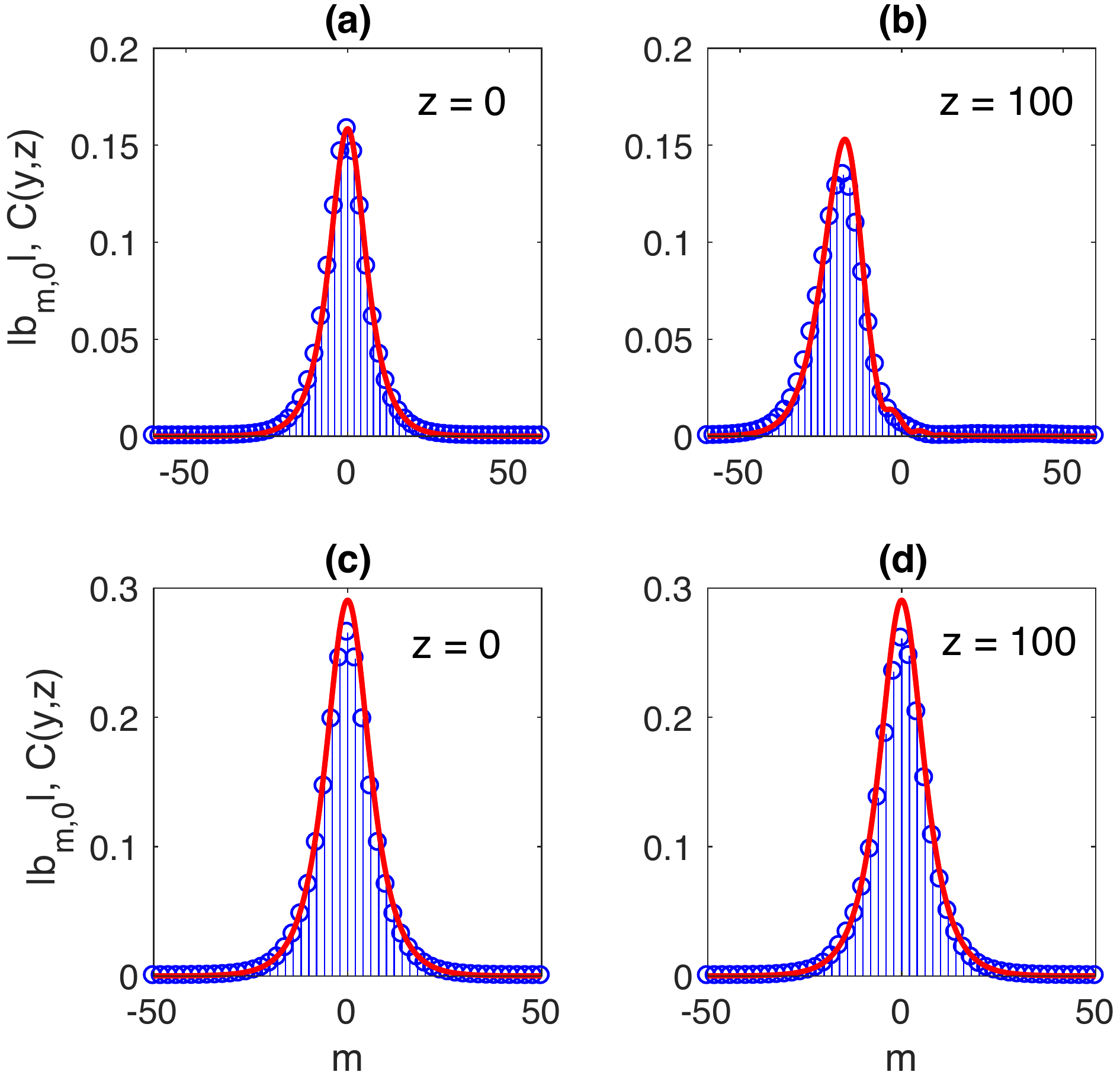}
\caption{Profile comparison between the discrete solution (blue {circles}), $b_{m,0}(z)$, and envelope (red curve), $C(y,z)$. The top row shows an envelope in a parameter regime described by the third-order NLS equation (\ref{higher_order_NLS}); the bottom row is a stationary bright soliton governed by (\ref{NLS}). Reprinted figure with permission from \cite{abjc2017}, copyright (2017) by the American Physical Society.}
\label{nonlinear_profiles}
\end{figure}

Through the careful selection of physical parameters, it is possible to engineer so-called Floquet edge solitons.
A slowly-varying envelope $B(y,Z)$  means that the higher-order dispersion terms (beyond third-order) in Eq.~(\ref{gen_NLS_eqn}) are {typically negligible}. 
Moreover, at moderate power levels the higher-order nonlinearity terms (beyond cubic Kerr term) can be neglected too. By judiciously picking the wavenumber, {$k_0$}, certain linear terms can {be} effectively eliminated. 
For example, near a critical point of an edge band (see Fig.~\ref{band_diagrams_floq}(a)), $\alpha_*'' \not= 0$ and $\alpha_*''' \approx 0$. As a result, the governing equation of the envelope  is the traveling NLS equation 
\begin{equation}
\label{NLS}
i \frac{\partial B}{\partial Z}  + \alpha_*B + i \alpha_*' \frac{\partial B}{\partial y}  + \frac{\alpha_*''}{2} \frac{\partial^2 B}{\partial y^2} + \alpha_{\rm nl} |B|^2 B = 0  ,
\end{equation}
for which  $\alpha_*'' > 0$  admits the  bright soliton solution
\begin{equation}
\label{soliton_define}
B(y,Z) = \nu \sqrt{\frac{\alpha_*''}{\alpha_{\rm nl}}} {\rm sech}\left[ \nu \left( y - \alpha_*' Z \right) \right] e^{i \left( \frac{\alpha_*'' \nu^2}{2} +  \alpha_* \right) Z}  ,
\end{equation}
with $\nu \in \mathbb{R}$. 
Experimentally,  Floquet solitons have been observed in the bulk \cite{Mukherjee2020} and along the edge \cite{Mukherjee2021}.
In the case of $\alpha_*'' < 0$, this equation  admits dark solitons of the form 
\begin{align}
& B(y,Z) = \\ \nonumber & \nu \sqrt{-\frac{\alpha_*''}{\alpha_{\rm nl}}} \Big[ \cos \alpha \\ \nonumber & + i \sin \alpha ~{\rm tanh}\left( \nu \left[ y - \left(\alpha_*'   - \nu \alpha_*'' \cos \alpha \right) Z \right] \right)  \Big] e^{i(\alpha_* - \nu^2 \alpha_*'' )Z} ,
\end{align}
where $\nu$ and $\alpha$ are real parameters.

On the other hand, if one considers modal values near the inflection point of Fig.~\ref{band_diagrams_floq}(a), then $\alpha_*'' \approx 0$ while 
$\alpha_*'''  \not=  0$ and (\ref{gen_NLS_eqn}) reduces to the third-order NLS equation 
\begin{equation}
\label{higher_order_NLS}
i \frac{\partial B}{\partial Z}  +\alpha_*B + i \alpha_*' \frac{\partial B}{\partial y} - i\frac{\alpha_*'''}{6} \frac{\partial^3 B}{\partial y^3} + \alpha_{\rm nl} |B|^2B = 0 ,
\end{equation}
for which no stable solitons are known{; in this case there is considerable dispersion. Theoretically, one expects solitons away from the zero dispersion point to propagate more effectively over long distances than  modes at the zero dispersion point.}

\begin{figure}
\centering
\includegraphics[scale=.3]{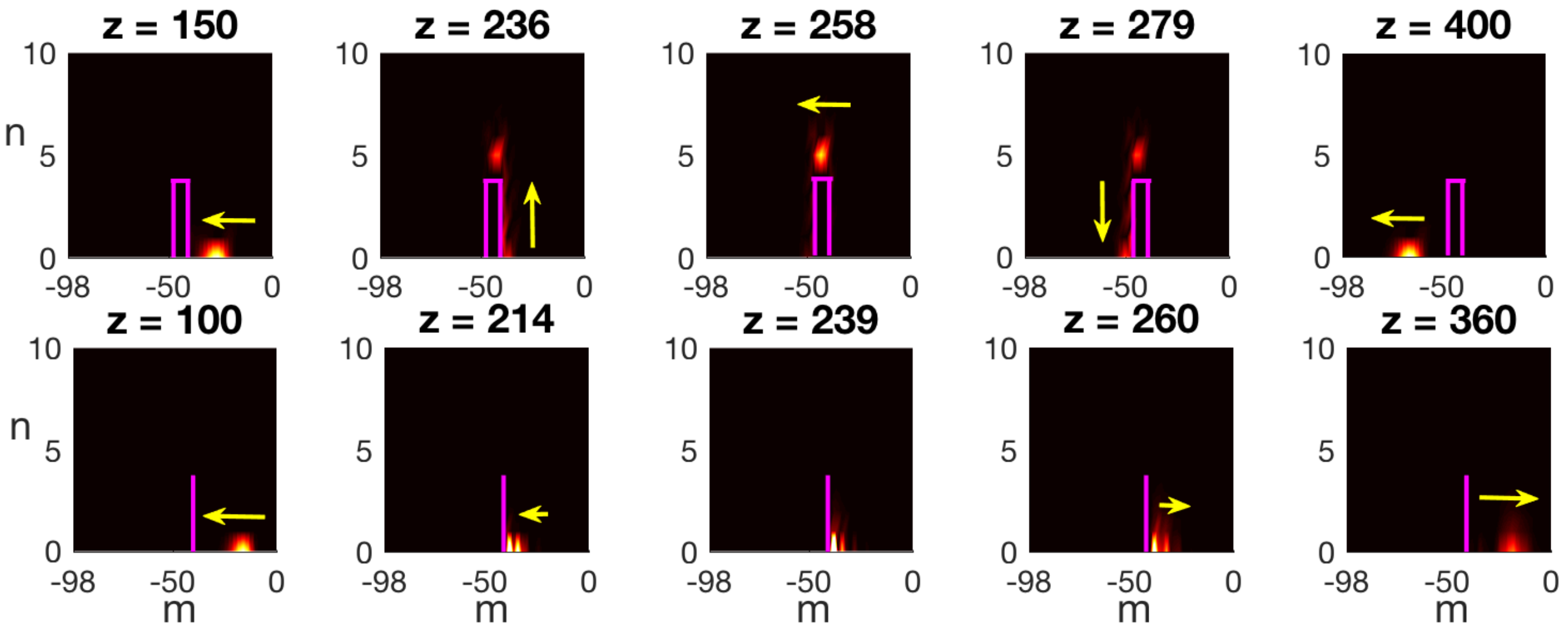}
 \caption{Intensity snapshots, $|b_{mn}(z)|^2$, for a (top row) topologically protected mode and (bottom row) non-topologically protected mode. The defect barrier is located in the region $[-46, -40] \times [0 , 4].$ Reprinted figure with permission from \cite{abjc2017}, copyright (2017) by the American Physical Society. \label{honey_floq_defect_evolve}}
\end{figure}

A comparison of the envelope approximation in (\ref{nonlinear_soln_asym}) with the full numerical solution of (\ref{HC_TB_eqn1})-(\ref{HC_TB_eqn2}) is shown in Fig.~\ref{nonlinear_profiles}. In the case of the soliton being described by the higher-order NLS equation (\ref{higher_order_NLS}), the envelope and discrete model both are seen to develop dispersive tails at large $z$. On the other hand, the stationary bright soliton profile is seen to maintain its form over long distances.

The final consideration is the effect of the topological protection on the edge envelope evolution. A defect barrier is introduced along the boundary wall. Physically, this defect corresponds to an absence of waveguides, so $a_{mn} = 0$ and $b_{mn} = 0$ is imposed in that region.
The evolution of a linear edge envelope with an associated nontrivial Chern number is displayed in the top row of Fig.~\ref{honey_floq_defect_evolve}. The envelope encounters the defect barrier, and rather than backscatter, propagates around and with virtually no loss in intensity. On the other hand, if one considers a non-topological edge envelope, the contrast is stark (see bottom row of Fig.~\ref{honey_floq_defect_evolve}). The envelope propagates into the barrier, reflects backward and loses a substantial amount of energy.

A similar evolution follows for the edge solitons described above (see \cite{abjc2017}). A potential advantage of incorporating nonlinearity is the reduction or removal of dispersion in envelopes. As was seen in Fig.~\ref{nonlinear_profiles}, dispersive degradation of modes is possible over long distances. In theory, a soliton is a perfect balance of dispersive broadening and self-focusing nonlinearity. Floquet edge solitons have the potential to combine the robust unidirectional propagation of topological edge modes with a stable soliton balance.

Finally, we remark on the case of narrow (in $y$) envelopes for this system. In the absence of driving (${\bf A}(z) = 0$), 
{generally} traveling solitary waves are not supported by the discrete NLS  equation \cite{Flach1999,Kevrekidis2009}. This effect is due to discretization of the original PDE, and is commonly known as the Peierls-Nabarro energy barrier {\cite{Jenkinson2015}}. Recently, it was shown that the topological nature of these systems {does} not allow highly localized modes to stop \cite{Ablowitz2021c}{, i.e. traveling modes exist}. However, 
{the solitary wave 
sheds} energy until it widens it's profile and is effectively continuous and described by the envelope in Eq.~(\ref{soliton_define}).
{The need to carefully prepare nonlinear edge states was also observed in \cite{Hu20}, albeit in a physically different system. In that latter work, solutions of the linear system were found to suffer from decoherence in the fully nonlinear system. Care must me taken when preparing coherent nonlinear modes.}

}

{\subsection{Other Lattice Models}

Longitudinally-driven lattice models can be {constructed} 
for other lattice types and with them {their own} unique band diagrams. Examples include staggered-square \cite{Maczewsky2017,abjc2017}, Lieb \cite{Guzman2014}, and kagome lattices \cite{Zong2016,abjc2019}. Each case 
{allows} topologically protected, unidirectional edge mode propagation. The principles used to derive a set of governing tight-binding models are similar to those used to obtain the honeycomb lattice above.

\begin{figure}
\centering
\includegraphics[scale=.5]{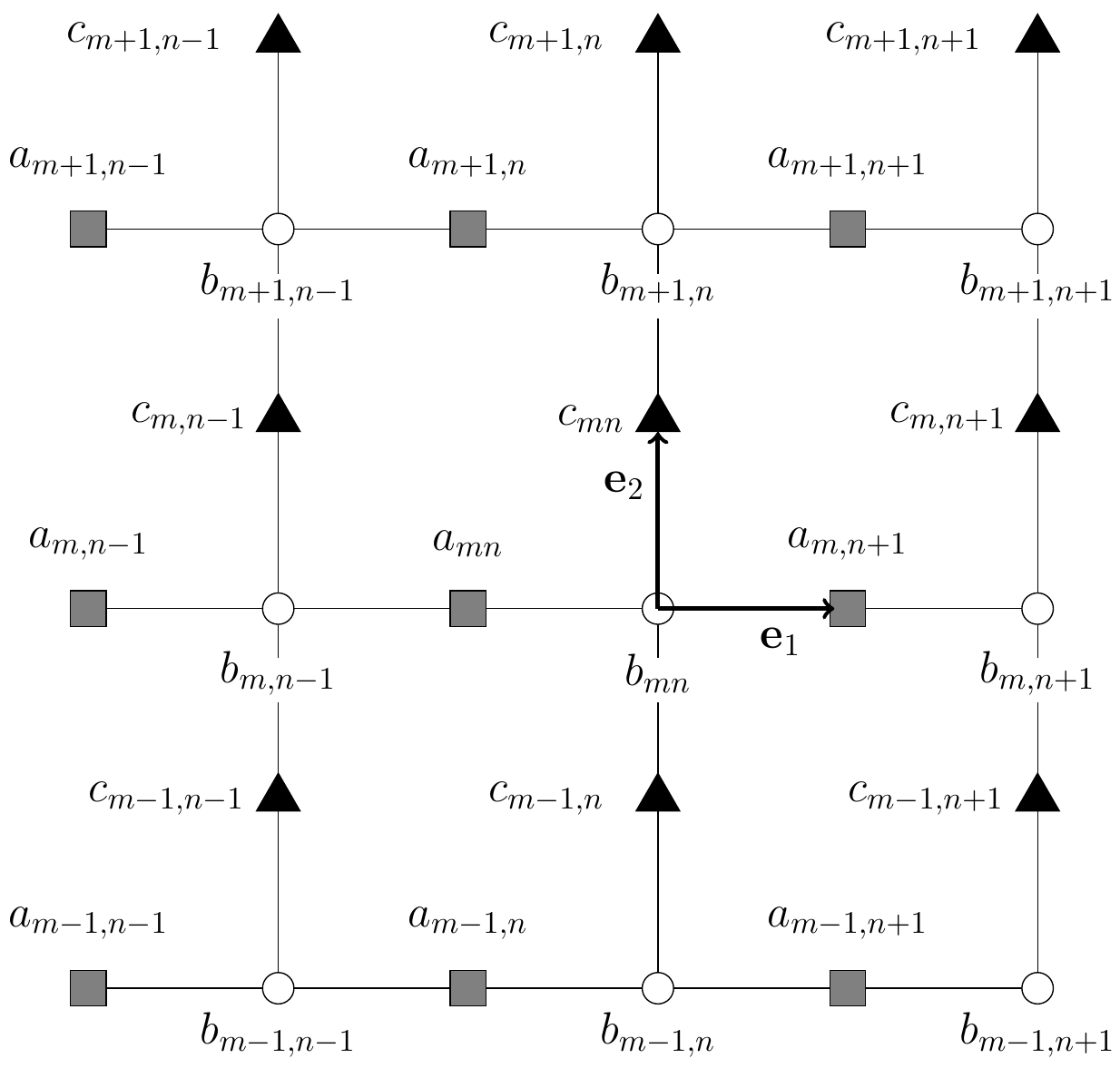}
 \caption{The Lieb lattice consists of three interpenetrating square sublattices $V_a({\bf r})$ (square site, ${\bf a}$), $V_b({\bf r})$ (circle site, ${\bf b}$) and $V_c({\bf r})$ (triangle site, ${\bf c}$). The lattice vectors are ${\bf e}_1 = (l,0)$ and ${\bf e}_2 = (0,l)$. Lines denote nearest neighbor interactions. Shown is a bearded (straight) boundary condition on the left (right) edge. Reprinted figure with permission from \cite{abjc2019}, copyright (2019) by the American Physical Society.  \label{lieb_fig}}
\end{figure}

An interesting example is the Lieb Floquet lattice (see Fig.~\ref{lieb_fig}) which contains three lattice sites per unit cell. The central (b)-site couples to the four nearest neighbor sites: two (a)-sites and two (c)-sites. The (a) and (c) sites do not directly couple to each other, that is a next-nearest neighbor interaction. The governing tight-binding model is given by the system of three equations
\begin{align}
\nonumber
 i \frac{d a_{mn}}{dz} & + C \left[ e^{i {\bf e}_1 \cdot {\bf A}(z)} b_{mn} + e^{- i {\bf e}_1 \cdot {\bf A}(z)} b_{m,n-1} \right] \\  \label{lieb_TBA_eq1} & +  \sigma g  |a_{mn}|^2  a_{mn}  = 0  ,
\end{align}
\begin{align}
\nonumber
& i \frac{d b_{mn}}{dz} 
 + C \Big[ e^{i {\bf e}_1 \cdot {\bf A}(z)} a_{m,n+1} +e^{- i {\bf e}_1 \cdot {\bf A}(z)}a_{mn} \\  \label{lieb_TBA_eq2}
  & +e^{i {\bf e}_2 \cdot {\bf A}(z)}c_{mn} +e^{- i {\bf e}_2 \cdot {\bf A}(z)} c_{m-1,n} \Big] + \sigma g  |b_{mn}|^2 b_{mn} = 0 ,
\end{align}
\begin{align}
\nonumber
  i  \frac{d c_{mn}}{dz} & + C \left[ e^{i {\bf e}_2 \cdot {\bf A}(z)} b_{m+1,n} + e^{-i {\bf e}_2 \cdot {\bf A}(z)} b_{mn} \right] \\ \label{lieb_TBA_eq3} & +  \sigma g  |c_{mn}|^2 c_{mn}   = 0  ,
\end{align}
such that $g = \int \phi_A^4({\bf r}) d{\bf r} = \int \phi_B^4({\bf r}) d{\bf r} =  \int \phi_C^4({\bf r}) d{\bf r}$.

The corresponding bulk and edge dispersion bands can be computed in manner similar to the honeycomb lattice in Sec.~\ref{HC_disp_bands_sec}. In the absence of driving, the bulk dispersion surfaces are characterized by the single Dirac point in the reciprocal unit cell where all three bands meet \cite{Ablowitz2021b}. The top and bottom bands exhibit locally conical structure near the Dirac point while the middle band is completely flat. Helically driving the waveguide opens a band gap between the top, bottom, and (flat) middle bands. The eigenmodes of the top and bottom bulk bands can acquire nontrivial Chern numbers.

A typical edge band diagram for the Lieb lattice is shown in Fig.~\ref{lieb_edge_bands_markup}. Driving the lattice opens two gaps: between the top and middle bands and between the bottom and middle. Within each gap is a single chiral edge state that spans the gap. As a result, the central band has a Chern number of zero ($1 - 1 = 0$). Moreover,  flat band edge modes  are stationary and do not  suffer from dispersion/diffractive effects \cite{Mukherjee2015,Vicencio2015}. Similar to the honeycomb lattice, the gapless edge modes propagate unidirectionally around lattice defects, scatter-free (see \cite{abjc2019} for details). Edge solitons have been predicted for the Lieb lattice in \cite{Ivanov2020}, meanwhile dipole solitons have been theorized in the kagome lattice \cite{Ivanov2021}.

\begin{figure}
\centering
\includegraphics[scale=0.3]{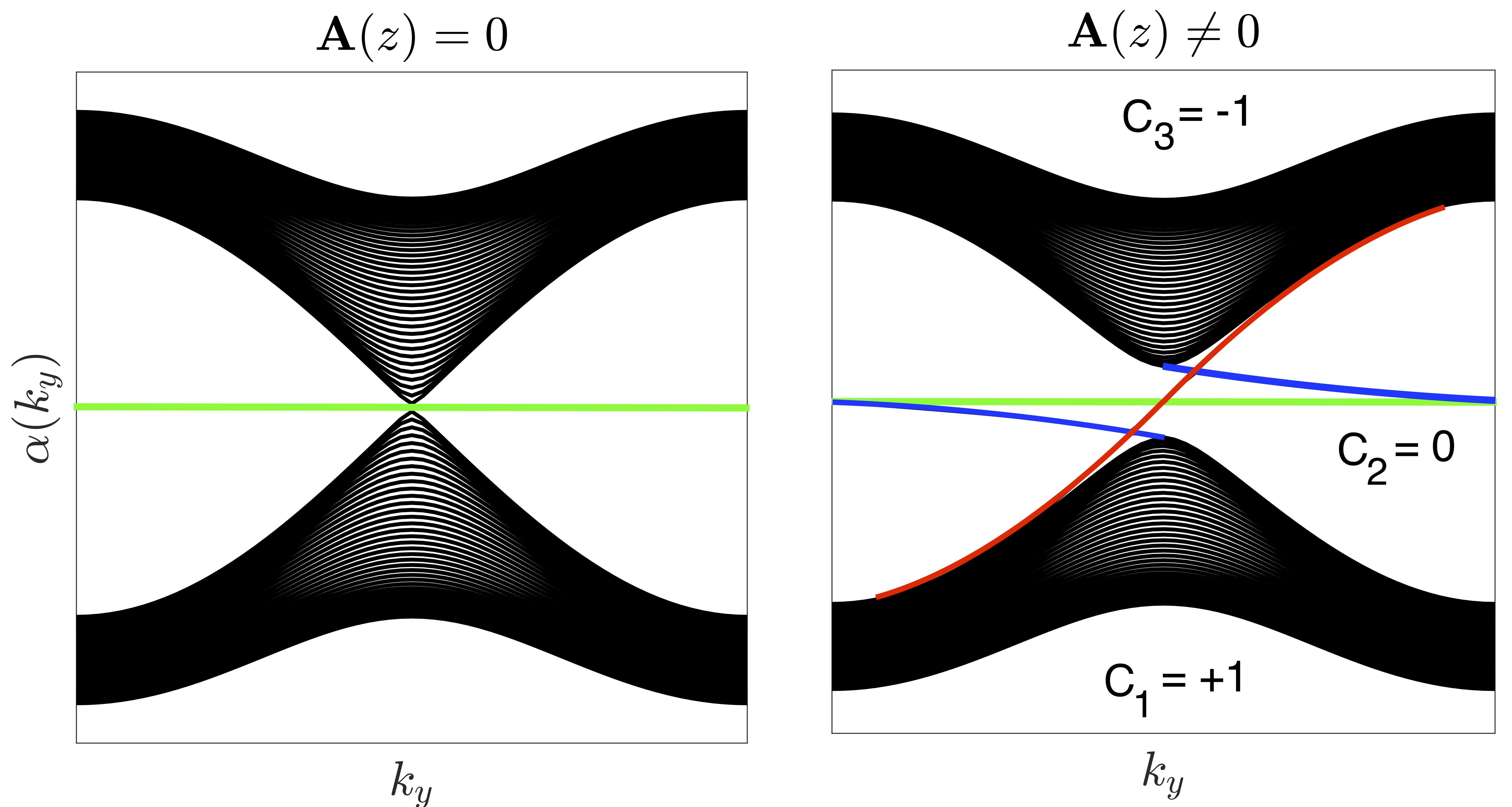}
 \caption{Lieb lattice edge band diagrams for bearded (straight) boundary conditions on the left (right) boundary.  Red curves indicate edge modes on the right edge, blue curves denote left edge modes, and green curves designate flat band modes on both edges. When driven, the Chern numbers for the corresponding bulk bands are shown. \label{lieb_edge_bands_markup}}
\end{figure}

}

{\section{Conclusions}
\label{conclude_sec}

Photonic waveguide arrays are a versatile platform for realizing  interesting physical phenomena. This article 
{discusses} some of the important experimental and theoretical work in the field. 
{The} field of optical waveguides is 
{vast;  the} focus of this article was primarily on work  done in the last decade.

The history of early experiments and their mathematical models was reviewed. Most experimentally realizable parameter regimes correspond to 
strong waveguide attraction {which are effectively} modeled by deep lattice potentials. As a result, the derivation of and study of various tight-binding models is a useful consideration. Here, tight-binding models in one and two spatial dimensions were examined. Emphasis was placed on the orbital expansion technique due to its ability to yield analytical descriptions of coupling coefficients. Numerous physical phenomena {are} found, including: Dirac cones, conical diffraction, gap solitons, topologically protected modes, and Floquet  {linear edge mode and nonlinear edge} solitons.

{Importantly, photonic waveguide arrays 
can be} experimentally realized. {Such waveguides are usually constructed in} the paraxial regime {and are} governed by the Schr\"odinger equation with a periodic potential. {The interesting phenomena discussed in this paper 
makes it likely that the study of photonic waveguide arrays and associated topological waves will continue to draw considerable research interest in physics, engineering and applied mathematics for many  years.}


}


\section*{Acknowledgements} This work was partially supported by AFOSR under grant No. FA9550-19-1-0084  and NSF under Grant DMS-2005343.

\clearpage

\bibliographystyle{plain}
\bibliography{biblio6_25_22.bib}\label{refs}

\end{document}